\documentclass[11pt,a4paper]{JHEP3}

\usepackage{graphicx}
\usepackage{amsmath,epsfig}
\usepackage{amsfonts}
\usepackage{xcolor}

\usepackage{epsfig}
\usepackage{subfigure}
\usepackage{subfig}

\def\eqc{\;,}

\relax
\renewcommand{\theequation}{\arabic{section}.\arabic{equation}}
\def\be{\begin{equation}}
\def\ee{\end{equation}}

\newcommand{\ha}{{1 \over 2}}

\newcommand{\bear}{\begin{eqnarray}}
\newcommand{\bea}{\begin{eqnarray}}
\newcommand{\eear}{\end{eqnarray}}
\newcommand{\eea}{\end{eqnarray}}
\def\hri#1#2{\href{http://arxiv.org/abs/#1}{[ArXiv:#1]#2}}
\def\hre#1#2{\href{http://arxiv.org/abs/#1/#2}{[ArXiv:#1/#2]}}

\newbox\pippobox

\def\II{\relax{\rm I\kern-.18em I}}

\def\cO{{\cal O}}

\def\e{\epsilon}

\def\m{\mu}
\def\n{\nu}
\def\mn{{\mu\nu}}
\def\r{\rho}

\def\s{\sigma}
\def\pa{\partial}

\def\sp{\;\;\;,\;\;\;}

\def\p{\partial}

\def\f{\varphi}

\def\b{\beta}

\def\le{\left}
\def\ri{\right}

\def\Vg{\sqrt{-g}} 

\newcommand{\eql}[2]
{ \begin{equation} \label{#1}
 #2
\end{equation}}

\title{de Sitter versus Anti de Sitter flows and the (super)gravity landscape }

\author{Elias Kiritsis$^{1,2}$, Alexandros Tsouros$^2$\\
 ~\\
 $^1$
 \href{http://www.apc.univ-paris7.fr}{Universit\'e de Paris, CNRS, Astroparticule et Cosmologie,  F-75013 Paris, France}\\
~\\
 $^2$
 \href{http://hep.physics.uoc.gr/}
 {Crete Center for Theoretical Physics}, Institute for Theoretical and Computational Physics, Department of Physics, University of Crete
 71003 Heraklion, Greece.
}

\abstract{
Generic solutions are studied in Einstein-scalar gravity in an ansatz that can interpolate between  de Sitter and Anti-de Sitter  regimes. The classification of regular solutions of \cite{exotic} is first extended to the dS regime. This implies, among others, the existence of cosmic clocks that reverse direction without passing through a curvature singularity.
We then consider an ansatz  for solutions that interpolate between the dS and AdS regimes. The structure of such more general solutions and their singularities are studied. It is  shown that there are no regular solutions that interpolate between dS and AdS extrema for generic potentials. This is unlike the Centaur solutions that were shown to exist in two bulk dimensions.
We also comment on the potential interplay with recent dS conjectures and the dS BF bounds.
}
\preprint{CCTP-2018-17\\  ~\hfill ITCP-IPP-2018/13}

\begin{document}

\maketitle 

\section{Introduction}
\label{sec:gen}

Quantum field theory and string theory have enjoyed, for some time, a certain hierarchical structure, with string theory encompassing quantum field theory (QFT) as a low energy theory. Moreover, string theory provided the very sought-after inclusion of quantum gravity to low-energy QFT interactions, and to a substantial extent,  ``unified" gauge theories and gravity.

The holographic correspondence, \cite{Malda,GKP,Witten98}, has restored a certain kind of democracy in theory space and provided a contrasting view of string theory, and the associated quantum gravity. It now looks plausible that string theory as we know it, and QFT as we know it, are collections of local patches in a construct that may be as big as QFT or bigger.
It is usually the case that the string theory behavior or (weakly coupled) QFT behavior are mutually exclusive and this amounts to a large extent for the presence of patches in the overall construct.
It is also clear that there may be many string theories missing from this, probably because they do not have a weakly-coupled, continuum limit\footnote{Some ideas in this direction can be found in \cite{nstring} but they are by no means unique.}.

A major restructuring of the conceptual view of string theory has emerged with the holographic correspondence: the notion of the landscape of string theory vacua has undergone a major reinterpretation. The string theory landscape has looked rather formidable and unwieldy, \cite{schelle}, but holography forced us to contrast it to the QFT landscape, which,  with the advent of non-perturbative techniques,  turned out to be as daunting as the stringy one.

The correspondence/map between the two landscapes may lie in the heart of understanding the emergence of gravity and space-time, \cite{Seiberg,Malda2,smgrav},  from purely QFT-based concepts.
It also seems relevant to the several deep and fundamental problems that plague the mariage of gravity and quantum mechanics, like the black-hole information paradox, \cite{Harlow,polpar,Marolf}, the cosmological constant problem, \cite{Polchinski} and to some extent the hierarchy problem.

In the case of asymptotically AdS space-times, holography provides a rather credible picture of the (structure of the) space of theories and their connections via string theory/supergravity solutions that correspond to QFT RG flows, \cite{GPPZ}-\cite{curved}. Moreover, there is a concrete framework to understand the mapping from QFT, \cite{Go}-\cite{Dou}.

The case of asymptotically flat space-times is the standard arena of critical (super)-string theory\footnote{This is not however the case away from the critical theory where, depending on sub-criticality or super-criticality, a negative or positive curvature is preferred. In such cases however, the zero mode/gravity approximation is mostly not a good guide.}
However, the holography of such space-times is not very well understood, \cite{Susskind}-\cite{petro}.

An even more exotic case, from the holographic point of view  is that of asymptotically de Sitter (dS) space-times.
The reason is because of a clear dichotomy that emerges: on one hand we have clear experimental/observational evidence that near dS space has been  relevant to the early universe as well as the current and near future one.
On the other, dS space and the associated quantum physics, presents several puzzles, one of which is of course the issue of the size of the cosmological constant and the fact that it seems to be dynamically unstable to quantum corrections, \cite{tw,br}.

To add insult to injury, weakly coupled, weakly curved string theory seems to be at odds with dS solutions, \cite{DaVa}.
Many attempts rely on general structure that is difficult to control quantitatively, \cite{KKLT}, and with further  difficulties associated to
(holographically) controlling anti-branes.

Other dS attempts  are direct,  but work in a regime where the dS solutions require ${\cal O}(1)$ stringy or loop corrections, \cite{DaSh,Wrase}, in consonance with a general (not without loopholes) argument of Dine-Seiberg \cite{DS}.
This difficulty, following recent trends, has been elevated to a time-evolving swampland conjecture, \cite{ov}-\cite{and2}.
It currently states that there are no dS extrema without directions in field space that are ``unstable".

The effort of finding controllable dS solutions in string theory is fully active, \cite{tom}.
New possible realizations have been proposed based on the brane-world idea\footnote{A simplified version of this was proposed a bit earlier in \cite{dSe}. In this realization, a bulk RG flow is approximated as an abrupt domain wall between the UV and IR CFT extrema. This approximate realization falls into the fine-tuned category of \cite{dS}.}, \cite{dS}. The idea is motivated by the self-tuning mechanism of the cosmological constant\footnote{The cosmology of moving branes has had a longer history in string theory and was related to holography, \cite{kraus}-\cite{holo}.} \cite{selftuning}. It says  that the bulk space can be negatively curved, but the geometry on a collection of branes (that must eventually carry, among other things,  the Standard Model) can be de Sitter.
It was shown in \cite{dS} that if the bulk string theory is holographically related to a boundary QFT, then this is not {\em generically} possible if the
QFT is defined on flat Minkowski space, but it is generically possible if the boundary  QFT is defined on de Sitter space. Moreover, it was shown that interesting hierarchies can appear between the de Sitter scale of the boundary QFT and of the de Sitter scale on the brane universe.
However, the implementation of this idea in a controllable bulk string theory framework needs to be worked out.

There is a putative holographic side of the de Sitter story. Since the advent of AdS/CFT it has been suggested that there maybe a de Sitter analogue, called dS/(pseudo)CFT correspondence, \cite{SdS}-\cite{Strom}.
There are several ideas on how this correspondence might work. At the scale invariant points, associated to de Sitter, the bulk gravitational theory is expected to be dual to (pseudo)CFT. The precise rules for this (pseudo)CFT were spelled out in \cite{MS}.

The cosmological evolution starting and ending in dS was analyzed from a holographic viewpoint in \cite{afim} where it was associated to (pseudo)RG flows, and where Wilsonian ideas were used to classify inflationary theories\footnote{For a recent discussion from a different starting point, see \cite{We}.}. A sharp contrast can be drawn on the conventional view of cosmological solutions and their fine tuning problems on one hand and the holographically dual picture on the other. In particular it was argued that holographic ideas may be crucial in resolving several fine tuning problems in cosmology.

In \cite{exotic}, a program was started aiming at a systematic study of holographic solutions in the AdS part of the landscape of gravitational theories, with the aim of producing a precise map with similar RG flows on the QFT side.
The intricate questions associated with the dS regime imply that an extension of this study to the dS regime is important, and the purpose of this paper is to start such a systematic study.
Moreover, the study of semiclassical, regular solutions that interpolate between the two regimes is also important, in order to understand the transition region.

Studies of dS solutions in the AdS context have been performed in \cite{hub,Lowe} that have used one of the two ansatze we will be using in this paper, although they have discussed the solutions in the thin wall approximation.
One of the motivations were to go beyond the results of \cite{Farhi} that argued that all such solutions are singular in the past.

A recent paper, \cite{AnHo}  addressed the latter question in two space-time dimensions, and found interpolating solutions that contain an AdS$_2$ boundary and an dS$_2$ part in the ``IR" geometry.

In this paper we shall study possible interpolating solutions, between the AdS and the dS regimes,  in higher than two dimensions, by choosing convenient metric ansatze.
Although, we shall not study all possible solutions, we shall manage to study a prominent ansatz and be able to present general results.

\subsection{Results and outlook}

In this paper we shall assume that the scalar potential does not have Minkowski extrema, but only dS or AdS extrema. We also assume that it is everywhere regular except at the boundaries of field space in accordance with the what is known to happen in string theory, \cite{book}.

Our results can be summarized as follows

\begin{itemize}

\item We extend our local and global analysis of AdS flows (performed in \cite{exotic}) to flows that are entirely in the de Sitter regime\footnote{In this paper we call ``de Sitter regime" the part of scalar field space where $V(\phi)>0$. We call ``AdS regime" the part where $V(\phi)<0$. We also, do not consider extrema in this paper where the potential vanishes.} Such metrics are sliced with flat slices and the coordinate system is in the Poincar\'e patch. We develop the superpotential formalism for such flows and match to previous preliminary results in \cite{afim}.

    The superpotential formulation is convenient in order to separate initial conditions that are innocuous (for example an initial size of space or the boundary value of the scalar field) from initial conditions that are crucial for regularity. The first kind appear in the first order flow equations whereas the second appear in the non-linear equation that relate the (fake) superpotential to the bulk scalar potential.

\item    We find that near de Sitter minima of the potential, we have  large/diverging  scale factors that correspond to a large universe. In this regime, the minus (-) branch\footnote{The plus and minus branches of solutions near potential extrema have been discussed in detail in \cite{exotic} in the AdS regime. They are also discussed here in section \ref{cripo} in the dS regime.} of solutions for the superpotential  has a free parameter (integration constant) that indicates a continuous family of locally regular solutions\footnote{This parameter is interpreted as the dual scalar vev in the AdS case.}. The plus (+) branch gives a unique solution as usual, corresponding to a moduli space of vevs.

On the other hand, maxima of the potential in the dS regime are associated with a universe that shrinks to zero size in Poincar\'e coordinates and therefore signal the ``beginning" of the universe. In this regime, both the + and - branches of solutions for the superpotential are unique, after imposing regularity, and therefore have no adjustable parameters.

\item We also find that in the dS regime there are ``bouncing solutions"\footnote{The notion of  a bounce here is different from the notion of bouncing universes common in cosmology, \cite{bounce1}.} as those found in the AdS regime in \cite{exotic}.
    Such bounces are places in the cosmic evolution where the scalar field changes direction without any singularity. They can appear for relatively steep potentials. As discussed in section \ref{clocks}, such solutions correspond to cosmic clocks that change direction in time, in a fully regular fashion.

  \item Like in the AdS regime, regular flows in dS start and end at neighboring extrema of the bulk scalar potential.
      A way to understand such regular flows as solutions of the associated differential equations is that the unique regular solution starting from a dS maximum, can be evolved to end into one of the many solutions ending near a minimum. The existence of the single free parameter near dS minima
is crucial for the existence of generic globally regular solutions.

\item As is known from the AdS regime, there can be isolated regular solutions for special (tuned) bulk potentials. Such solutions are such that in the dS regime, they start at a maximum of a potential with a + solution which then  ends up at another maximum of the potential. An example was discussed in \cite{Libanov14}. The analogous flows in the AdS regime are flows from minimum to minimum.
    In that case the region near the first maximum describes a large expanding univers, while the second a (regular) big-bang.
    In appendix \ref{exam} we also present a similar example partly in the AdS regime, going from a minimum to a minimum of the potential,   without any AdS boundary.

    \item There is an analogous BF bound in the dS regime as in the AdS regime. In the AdS regime the BF-violating extrema are unstable under small perturbations, \cite{BF82}.
        This is associated with the fact that there are operators associated to the dual CFT that have complex dimensions. The associated CFTs have been studied recently in \cite{Rychkov}.
There are flows that end up at BF-violating extrema, both in the AdS regime and the dS regime. For such flows, the scalar reverses direction an infinite number of times before arriving at such extrema.

A particular property in the AdS regime is that flows starting at BF-violating maxima are always subleading compared to ones where vevs are turned on, avoiding such maxima. This is a kind of ``censorship" of    BF violating maxima, in the holographic landscape.

In the AdS regime, the BF bound forbids scalar directions with sufficiently negative masses. This implies that the bound is relevant for maxima of the potential.
The situation in dS is the inverse. Sufficiently massive scalars violate the BF bound. This implies that here it applies to minima of the potential.
It should be stressed that in the dS regime, the definition of the BF bound is by analogy with the AdS case: it delineates extrema where $\Delta_{\pm}$ are complex, or where a scalar changes direction an infinite number of times before reaching an extremum.

However, unlike AdS, it is not clear whether violations of the BF bound in dS implies instabilities or the necessity to turn on the relevant scalar solutions. An answer to these issues is a very interesting problem which may have consequences for theoretical cosmology.

\item We investigate then, flows that interpolate between the dS and the AdS regime. We use an ansatz with flat slices, given in  (\ref{c38}), that is capable of interpolating between AdS and dS solutions and also starts and ends at extrema of the potential. It includes a blackness function that must change sign along the flow. Therefore, the relevant solutions must have a horizon along the flow as was found in earlier works, \cite{hub,Lowe}.

    We develop the first order (superpotential) formalism for such solutions, and investigate their generic regularity.
We also investigate the properties of flows between dS and AdS regimes.

We deduce, by counting free integration constants and other properties, that potentially regular solutions that exist for {\it generic potentials} must start at a dS minimum and end at an AdS maximum (in order to interpolate between dS and AdS).
We then show that the null energy condition in the bulk implies that no such solutions exist.

To the contrary, the null-energy condition implies that only solutions that  interpolate between a maximum in dS (where the scale factor vanishes) and a minimum in AdS (where again the scale factor vanishes) could exist. However, according to our general analysis, in the dS regime, a vanishing scale factor implies that the solution must be a minus solution.

We can show however, that minus solutions with vanishing scale factor and a non-constant blackness function  are singular at the extremum.
Therefore, regular solutions,  even for tuned potentials that interpolate between AdS and dS in this ansatz
{\it do not exist}.

\end{itemize}

An intriguing issue is whether recent conjecture dS swampland constraints \cite{ov}-\cite{and2} can be intertwined with the results of this work. In particular, whether they allow even tuned potentials that have interpolating solutions between dS and AdS vacua.

A final question concerns the impact of this analysis for real cosmology.
We hope to address these issues in future work.

\section{Holographic RG Flows in Einstein-dilaton gravity} \label{section:2}

As a gravitational theory we study  Einstein-scalar theory, whose action consists of an Einstein-Hilbert term and a minimally coupled scalar field with a potential. This is the most general two-derivative theory of a metric and a scalar field and it is a proxy for the more general multiscalar theory. As was argued in detail in \cite{exotic}, the general solutions of the multiscalar theory can be always mapped to solutions of the single scalar theory.
The general analysis was done in \cite{exotic}. We will repeat some of it here, so that we setup notation.

The connection between the bulk gravitational setup and the boundary QFT is made by mapping  the bulk metric with the stress energy tensor, $T_{\m\n}$,  of the boundary theory and the scalar field with the single-trace scalar operator, $O(x)$.

 In this section we review how the second-order equations that result from the above action can be transformed into a system of first-order equations in terms of a \emph{superpotential} and how this formalism can be used to classify solutions in the AdS regime\footnote{We shall define the AdS regime are the region in field space where the scalar potential is negative. When it is positive we shall label it as the dS regime.}. This function is used to make contact with the holographic renormalization group flow of the boundary QFT  \cite{dVV}-\cite{rg2}.

\subsection{The setup}
\label{ssec:setup}

We define the Einstein-scalar theory in $d+1$ dimensions, with signature $(-, +\ldots +)$. The action we consider throughout is of the form:
\eql{i1}{
			S\le[g,\f\ri]= \int d^{d+1}x \Vg \le(
			 R
			-\ha\p_a\phi \p^a\phi-V(\phi)  \ri)+S_{GHY}.
			}
where $S_{GHY}$ is the Gibbons-Hawking-York term. To arrive to this from the most general two derivative action a Weyl rescaling of the metric as well as a redefinition of the scalar are necessary.

 The dual QFT is defined in $d$-dimensional Minkowski space-time, which is the boundary of the $d+1$ space-time on which the Einstein-scalar theory is defined. The ground state saddle point of the QFT is related via holography to  Poincar\'e-invariant solutions of the Einstein-scalar theory.

Such solutions can be always  put in the so-called domain-wall coordinate system:
\eql{i2}{
		 \phi=\phi(u),\qquad ds^2=du^2+e^{2A(u)}\eta_{\mn}dx^\m dx^\n
		}
where $u$ is  the holographic coordinate. The metric has manifest ISO(d) invariance and the only remaining dynamical variable is the scale factor $e^A$ of the Minkowski slices.

We denote by dot derivatives with respect to the holographic coordinate, whereas derivatives with respect to the scalar field are indicated with a prime.

Using \eqref{i2} and by varying the action \eqref{i1}, we arrive at the following equations of motion:
\begin{subequations}\label{i3}
\begin{align}
	&2(d-1)\ddot{A}(u)+\dot{\phi}^2(u)=0,\label{i3_1}\\
	&d(d-1)\dot{A}(u)^2-\ha\dot{\phi}^2(u)+V(\phi)=0.\label{i3_2}
\end{align}
\end{subequations}
We deduce  from \eqref{i3_1}  that $\dot A(u)$  cannot increase. In the holographic RG, this is related to the holographic c-theorem \cite{GPPZ,Freedman}.

From (\ref{i3}) we may also deduce the Klein-Gordon equation for the scalar, which is given by
\eql{i4}{\ddot{\phi}+d\dot A\dot{\phi}-V'(\phi)=0,}

The  (dual) QFT energy scale is taken to be
\eql{i5}{
\m\equiv\m_0 e^{A(u)},
}
with some arbitrary mass scale $\m_0$. This choice is justified in that it gives the correct trace identities for the dual stress tensor, if the running coupling is identified with $\phi(\mu)$ \cite{rg2}. Further, it is monotonic along the RG Flow in view of (\ref{i3_1}).

In the context of the gauge/gravity duality, the fixed points of the holographic renormalization group flow are connected with asymptotically AdS geometries. Therefore, in this section,  we shall consider only the cases where the extrema of the potential in (\ref{i1}) occur at negative values of the potential.

This restricts our solutions to not include time-dependent or asymptotically flat space-times. We further assume that $V(\phi)$ is analytic for all finite $\phi$ as this is standard property of string theory effective potentials, \cite{book,Trigiante}.

We also define the radial coordinate in a way such that $u$ increases along the flow, which is equivalent to take $A(u)$ to be monotonically \emph{decreasing}. This is permitted by (\ref{i3_1}),  which states that $\ddot{A} \leq 0$.
Therefore,  our solutions in the AdS regime will have an AdS boundary at $u\to-\infty$,  while they will have a Poincar\'e horizon at $u\to +\infty$.


\subsection{The superpotential}
\label{ss:i10}

 In this section, we convert the second-order Einstein equation (\ref{i3_2}) in two first order equations. Through this formalism, we connect the notion of the RG flow in $d$ space-time dimensions, and the Einstein equations in $d+1$ dimensions with solutions that respect $d$-dimensional Poincar\'e invariance ( see \cite{ceresole},\cite{dVV}-\cite{rg2}).

In the context of the gauge/gravity duality, the RG flow of the scalar coupling is governed by the Einstein equations, (\ref{i3_1}-\ref{i3_2}).
 The unusual fact is that these equations are second order, as opposed to the RG flow equation of QFT, which is first order. Following \cite{Freedman}, we shall rewrite the Einstein equations as a system of first order equations. This will enable us to make contact with the RG flow of the boundary theory. We define the function $W(\phi)$ such that:
\eql{i8}{W(\phi(u))=-2(d-1)\dot A(u),}
with the proportionality constant chosen for future convenience.
Notice that equation \eqref{i3_1} is automatically satisfied if $\phi$ obeys:
\eql{i9}{W'(\phi)=\dot \phi (u).}
Substituting \eqref{i8} and \eqref{i9} into \eqref{i3_2} we obtain an equation for $W(\phi)$:\footnote{In supergravity theories, $W(\phi)$ is known as the \emph{superpotential}, and the first order equations define the BPS flows. Even though we do not have supersymmetry in our setup, we shall still use this name. We shall further refer to\eqref{i10} as the superpotential equation. Sometimes $W(\phi)$ is also called the ``fake superpotential".}
\eql{i10}{V(\phi)=\ha W'^2(\phi)-\frac{d}{4(d-1)}W^2(\phi).}

We may now, using \eqref{i8} and \eqref{i9},  write the $\b$-function of the boundary QFT in terms of the coupling $\phi$:
\eql{i11}{{d\phi\over d\log \m}=\b(\phi)=-2(d-1){W'(\phi)\over W(\phi)}.}

If a solution $W(\phi)$ of equation \eqref{i10} is known,  one can use it to solve for $\phi(u)$ and $A(u)$ using \eqref{i8} and \eqref{i9}. {\it Hence from now on, for a given $V(\phi)$, we solve the superpotential equation \eqref{i10} and find $W(\phi)$. From there,  we can find the $\beta$-function of the boundary theory, together with $\phi(u)$ and $A(u)$.}


\subsection{General properties of the superpotential}
\label{ss:gen_prop}

In this section we list some important properties of solutions $W(\phi)$ of equation \eqref{i10} for potentials $V(\phi)$ which are taken to be analytic\footnote{
By analytic we mean that around any finite $\phi$ the potential is given by a convergent power series.} excepts at the boundaries of field space, $\phi\to \pm\infty$. Also,  we assume that we are in the AdS regime ($V<0$).

Next, we note some general properties of the superpotential $W(\phi)$.

\begin{enumerate}

\item {\it W(u)$\equiv$W$(\phi(u))$ is monotonically increasing along the flow}
\label{it:i12}

From equation \eqref{i9} we can show that,
\eql{i12}{
	{dW(u)\over du}={d\phi(u)\over du}{d W(\phi)\over d\phi}=W'^2\geqslant0
}


\item {\it The $(W,\phi)$ plane has a forbidden region where no solution of \eqref{i10} exists. This is the region bounded by the curve:}
\label{it:bound}
\be
B(\phi) \equiv \sqrt{-\frac{4(d-1)}{d}V(\phi)}\label{i13}.
\ee

This property  is easily seen from \eqref{i10}, from which we obtain:
\eql{i14}{|W(\phi)|= \sqrt{\frac{4(d-1)}{d}\le(\ha W'^2-V(\phi)\ri)}\geqslant\sqrt{-\frac{4(d-1)}{d}V(\phi)}.}

This property in turn implies that $W(\phi)$ \emph{cannot change sign} in the AdS regime.

\item {\it We can assume $W(\phi)>0$ without loss of generality}.

This follows by noting that \eqref{i10} is invariant under the transformation

\be
W\to -W.
\label{w5}\ee

Also, the following transformation:

\eql{i14_1}{
(u,W)\to-(u,W)
}
is a symmetry of the full set of equations (\ref{i8} - \ref{i10}).

\item {\it For a generic point in the allowed region of the $(\phi,W)$ plane with $W(\phi)> B(\phi)$,
    there exist two and only two solutions, $W_{_{\uparrow}}(\phi)$ and $W_{_{\downarrow}}(\phi)$, passing through that point}. See \cite{exotic} for a justification.
\label{it:updown}

\item{\it The geometry is regular if the potential and the superpotential are finite along the flow. This is equivalent to the scalar field staying finite along the flow, \cite{Bourdier13,exotic}}
\label{it:reg}


\item {\it Extremal points of $W(\phi)$ lie on the critical line $B(\phi)$, } since there $W'=0$ by equation (\ref{i10}).

There are two kinds of critical points:
\begin{enumerate}
\item {\em fixed points}, which are points on the critical line where $B'=0$. These are the  extrema of $V(\phi)$, and correspond to the usual UV and IR  fixed point in the RG flow. The geometry is near AdS there, and the boundary QFT is a conformal field theory.
\item  {\em bounces}, which are generic points on the critical line where $V'\neq 0$.  In this case the superpotential becomes multi-branched but the geometry is regular. They correspond to points where $\phi(u)$ reaches a maximum or a minimum and the flow reverses its direction.
\end{enumerate}

In the next subsection, we shall study the critical points in detail.

\end{enumerate}

\subsection{Critical points}
\label{ssec:crit}

Points at which $W'(\phi)=0$ are called \emph{critical points}, as these are the only points where the solutions of the superpotential equation can have singularities.
These correspond to extrema of the coupling $\phi(u)$, as can be seen from  \eqref{i9}.
The behavior of $W(\phi)$ at a critical point is dictated by whether $V'(\phi)$ vanishes or not, at the critical point. Consider the derivative of equation \eqref{i10} with respect to $\phi$,
\eql{i19}{
V'(\phi)=W'(\phi)\le(W''(\phi)-\frac{d}{2(d-1)}W(\phi)\ri),
}
and let $W'(\phi_*)=0$. Since $W$ is taken to be finite for finite all finite $\phi$,  there are only two possibilities:
\begin{enumerate}
\item $V'(\phi_*)\neq0$ with  $W''(\phi_*)$  divergent. \label{i20}

{\bf or}

\item $V'(\phi_*)=0$; in this case  $W''(\phi_*)$ is finite.
\label{i21}
\end{enumerate}

The latter statement is not entirely obvious but it  is shown in \cite{exotic}.

The first case corresponds to ``bounces"\footnote{Bounces were defined in \cite{exotic}.} as we shall see.
 Bounces are points where two different branches of $W(\phi)$ (one growing, one decreasing) are glued together. This generic situation will be treated in (\ref{bounces}).

The second case corresponds to  fixed points of the holographic RG flow. In what follows, we discuss the extrema of $V(\phi)$. The inflection points are also studied in \cite{exotic}.

\subsubsection{Local maxima of the potential}
\label{sssec:max}

In general, we consider the extrema to be situated at $\phi=0$, which can always be done by a translation in scalar field space.  Near a maximum  of the potential,we may expand as:
\be
	V=-{d(d-1)\over \ell^2}+{m^2\over 2}\phi^2 + \cO(\phi^3), \qquad m^2 <0 \label{i21_1}
\ee
As is shown in \cite{papa1}  equation \eqref{i10} has two distinct classes of solutions that we shall label with $W_{\pm}$,
\begin{subequations}			\label{i22}
\begin{align}
	W_+(\phi)
			=&{1\over \ell}
			\le[
				{2(d-1)} + {\Delta_+\over2}\phi^2+\cO(\phi^3)
			\ri], 			\label{i22_a}\\
	W_-(\phi)
			=&{1\over \ell}
			\le[
				{2(d-1)} + {\Delta_-\over2}\phi^2+\cO(\phi^3)
			\ri]+{C\over \ell}|\phi|^{d/\Delta_-}\le[1 +\cO(\phi)\ri]
			+\cO(C^{2})
			\label{i22_b},\\
	\Delta_\pm =&\ha \le( d \pm \sqrt{d^2 +4 m^2\ell^2}  \ri) \quad \text{with}\quad -{d^2\over 4\ell^2}<m^2<0,
	\label{i23}
\end{align}
\end{subequations}
where $C$ is an integration constant.

The two solutions are characterized by the value of $W''(\phi)$ at the extremum: the $W_-$ branch is a continuous family of solutions parameterized  by the constant $C$; $W_+$ is instead an isolated solution of the equation. The $W_+$ solution can be obtained from the continuous family $W_-$ in the limit  $C\to +\infty$, as was shown in \cite{papa1}. Notice that the symbol $\cO(C^{2})$  in equation (\ref{i22_b}) is not to imply that $C$ itself is infinitesimal, but it is used to mean that the subsequent  terms in the expansion are accompanied by higher powers of $\phi$ and are therefore sub-leading as $\phi$ approaches $0$. The general structure of the $W_-$ solution is shown in \cite{rg1,papa1} to be of the form
\be
\ell ~W_-=\sum_{m=0}^{\infty}\sum_{n=0}^{\infty} ~A_{m,n} \left(C~|\phi|^{d/\Delta_-}\right)^m~\phi^n
\label{i24}
\ee

In (\ref{i24}), the coefficients $A_{m,n}$ depend on the potential, and the expansion is valid if $\Delta$ is irrational.  The expansion in (\ref{i24}) has poles for rational values of $\Delta$, which implies that in such cases there are also logarithmic terms that go into (\ref{i24}), for further details see \cite{Martelli,Aharony}.

A word on the geometry that corresponds to the superpotential solutions, (\ref{i22_a}) and (\ref{i22_b}), is now in place. From \eqref{i9} we have:
\begin{subequations}\label{i25}
	\begin{align}\label{i25a}
		&\dot\phi(u)=W_+'(\phi)
			\implies
		\phi(u)=\phi_+ e^{\Delta_+ u/\ell}+\cdots\\
		&\dot\phi(u)=W_-'(\phi)
			\implies
		\phi(u)=\phi_- e^{\Delta_- u/\ell}
			+{
				d~C
					\over
				\Delta_-(2\Delta_+-d)}~\phi_-^{\Delta_+/\Delta_-}~
			e^{\Delta_+u \ell}+\cdots\label{i25b}
	\end{align}
\end{subequations}
where $\phi_+$ and $\phi_-$ are integration constants.

By integrating (\ref{i8}), one finds the corresponding scale factors:
\begin{subequations}\label{i26}
	\begin{align}
		&\dot A(u)=-{1\over 2(d-1)}W_+'(\phi)
			\implies
		A(u)=-{u-u_*\over \ell}+{\phi^2_+ \over 8(d-1)} e^{2\Delta_+ u/\ell}
		+\cO\le( e^{3\Delta_+u/\ell}\ri)
		\label{i26a}\\
		&\dot A(u)=-{1\over 2(d-1)}W_-'(\phi)
			\implies
		A(u)=-{u-u_*\over \ell}+ {\phi^2_- \over 8(d-1)} e^{2\Delta_- u/\ell}
		+\cO\le(C e^{ud/\ell}\ri)
		\label{i26b}
	\end{align}
\end{subequations}
where $u_*$ is an integration constant and the sub-leading terms come from the $\cO(\phi^2)$ terms in $W_\pm$.

Since we are expanding near $\phi=0$, we must take $u\to -\infty$. In that limit, the scale factor is divergent, and therefore these solutions describe the near-boundary regions of an  asymptotically $AdS$ space-time,  with  AdS length $\ell$.  As $u$ increases away from the boundary the scale factor decreases, and the solutions correspond to a flow leaving a UV AdS fixed-point. We have shown that {\it local maxima of the potential $V(\phi)$ and, hence, local minima of $B(\phi)$, correspond to UV fixed points}.

Each superpotential solution corresponds to {\em two} disconnected geometries: one with $\phi>0$,  the other with $\phi<0$,  because as $\phi$ approaches the critical point from each side the geometry is geodesically complete.

This analysis was done in the context of the bulk gravitational theory. We now look at the boundary QFT. {The $W_-$, branch is interpreted by associating with $\phi(u)$ a source\footnote{
This choice of source is called the standard quantization. Other quantizations exist \cite{BF82,KlebWit,Witten02}, as for example, the so-called alternative quantization where the roles of the source and the VEV are reversed. Different quantizations, however, only exist when $\Delta_-$ is between $d/2-1$ and $d/2$.
} $J=\phi_-~\ell^{-\Delta_-}$ of an operator ${\cal O}(x)$. The scaling dimension of $\cO$ is $\Delta_+$, \cite{Witten98}.
The vacuum expectation value (VEV) of the scalar operator $\cal O$ is given by
\eql{i27}{\le<\cal O\ri>_{W_-}=C{
				 d
					\over
				\Delta_-}
				\phi_-^{\Delta_+/\Delta_-}\ell^{-\Delta_+} = C{
				 d
					\over
				\Delta_-}
				J^{\Delta_+/\Delta_-}   }\;.
Therefore, for a given source $J$, the integration constant $C$ fixes the VEV \footnote{If we want to  interpret \eqref{i27} as the renormalized VEV,  the appearance of the same $C$ in \eqref{i22_b} and in \eqref{i27} corresponds to a specific choice of holographic renormalization scheme \cite{papa1}.}.  The terms of order $C^2$ and higher do not contribute to  equation \eqref{i27} as they vanish at the boundary \cite{rg2,papa1}.

The $W_+$ solution in equation (\ref{i22_a})  also corresponds to a boundary operator $\cal O$ with scaling dimension $\Delta_+$ and  a non-vanishing VEV given by:
\eql{i28}{\le<\cal O\ri>_{W_+}=(2\Delta_+-d)\phi_+\ell^{-\Delta_+}.}
The source of $\cal O$ however is set to zero, and the vev is arbitrary. We have a moduli space of vacua in this case, and the flow is driven by the operator vev.

Once the sign of the source $\phi_-$ in equation \eqref{i25a} is fixed, close to the UV fixed point, the solution will correspond either to $\phi(u)\geqslant0$ or to $\phi(u)\leqslant0$. For either choice of sign, the geometry will approach an AdS boundary, meaning that the positive source and the negative source solutions correspond to disconnected geometries. Similarly, a $W_+$ solution associated with a positive VEV in \eqref{i28} is disconnected from the geometry corresponding to a negative VEV. This corresponds to the sign of $\phi_+$ in \eqref{i25a}.}

The $W_{+}$ solution is the upper envelope of all $W_-$ solutions. To see this, from (\ref{i25b}), take $\phi_-\to 0$ and $C\to \infty$ while keeping $C\phi_-^{\Delta_{+}/ \Delta_{-}}$ fixed, we obtain the $W_+$ solution.
The marginal case with $\Delta_+=\Delta_-={d\over 2}$ is considered  in \cite{papa1}.

The BF bound $m^2\geqslant-{d^2\over 4\ell^2}$ is required in order to have stability of small perturbations close to a maximum of $V$ \cite{BF82, exotic} . On the boundary CFT, it corresponds to a reality condition on the dimension of the operator dual to $\phi$.
The BF bound is also required, as a reality condition for the superpotential.

Rewrite, now, the squared mass and the AdS length in terms of the potential with maximum at $\phi=\phi_*$ and replace the results into the BF bound inequality:
\eql{i31}{
1\geqslant {4(d-1)\over d}{V''(\phi_*)\over V(\phi_*)}
}
If there is a solution of \eqref{i10} with $W'(\phi_*)=0$, then \eqref{i31} becomes an identity:
\eql{i32}{\le(1-{4(d-1)\over d}{W''(\phi_*)\over W(\phi_*)}\ri)^2\geqslant0,}

and hence, having a (real) critical point  of the superpotential is incompatible with a violation of the BF bound.

This shows that \emph{a local maximum of the potential which violates the BF bound does not correspond to a critical point for the  superpotential.}

This can be justified in another way as well : notice that the expansion \eqref{i22} around a BF bound-violating point leads to a  complex superpotential, as the dimensions $\Delta_\pm$  of \eqref{i23} become complex. This  represents a breakdown of the first order formalism.

However, a UV-regular solution $(\phi(u), A(u))$ which reaches the  AdS boundary with vanishing $\dot \phi(u)$ does exist. This solution is unstable against linear perturbations and corresponds to a non-unitary CFT.

We conclude this part by indicating that solutions near a maximum of the potential in the AdS regime have a single free parameter in the $W_-$ branch, and no free parameter in the $W_+$ branch. This fact will be crucial in order to understand globally the generic existence of regular solutions.


\subsubsection{Local  minima of the potential}
\label{sssec:min}

We now assume the potential has the form \eqref{i21_1},  but  with $m^2>0$. Then, as shown in \cite{exotic},  solutions of the superpotential equation close to the critical point $\phi=0$ have the  regular power series expansion:
\begin{subequations}\label{i33}
\begin{align}
	W_\pm(\phi)
			=&{1\over \ell}
			\le[
				{2(d-1)} + {\Delta_\pm\over2}\phi^2+\cO(\phi^3)
			\ri],
			\label{i33_a}\\
\Delta_\pm =&\ha \le( d \pm \sqrt{d^2 +4 m^2\ell^2} \ri)\quad \text{with}\quad m^2>0. \label{i33_b}
\end{align}
\end{subequations}
Note that now we have necessarily $\Delta_-<0$  and $\Delta_+>0$.
Also it is important to know there are no free parameters in either of the two branches. The reason is that the turning on of free parameters leads to singularities.

The $W_+$ solution has a local minimum at $\phi=0$ while  $W_-$ has a local maximum. This implies a different geometrical and holographic interpretation. To see this, we solve for $\phi(u)$ using \eqref{i9} and for $A(u)$ using \eqref{i8} with $W_\pm$ from \eqref{i8},\eqref{i9} :
\begin{subequations} \label{i34}
\begin{align}
		&\dot\phi(u)=W_+'(\phi)
			\implies
		\phi(u)=\phi_+ e^{\Delta_+ u/\ell}+... \label{i34_1}\\
		&\dot\phi(u)=W_-'(\phi)
			\implies
		\phi(u)=\phi_- e^{\Delta_- u/\ell}
			+...  \label{i34_2}\\
	&A_\pm(u)=-{u-u_*\over \ell}
				- {1\over 8(d-1)}{\phi_\pm^2}e^{2\Delta_\pm u/ \ell}+...
\label{w6}\end{align}
\end{subequations}
where $\phi_\pm$ and $u_*$ are integration constants.
Equations \eqref{i34} are valid for small $\phi$ (near the critical point). Because $\Delta_-<0$, small $\phi$ in (\ref{i34_1}) requires  $u \to +\infty$,  and $\m=\m_0 e^{A_-(u)} \to 0$. Therefore, {\em a $W_-(\phi)$ solution with a critical point at a local minimum of the potential corresponds to a flow that arrives at an infra-red (IR) fixed point}.

We can use again eq. \eqref{i27} with $C$ set to zero,  to see that the $W_-$ solution with positive $m^2$ represents a RG flow reaching an IR fixed point where the operator $\cal O$ of the dual field theory has dimension $\Delta_+>d$ and a vanishing VEV.

On the other hand,  because $\Delta_+>0$, small $\phi$ in (\ref{i34_1})  requires $u \to -\infty$. In this case the scale $\m=\m_0 e^{A_+(u)}$ diverges, therefore,  the $W_+$ solution corresponds to a flow {\it leaving a UV fixed point}. Here however, the source for the operator $\cal O$ vanishes,  so it is a flow driven purely by a VEV.

We summarise by stating that all regular solutions near a minimum of a potential have no adjustable parameters. The $W_+$branch describes a the UV part of a flow driven by the vev of an irrelevant operator. The $W_-$ part describes the IR part of a flow driven by the source of an irrelevant operator.

There are subtler cases of extrema that are inflection points. We shall not considered them here. They have been considered in \cite{Bourdier13,exotic}.

\subsubsection{Bounces: $W'(\phi)=0$ and $V'(\phi)\neq0$}
\label{bounces}

If we choose an arbitrary point $\phi_*$ on the critical curve $W(\phi)=B(\phi)$, $V'(\phi_*)$ will generically be non-zero. Despite this fact, $W(\phi)$ still has a critical point at $\phi_*$, as follows from equation \eqref{i10}. In fact, generic critical points occurs where $V'(\phi_*)\neq 0$.

As we shall see,  such critical points are reached  at  finite values of the scale factor, unlike critical points  at  extrema of $V$, which instead are reached as $A \to \pm \infty$.

We denote a generic critical point by $\phi_B$ where by definition $W'(\phi_B)=0$. The  superpotential equation also implies that $W(\phi_B)=B(\phi_B)$ and both are finite.
Then, equation (\ref{i19}) implies that  $W''$ must necessarily diverge at $\phi_B$. Therefore,  we can approximate equation \eqref{i19} by:
\eql{i35}{
 W'(\phi)W''(\phi) \simeq V'(\phi) \qquad \phi \simeq \phi_B,
}
which can be integrated once to give, close to $\phi_B$:
\be\label{i36}
W'(\phi) \simeq \pm \sqrt{2(\phi-\phi_B) V'(\phi_B)},
\ee
where we have used the assumption that $V$ has a regular power series expansion around $\phi_B$.
We conclude that a real superpotential exists only on the right (left) of $\phi_B$ for $V'(\phi_B)$ positive (negative). In either case, there are two solutions $W_\uparrow$ and $W_\downarrow$ terminating at $\phi=\phi_B$ from the right (left), which correspond to the two choices of sign in equation (\ref{i36}).  After one more integration, we obtain the approximate form of the two solutions close to $\phi_B$.  For $V'(\phi_B)>0$,  they are:
\bea
 W_{\uparrow}(\phi)\simeq  W_B+\frac{2}{3} \sqrt{2V'(\phi_B)}(\phi-\phi_B)^{3/2}, \nonumber\label{w7} \\
&& \quad \phi > \phi_B, \label{bW2}\\
 W_{\downarrow}(\phi)\simeq  W_B-\frac{2}{3} \sqrt{2V'(\phi_B)}(\phi-\phi_B)^{3/2},  \nonumber \label{w8}\\
\eea	
where
\eql{i37}{W_B=\sqrt{-{4(d-1)\over d}V(\phi_B)}.}
The two-branched  solution in equation (\ref{bW2}) was called a {\em bounce} in \cite{exotic}. In  particular, equation  (\ref{bW2})  describes the increasing and decreasing branches at a left bounce. For $V'<0$  the solutions have the same  expression,   but now   $\phi<\phi_B$, with $W_\uparrow$ and $W_\downarrow$ interchanged. At $\phi=\phi_B$, both solutions  (\ref{bW2}) have vanishing first derivative  and infinite second derivative,  as expected from equation (\ref{i19}). As shown in \cite{exotic}, bounces do not admit continuous deformations, i.e. there are only two solutions reaching the critical curve at a given generic point $\phi_B$ which is not an extremum of $V$. This property is also generalized to the other ansatze considered in this paper, later on, see appendix \ref{flats}.

To obtain a complete geometry  we must glue the two solutions $W_{\uparrow}(\phi)$ and $W_{\downarrow}(\phi)$ into a single solution with multi-valued superpotential. Although the superpotential is non-analytic at $\phi_B$, the resulting geometry  is smooth.


To obtain the solution in terms of $A(u)$ and $\phi(u)$, we first rewrite equation \eqref{i9} for each of two  branches in equation (\ref{bW2}). Denoting by  $\phi_\uparrow(u)$ and $\phi_\downarrow(u)$ the two corresponding scalar field profiles, we have:
\be
\dot\phi_\uparrow \simeq \sqrt{2\le(\phi_{\uparrow}-\phi_B\ri) V'(\phi_B)} \label{eqphiB1}
\ee
\be \dot\phi_\downarrow \simeq -\sqrt{2\le(\phi_{\downarrow}-\phi_B\ri) V'(\phi_B)} \label{eqphiB2}
\ee
where for definiteness we are considering the case a left bounce with $V'>0$ and $\phi>\phi_B$.

Equations (\ref{eqphiB1}-\ref{eqphiB2})  integrate to the two halves $u>u_B$ and $u<u_B$ of a solution $\phi(u)$ which is analytic at the location of the bounce $u=u_B$

\begin{align}
	\phi(u)=\phi_B+{V'(\phi_B)\over 2}(u-u_B)^2+\cO(u-u_B)^3=
		\begin{cases}
			\phi_\uparrow(u)\text{ for } u>u_B,\\
			\phi_\downarrow(u)\text{ for } u<u_B.
		\end{cases}
	\label{b26}
\end{align}

Using equation  (\ref{phiB}) we can write  both superpotentials \eqref{bW2} as functions of the holographic coordinate $u$:
\eql{b26_a}{W(u)=W_B+{\le(V'(\phi_B)\ri)^2\over 3}(u-u_B)^3+\cO(u-u_B)^4
=
		\begin{cases}
			W_\uparrow(\phi(u))\text{ if } u > u_B\\
			W_\downarrow(\phi(u))\text{ if } u< u_B
		\end{cases}
}
As for the scalar field, the two solutions $W_{\uparrow}(\phi)$ and $W_{\downarrow}(\phi)$  combine into a  a single-valued function of $u$. We can integrate \eqref{i8} using \eqref{b26_a} to obtain the scale factor:
\be \label{Ab}
A(u)=A_B-\sqrt{V(\phi_B) \over d(d-1)}(u-u_B)-{\le(V'(\phi_B)\ri)^2\over 4!(d-1)}(u-u_B)^4+\cO(u-u_B)^5,
\ee
which is also regular at the bounce. In particular, keeping only one branch of the bounce would result in an incomplete geometry which terminates at an ``artificial'' boundary at $u=u_B$.

As mentioned already, the geometry at the bounce is non-singular: all curvature invariants are finite . The curvature invariants of the geometry are functions of $V(\phi), W(\phi)$, and their derivatives \cite{Bourdier13}.  The fact that the geometry is regular at the bounce is clear when invariants are  expressed in terms of $(A(u),\phi(u))$ and their derivatives, \cite{Bourdier13,exotic}.

To summarize, the full solution  is regular as it goes  through the bounce, which is characterized by a turning point for $\phi(u)$.  The non-analyticity (and multi-valuedness) of $W(\phi)$ is a consequence of the fact that $\phi(u)$ is not monotonic, therefore,  it is not a good coordinate in a neighbourhood of $\phi_B$.  The same considerations hold for a bounce to the right, i.e. such that $\phi<\phi_B$: in this case,  $V'(\phi_B)<0$ and $\phi(u)$ goes through  a relative maximum.

\section{Asymptotically de Sitter flows\label{dSflows}}

We now turn to regions of the scalar potential where it is positive, $V(\phi)>0$. We call such regions the ``dS regime".
As with the AdS regime, we expect to find here solutions that are asymptotically de Sitter\footnote{Such solutions were considered in \cite{afim} as cosmologies in a map between holographic and pseudo-holographic solutions.}.

Here, we use the action (\ref{i1}), and work in a coordinate system\footnote{Several useful coordinates for de Sitter can be found in appendix \ref{a}.} where the holographic coordinate $u$ is now time like,
\be
\phi=\phi(u),~~~ds^2=-du^2+e^{2A(u)}dx_{\m}dx^{\m}
\label{b1}\ee
where we imposed again ISO(d) invariance.
For $A(u)=H u$ this is de Sitter space in Poincar\'e coordinates, as seen in (\ref{a26}) .The resulting Einstein equations are, in this case,

\begin{subequations}\label{b2}
\begin{align}
	&2(d-1)\ddot{A}(u)+\dot{\phi}^2(u)=0,\label{b2a}\\
	&d(d-1)\dot{A}(u)^2-\ha\dot{\phi}^2(u)-V(\phi)=0.\label{b2b}
\end{align}
\end{subequations}
And they are related to those in (\ref{i3_1}) and (\ref{i3_2}) by $V\to -V$.

The Klein-Gordon equation is unchanged
\eql{b3}{\ddot{\phi}+d\dot A\dot{\phi}+V'(\phi)=0.}
As before, we observe that $\dot{A}(u)$ does not increase, due to equation (\ref{b2a}).

We define the function $\mu=\mu_0 e^{A(u)}$. For now, we consider the fixed points of the cosmological evolution to correspond to asymptotically dS geometries. Therefore, we shall be in the dS regime.

We take $A(u)$ to be monotonically increasing, an assumption consistent with (\ref{b2a}), and the timelike coordinate $u$ decreases with the flow. This convention matches with our cosmology of an expanding (spatially flat) universe.

\subsection{The superpotential}

The superpotential formalism in cosmological solutions was introduced a long time ago in \cite{Bond}. It was used in \cite{afim} in order to define the analogue of a $\beta$-function in cosmology and use Wilsonian intuition in the classification of inflationary theories\footnote{See also \cite{Adam}.}.

We define the superpotential as in (\ref{i8}),
\eql{b5}{W(\phi(u))=-2(d-1)\dot A(u).}
Proceeding similarly as in the case of the holographic flows, we obtain

\eql{b6}{W'(\phi)=\dot \phi(u),}
\eql{b6_1}{V(\phi)=\frac{d}{4(d-1)}W^2(\phi)-\ha W'^2(\phi)}.

The analogue to what in the holographic RG case was the $\beta$ function\footnote{Which for consistency we shall symbolize as $\beta$ in the case of cosmological flows as well.} is, once again, found to be

\eql{b8}{{d\phi\over d\log \m}=\b(\phi)=-2(d-1){W'(\phi)\over W(\phi)}.}

\subsection{General properties of the superpotential}
Using exactly the same technique that we used in section (\ref{ss:i10}), we went from one second order and one first order equation, to three first-order equations , namely,  (\ref{b5}-\ref{b8}). The properties of the superpotential in the case of asymptotically dS flows are now the following:

\begin{enumerate}

\item {\it W(u)$\equiv$W$(\phi(u))$ is monotonically decreasing along the flow}.

From equation \eqref{b6},
\eql{b9}{
	{dW(t)\over du}={d\phi(u)\over du}{d W(\phi)\over d\phi}=W'^2\geqslant0
}
Because we have taken $A(u)$ to increase with $u$, and the flow is toward what would in the AdS picture be lower energy, then (\ref{b9}) means that $W(u)$ is decreasing along the evolution.

Recall that in the case of holographic RG flows, $W(u)$ was taken to be monotonically \emph{increasing} with the flow.


\item {\it The $(W,\phi)$ plane has a forbidden region where no solution of \eqref{b6_1} exists. It is the region bounded  by the critical curve:}
\be
B(\phi) \equiv -\sqrt{\frac{4(d-1)}{d}V(\phi)}\label{b10}
\ee
This follows from the superpotential equation \eqref{b6_1}:
\eql{b11}{|W(\phi)|= \sqrt{\frac{4(d-1)}{d}\le(\ha W'^2+V(\phi)\ri)}\geqslant\sqrt{\frac{4(d-1)}{d}V(\phi)}.}

Again, as before, $W(\phi)$ cannot change sign. From the definition (\ref{b10}), we again allow for local minima of the potential to correspond to local maxima of $B(\phi)$, and vice versa.

This is juxtaposed with the previous case, where we had taken

\be
B(\phi) \equiv \sqrt{-\frac{4(d-1)}{d}V(\phi)}\label{b12}.
\ee

To give some intuition for the connection of the properties of holographic RG and cosmological flows, notice that results are similar in both cases; in particular, in our conventions, the cosmological flow is essentially acquired from the holographic RG flow by reflecting along the $\phi$-axis.
\item {\it We can assume $W(\phi)<0$ without loss of generality}.

This property generalizes naturally from the holographic superpotential, since the symmetry
\eql{b13}{
(u,W)\to-(u,W)
}
is still a symmetry of the equations (\ref{b5} - \ref{b6_1}). Recall that in the holographic case, we had taken $W(\phi)>0$. This choice justifies our assumption in the preceding section, that we can take $A(u)$ to be monotonically increasing.

It is noted, as will be shown later on, that with these choices for $W$ \emph{the situation is now reversed with respect to the RG (AdS) flows}. Here, local maxima (instead of minima) correspond to the big-bang point  of the cosmological evolution, and this means that depending that the space volume is near infinite around their neighborhood. This corresponds to the ${\cal I}^{+}$ boundary of global de Sitter (see appendix \ref{a}).

Local minima in the dS regime on the other hand serve as either departure or arrival points of the evolution (with the intuition acquired from holography, UV and IR fixed points respectively). We shall analyze this further soon.

\item {\it For a generic point in the allowed region of the $(\phi,W)$ plane with $W< B$,
    there exist two and only two solutions, $W_{_{\uparrow}}(\phi)$ and $W_{_{\downarrow}}(\phi)$, passing through that point}

Equation \eqref{b6_1}  can be separated in two equations as:
\begin{subequations}\label{b13_1}
\begin{align}
W'_{\uparrow}(\phi)=+\sqrt{{d\over2(d-1)}\big(W_{\uparrow}^2(\phi)-B^2(\phi)\big)}\label{b14a}\\
W'_{\downarrow}(\phi)=-\sqrt{{d\over2(d-1)}\big(W_{\downarrow}^2(\phi)-B^2(\phi)\big)}\label{b14b}
\end{align}
\end{subequations}

Notice that the equations (\ref{b14}) are exactly the same as the equations in the AdS case. Therefore, all properties discussed in the corresponding item 4 of the respective holographic discussion, are inherited here as well.

\item{\it The geometry is regular if the potential and the superpotential are finite along the flow. This is equivalent to the scalar field staying finite along the flow.}
The justification is identical to holographic RG case : $V(\phi)$ is analytic and hence finite for all finite $\phi$, and from (\ref{b11}), $W$ is also finite.


\item {\it Extremal points of $W(\phi)$ lie on the critical line $B(\phi)$, } since there $W'=0$ by equation (\ref{b14}).
These again correspond to fixed points, and bounces.
\end{enumerate}

\subsection{The analogue of the Breitenlohner-Freedman bound} \label{BFbound}

Let $\phi_*$ be a fixed point of the flow. Therefore, $V'(\phi_*)=0$ and $W'(\phi_*)=0$, and by a change of coordinates we set $\phi_*=0$.

The expansion of the potential about the fixed point is

\be
	V={d(d-1)H^2}+{m^2\over 2}\phi^2 + \cO(\phi^3),
\label{b14}\ee

At the critical point, it is obvious that $B(0)=W(0)=-2(d-1)H$, since we have taken both functions to be negative. Furthermore, from (\ref{b14}) we have

\eql{b15}{
B''(0)=-\frac{m^2}{d H}
}

Writing (\ref{b6_1}) as

\eql{b16}{
\frac{2(d-1)}{d} W''(\phi)=W^2(\phi)-B^2(\phi),
}

differentiating twice, setting $\phi=0$ and solving for $W''(0)$, we have

\eql{b17}{
W''(0)=-\Delta_{\pm} H,~~~\Delta_{\pm}= \frac{1}{2} \bigg( d \pm \sqrt{d^2-\frac{4 m^2}{H^2}} \bigg).
}

To have a real superpotential we must have,
\be
{m^2\over H^2} \leq {d^2\over   4}~~~\to~~~ {V''\over V}\leq {d\over 4(d-1)}
\label{b17a}\ee
This the analogue of the BF bound, for dS solutions.

It should be noted that the BF bound in the AdS regime has a dual interpretation. First, although $\Delta_{\pm}$ are complex, the solution for $\phi$ and $A$ are well defined near an extremum that violates the BF bound.
However, under the AdS/CFT correspondence this extremum corresponds to a complex and therefore non-unitary CFT\footnote{Such QFTs have been described recently in \cite{Rychkov}.} .
Moreover the solution for the field $\phi$ oscillates an infinite number of times until it arrives at the extremum. Small fluctuations around this solution have exponentially growing dependence on time suggesting that the (pseudo)-CFT is unstable.

There is another aspect of the violation of the BF bound. Consider a flow along a field direction, ending at a minimum of the potential along the same direction. Consider also the case where at the same extremum another direction violates the BF bound.
In all such examples, there is another flow with the same sources (but different vevs) and lower free energy, that bypasses the BF violating extremum. Moreover, the scalar along the BF-violating direction obtains a non-trivial vev (but has no source).
This is has been observed in many examples but there is no general theorem so far
guaranteeing the validity of this censorship property, against BF-violating solutions.

In the AdS regime, the BF bound forbids scalar directions with sufficiently negative masses. This implies that the bound is relevant for maxima of the potential.
As we have seen above the situation in dS is the inverse. Sufficiently massive scalars violate the BF bound. This implies that here it applies to minima of the potential.
It should be stressed that in the dS regime the definition of the BF bound is by analogy with the AdS case: it delineates extrema where $\Delta_{\pm}$ are complex, or where a scalar changes direction an infinite number of times before reaching an extremum.

However, unlike AdS, it is not clear whether violations of the BF bound in dS implies instabilities or the necessity to turn on the relevant scalar solutions. An answer to these issues is a very interesting problem which may have consequences for theoretical cosmology.

\subsection{The structure near critical points.\label{cripo}}

We shall now analyse the structure of the solutions in the dS regime near the extrema of the potential.

\subsubsection{Local minima of the potential}

The potential has the following series expansion around a local minimum at $\phi=0$,  in the dS regime:
\be
	V(\phi)={d(d-1)\over \ell^2}+{m^2\over 2}\phi^2 + \cO(\phi^3), \qquad 0<m^2<\frac{d^2 H^2}{4}
\label{b17_1}
\ee
The equation \eqref{b6_1} has solutions:
\begin{subequations}			\label{b18}
\begin{align}
	{W_+(\phi)\over H}
			=&-
			\le[
				{2(d-1)} + {\Delta_+\over2}\phi^2+\cO(\phi^3)
			\ri], 			\label{b18a}\\
	{W_-(\phi)\over H}
			=&-
			\le[
				{2(d-1)} + {\Delta_-\over2}\phi^2+\cO(\phi^3)
			\ri]+C|\phi|^{d/\Delta_-}\le[1 +\cO(\phi)\ri]
			+\cO(C^{2})
			\label{b18b},\\
	\Delta_\pm =&\ha \le( d \pm \sqrt{d^2 -4 {m^2\over H^2}}  \ri) \quad \text{with}\quad {d^2 \over 4}>{m^2\over H^2}>0,
	\label{b19}
\end{align}
\end{subequations}

The possible deformations of the solutions about the critical point are

\be\
\delta W_- = C  \phi^{d\over \Delta_-},  \quad \delta W_+ = C  \phi^{d\over \Delta_+}
\label{w9}\ee
Both deformations vanish as $\phi \to 0$. But since $0<\Delta_-< d/2$ and  $\Delta_+> d/2$, we have:
\be
	{d\over \Delta_-}>2, \quad \quad 0< {d\over \Delta_+}<2.
\label{w10}\ee
and $\delta W_-$ is the only acceptable deformation, since only then do we have that ${\delta W}$ is subleading to the $\phi^2$ term as $\phi \rightarrow 0$. This implies, like in the AdS case, that the $\delta W_-$ solution has an adjustable parameter, and therefore there exists a complete family of solutions parametrized by $C$.  This is the reason why we have included the extra term in (\ref{b18b}).

Like the AdS case the $W_-$ solution has the form of a resurgent series
\be
{W_-\over H}=\sum_{m,n}A_{m,n}~\phi^m~\left(C\phi^{d\over \Delta_-}\right)^n
\label{w11}\ee
The terms with $n=0$ are the perturbative series of the solution where as the $n\not=0$ terms are non-perturbative, and are driven by the subleading boundary condition of the scalar equations.

Solving the first order equations for the scalar field, we obtain,
\begin{subequations}\label{b20}
	\begin{align}\label{b20a}
		&\dot\phi(u)=W_+'(\phi)
			\implies
		\phi(u)=\phi_+ e^{-\Delta_+ H u}+\cdots\\
		&\dot\phi(u)=W_-'(\phi)
			\implies
		\phi(u)=\phi_- e^{-\Delta_- H u}
			+{
				d~C
					\over
				\Delta_-(2\Delta_+-d)}~\phi_-^{\Delta_+/\Delta_-}~
			e^{-\Delta_+ H u}+\cdots\label{b20b}
	\end{align}
\end{subequations}
where $\phi_+$ and $\phi_-$ are integration constants.

The scale factors are found by integrating equation (\ref{b5}):
\begin{subequations}\label{b21}
	\begin{align}
		&\dot A(u)=-{1\over 2(d-1)}W_+'(\phi)
			\implies
		A(u)=H(u-u_*)+{\phi^2_+ \over 8(d-1)} e^{-2\Delta_+ H u}
		+\cO\le( e^{-3\Delta_+ H u}\ri)
		\label{b21a}\\
		&\dot A(u)=-{1\over 2(d-1)}W_-'(\phi)
			\implies
		A(u)=H(u-u_*)+ {\phi^2_- \over 8(d-1)} e^{-2\Delta_- H u}
		+\cO\le(C e^{-u d H}\ri)
		\label{b21b}
	\end{align}
\end{subequations}
where $u_*$ is an integration constant and the sub-leading terms come from the $\cO(\phi^2)$ terms in $W_\pm$.

The expressions above are valid for small $\phi$, so we must take $u \to +\infty$. The scale factor diverges and these solutions describe the near-boundary regions of an   asymptotically $dS$ space-time,  with $dS$ length scale $1/H$.  As $u$ goes to past infinity,  the scale factor decreases, therefore,  these solutions correspond to a flow leaving a  $dS$ fixed-point corresponding to a large universe. We conclude that {\it local minima of the potential $V(\phi)$ and, hence, local maxima of $B(\phi)$, correspond to large universes and in particular the ${\cal I}^{+}$ boundary of dS.}

\subsubsection{Local maxima of the potential}

We now assume the potential has the form \eqref{b18},  but  with $m^2<0$. The two solutions of the superpotential are:
\begin{subequations}\label{b22}
\begin{align}
	W_\pm(\phi)
			=&-H
			\le[
				{2(d-1)} + {\Delta_\pm\over2}\phi^2+\cO(\phi^3)
			\ri],
			\label{b22a}\\
\Delta_\pm =&\ha \le( d \pm \sqrt{d^2 -4 {m^2\over H^2}} \ri)\quad \text{with}\quad m^2<0. \label{b22b}
\end{align}
\end{subequations}
Now, we have $\Delta_-<0$  and $\Delta_+>0$. The two possible continuous deformations are once again $\delta W_\pm(\phi) = C \phi^{d/\Delta_\pm}$. However, for the case $\Delta_+$ we have $2 / \Delta_+ < 2$. and hence $\delta W_+ /W_+$ does not vanish as $\phi\rightarrow 0$. Also, $d/\Delta_-<0$, and again, $\delta W_-/W_-$ blows up as we approach the fixed point. Therefore,  there are no acceptable regular deformations of the solutions (\ref{b22}).

The $W_+$ solution has a local maximum at $\phi=0$ while  $W_-$ has a local minimum. Solving for the field and scale factor in each case, we have:
\begin{subequations} \label{b23}
\begin{align}
		&\dot\phi(u)=W_+'(\phi)
			\implies
		\phi(u)=\phi_+ e^{-\Delta_+ H u}+... \label{b23a}\\
		&\dot\phi(u)=W_-'(\phi)
			\implies
		\phi(u)=\phi_- e^{-\Delta_- H u}
			+...  \label{b23b}\\
	&A_\pm(t)=H(u-u_*)
				- {1\over 8(d-1)}{\phi_\pm^2}e^{-2\Delta_\pm H u}+...
\label{w12}\end{align}
\end{subequations}
where $\phi_\pm$ and $t_*$ are integration constants.
Equations \eqref{b23} are valid for small $\phi$ (near the critical point).

Because $\Delta_-<0$, small $\phi$ in the minus solution in (\ref{b23b}) requires  $u \to -\infty$,  and $\m=\m_0 e^{A_-(u)} \to 0$. Therefore, {\em a $W_-(\phi)$ solution with a critical point at a local maximum of the potential corresponds to a flow that arrives at an initial big-bang apparent singularity (as explained, this is a point in the ${\cal I}^-$ boundary of dS).}

On the other hand,  for the plus solution, because $\Delta_+>0$, small $\phi$ in (\ref{b23a})  requires $u \to +\infty$. In this case the scale factor diverges, and therefore the $W_+$ solution near a maximum
approaches the ${\cal I}^+$ of dS.

\subsection{Bounces: $W'(\phi)=0$ and $V'(\phi) \neq 0$\label{bounce}}

{In this section  we consider critical points $W'=0$ that happen away from critical points in the potential, $V'\not=0$.
If such a point is at $\phi=\phi_B$ then $W(\phi_B)=B(\phi_B)$, with $V'(\phi_B) \neq 0$. For the case of asymptotically AdS geometries, we studied such solutions in section (\ref{bounces}). We now make a similar analysis, for asymptotically dS geometries.

A similar arguments indicates that $W''$ diverges at a bounce and that

\be \label{b25}
W'(\phi) \simeq \pm \sqrt{2 V'(\phi_B) (\phi_B-\phi)}.
\ee

If the superpotential is to be real, then the requirement on $V'(\phi_B)$ is

\begin{align}
&\phi > \phi_B \implies V'(\phi_B) <0\label{w1} \\
&\phi < \phi_B \implies V'(\phi_B) >0.\label{w2}
\end{align}

From (\ref{b25}) and choosing $V'(\phi_B)>0$ (the analysis is the same for $V'(\phi_B)<0$, but one needs to replace $\phi_B-\phi$ with $\phi-\phi_B$ in all proceeding expressions), we arrive at two solutions for $W(\phi)$, each corresponding to a choice of sign. Therefore,

\be \
W_{\uparrow}(\phi) \simeq W_B + \frac{2}{3} \sqrt{2 V'(\phi_B)}(\phi_B-\phi)^{\frac{3}{2}},
\label{w3}\ee
and
\be
W_{\downarrow}(\phi) \simeq W_B - \frac{2}{3} \sqrt{2 V'(\phi_B)}(\phi_B-\phi)^{\frac{3}{2}},
\label{w4}\ee

where we have defined

\be
W_B \equiv \sqrt{\frac{4 (d-1)}{d} V(\phi_B)}.
\label{w13}\ee

We have shown, as was expected, that in the asymptotically dS flows we also have bouncing solutions, in analogy with the asymptotically AdS solutions. The two solutions (\ref{c3}), (\ref{c4}) terminate at $\phi=\phi_B$.

Using (\ref{b6}) we may write

\be
\phi_{\uparrow} \simeq + \sqrt{2(\phi_B-\phi_{\uparrow}) V'(\phi_B)}
\label{w14}\ee

and

\be
\phi_{\downarrow} \simeq - \sqrt{2(\phi_B-\phi_{\downarrow}) V'(\phi_B)}.
\label{w15}\ee

These equations may be integrated, yielding $\phi(u)$ for each of the cases $u<u_B$ or $u>u_B$

\begin{align}
	\phi(u)=\phi_B-{V'(\phi_B)\over 2}(u-u_B)^2+\cO(u-u_B)^3=
		\begin{cases}
			\phi_\uparrow(u)\text{ for } u>u_B,\\
			\phi_\downarrow(u)\text{ for } u<u_B.
		\end{cases}
	\label{phiB}
\end{align}

With this result, we can write the superpotential in terms of $u$ as

\eql{b27}{W(u)=W_B+{\le(V'(\phi_B)\ri)^2\over 3}(u-u_B)^3+\cO(u-u_B)^4
=
		\begin{cases}
			W_\uparrow(\phi(u))\text{ if } u > u_B\\
			W_\downarrow(\phi(u))\text{ if } u< u_B
		\end{cases}.
}		

This allows us to integrate equation (\ref{b5}) in order to find the scale factor in terms of $u$, near the location of the bounce. We have

\be
A(u)=A_B-\sqrt{\frac{V(\phi_B)}{d(d-1)}}(u-u_B)-\frac{(V'(\phi_B))^2}{4! (d-1)} (u-u_B)^4 + \cO(u-u_B)^5,
\label{w16}\ee

where $A_B$ and $u_B$ are constants of integration.

As is evident, keeping only one of the solutions results in an incomplete geometry, since the the scale factor terminates at some arbitrary $u=u_B$.

A few comments regarding the regularity of the geometry are in order. The argument is essentially as the one produced in section (\ref{bounces}) for asymptotically AdS flows, but we repeat it here for convenience.

 All curvature invariants are finite at the bounce. Since, however, $W''(\phi)$ is singular at the bounce, this is not entirely obvious. Expressing the invariants in terms of $A(u)$ and $\phi(u)$, which were shown to be regular, the regularity becomes manifest. To see that there is no discrepancy, consider once more the example of the invariant $\nabla^\mu \nabla_\mu R^{\rho \sigma} R_{\rho \sigma}$. This invariant will be dependent on $W''$. Since covariant derivatives are with respect to $u$, the terms that contain $W''$ will not diverge since

 \be
 \ddot{W}=2 W'' (W')^2 \sim \sqrt{\phi_B-\phi},
\label{w17} \ee
 modulo some multiplicative constants. Hence the vanishing of $W'$ at the bounce, cancels the divergence of $W''$.

In conclusion, we have shown that, as in the asymptotically AdS case studied in (\ref{bounces}), there exist points of the flow where the superpotential, $W(\phi)$ is not analytic and multivalued, where the geometry however is regular. This is due to the lack of monotonicity of $\phi(u)$ at the point of the bounce.

}

\subsection{On regular scalar clocks that go back in time\label{clocks}}

There are many time coordinates in any physical system, and in a theory of gravity, diffeomorphism invariance allows us to use a grand variety of such coordinates.

It is clear from physical experience, that time has a very different nature from the space coordinates. One of its characteristic properties is that it is perceived to run only in one direction, (known as the ``arrow of time") a physical puzzle that has received a lot of attention from physicists and laymen alike.

Locally time is measured by comparing to periods of periodic phenomena (the analogue of distance sticks), but in cosmology, in most cases  one can define global time, like in the ansatz we are using in (\ref{b1}).
In our ansatz $u$ is the global time.
There are however an infinite number of other ``clocks" we can use in such a universe, including the scalar factor $e^A(u)$ and the scalar $\phi(u)$ (as well as many other possible scalar variables.

A conventional clock in cosmology is given by the scale factor $e^A$ which according to the gravitational equations it can either increase indefinitely or decease indefinitely. Generically, if matter satisfies the null energy condition, it is not possible to reverse the direction of $e^A$ without going through a singularity.  Over the years, many efforts have been made to avoid source ``bounce singularities", but it is clear that the assumptions must dwell in the realm of science fiction, \cite{bounce1}. In string theory there are frameworks where effective bounces can be produced, \cite{top} but no realistic fully controllable solution exists so far.

Another class of cosmic clocks are scalar clocks, where time is given by the value of the scalar field $\phi(u)$. The inflaton is the canonical example in this class, and can be identified with our scalar field, $\phi$.
This can be made precise by choosing $\phi$ as a time coordinate in our metric (\ref{b1})
\be
ds^2=-{d\phi^2\over \dot\phi^2}+e^{2A(\phi)}d x_{\m}dx^{\m}=-{d\phi^2\over W'(\phi)^2}+e^{2A(\phi)}d x_{\m}dx^{\m}
\label{w18}\ee
It is clear that when $\dot\phi=0$ there is a coordinate singularity in the metric in question.
This is precisely the place where the ``bounce" happens in the solutions we have described in the previous subsection. At such points, time as measured by $\phi$ changes direction. The question is whether such a point has a (curvature) singularity or not.
Our analysis however of the question in the AdS regime, \cite{Bourdier13} can be repeated verbatim here, and has been mentioned in the previous subsection. The answer is that that such points have all their curvature invariants finite.
Therefore, they correspond to regular points at which the direction of the time coordinate ($\phi$ here) changes but the geometry remains regular there.

As is already known in the AdS regime, \cite{exotic}, there are many globally regular bouncing solutions that exist without fine tuning the scalar potential.
Such solutions can be turned into cosmological solutions by a simple change of the sign of the relevant potential. There is also some intuition on when such solutions may exist: this happens when the potential is steep, although no precise estimates exist on the steepness.
On the other hand, this condition seems to move in the same direction as the recently proposed swampland constraints, \cite{ov}-\cite{wrase}, which in their current form, for the single scalar case we are using and in supergravity units (where the scalar is dimensionless) it reads
\be
{|V'|\over V}\geq c ~~~{\rm or}~~~ {\rm min}~{V''\over V}\leq -c'
\label{w19}\ee
where $c$ and $c'$ are supposed to be ${\cal O}(1)$ positive numbers\footnote{There are also modifications of this, as the one in \cite{and2}.}.
We may therefore conclude that with current knowledge, the presence of scalar clocks in cosmology that can switch ``time" back and forth in a regular fashion seems to be allowed.

A final comment on such scalar clocks relates them to dS extrema that violate the analogue of the BF bound, as in (\ref{b17a}).
According to (\ref{b20b}) the solution for $\phi$ near a BF violating dS minimum is
\be
\phi(u)=\phi_-~e^{-{d\over 2}{Hu}}\left(\cos\left[\nu{Hu}+\theta_0\right]+{\cal O}\left(e^{-Hu}\right)\right)\sp t\to+\infty
\label{w72}\ee
 with
 \be
 \nu=\sqrt{{m^2\over H^2}-{d^2\over 4}}
\label{w73} \ee
and $\phi_-$,$\theta_0$ constants.
It is clear from (\ref{w72}) that $\phi$ changes sign an infinite amount of times before it reaches the dS fixed points at $u\to+\infty$, and $\phi\to 0$.
It is nor clear whether such extrema are allowed in consistent gravitational theories.

\section{Interpolating Ans{\"a}tze\label{inter}}

{
One of the main  purposes of this work is to study solutions that interpolate between asymptotically de Sitter and asymptotically Anti de Sitter space-times. Up until now we have only studied metric ans{\"a}tze that do not permit such behaviour. Indeed, the ans{\"a}tze,  (\ref{i2}) and (\ref{b1}), while having manifest $\text{ISO}(d)$ invariance, only contain one dynamical variable, namely the scale factor $A(u)$.

If we wish to allow for interpolating solutions, we must use a different coordinate system that allows for the coordinate $u$ to change from being spacelike (as is the case in the asymptotically AdS space-times) to timelike (as is the case in the asymptotically dS space-times). There is also an interpolating ansatz with a single scale factor that will be discussed below.

 We shall therefore introduce an additional dynamical variable, the \emph{blackness function} $f(u)$, which will give us this liberty. The general idea is the following : the vanishing of $f(u)$ yields  a horizon, on either side of which $f$ has a different sign. Therefore, a solution that passes through a horizon, for appropriately placed $f$ in the ansatz, exchanges $u$ from space-like to time-like and vice-versa.

In the rest of this section, we shall explore  three classes of ans{\"a}tze, one corresponding to a flat slicing, one corresponding to a spherical slicing, and finally, one corresponding to  dS slicing.

\bigskip

$\bullet$ \emph{The black hole ansatz with a flat slicing.} The corresponding metric is given by

\be
ds^2={du^2\over f(u)}+e^{2A(u)}\left[-f(u)dt^2+dx_{i}dx^{i}\right].
\label{c38}\ee

Notice that when $f=1$ and $e^{A}=e^{-{u\over \ell}}$, (\ref{c38}) reduces to AdS space in Poincar\'e coordinates (where $u$ is space-like), while with $f=-1$ and  $e^{A}=e^{-{u\over \ell}}$ it reduces to dS space in Poincar\'e coordinates (where $u$ is now time-like).

\bigskip

$\bullet$ \emph{The black hole ansatz with a spherical slicing.} This is similar to the previous ansatz, with the substitution $dx_{i}dx^{i} \rightarrow d\Omega^2_{d-1}$. Hence, the corresponding metric is given by

\be
ds^2={du^2\over f(u)}+e^{2A(u)}\left[-f(u)dt^2+R^2~d\Omega_{d-1}^2\right]
\label{c39}\ee

The radius of the sphere $R$ is not of physical significance, as it may be changed at will, by a shift in $A$ and a rescaling of $t$.

To see that AdS space-time can be obtained from this metric, set
\be
e^A=e^{-{u\over \ell}}\sp f=1+e^{2{u\over \ell}}\sp R=\ell.
\label{w22}\ee

Making the coordinate transformation $r=\ell~e^{-{u\over \ell}}$, this is mapped to the static patch metric of AdS in  (\ref{a35}).
We can also write AdS in global coordinates in this ansatz.
If we set
\be
e^{A}=\sinh{\rho}\sp f=\coth^2\rho\sp \coth\rho d\rho=du
\ee
we obtain AdS in global coordinates, as in (\ref{a30_1})

To obtain a dS space-time instead, we must set
\be
e^{A}=e^{Hu}\sp f=-1+e^{-2Hu}\sp R={1\over H}
\label{w23}\ee
 which can be mapped to the static patch metric in (\ref{a27}) by the coordinate transformation $Hr=~e^{{Hu}}$. This ansatz is dual to thermal states in QFTs defined on $S^{d-1}\times R$.
 We shall not study solutions related to  this ansatz in the present paper.

\bigskip

$\bullet$ \emph{The dS sliced ansatz.} This ansatz is given by the metric
\be
ds^2=du^2+e^{2A(u)}\left(-dt^2+{\cosh^2(Ht)\over H^2}d\Omega_{d-1}^2\right).
\label{w24}\ee

By choosing $e^A=\sinh{u\over \ell}$ we obtain AdS as in (\ref{a37}), whereas setting $e^A=\sin{Hu}$ we obtain dS as in (\ref{a29}). We shall not study solutions that arise from this ansatz in this paper.

In the next two sections, we shall study solutions and their properties that arise from the first two ansatz classes, for flat and for spherical slices.
}
\section{Black-hole ansatz with a flat slicing\label{flat}}

{ We now turn our attention to the case of the flat slicing. As seen in the previous section, this corresponds to the ansatz (\ref{c38}), that we reproduce here,
\be
ds^2={du^2\over f(u)}+e^{2A(u)}\left[-f(u)dt^2+dx_{i}dx^{i}\right].
\label{e1}\ee

Using the action (\ref{i1}), we derive the corresponding Einstein equations, which are found to be}

\begin{subequations}\label{e2}
\begin{align}
	&2(d-1)\ddot{A}(u)+\dot{\phi}^2(u)=0,\label{e2a}\\
	&\ddot{f}(u)+d\dot{f}(u)\dot{A}(u)=0\label{e2b}\\ &(d-1)\dot{A}(u)\dot{f}(u)+f(u)\left[d(d-1)\dot{A}^2(u)-\frac{\dot{\phi}^2}{2}\right]+V(\phi)=0.\label{e2c}
\end{align}
\end{subequations}
\be
\ddot\phi(u)+\left(d\dot A(u)+{\dot f(u)\over f(u)}\right)\dot\phi(u)-{V'\over f(u)}=0
\label{e2d}\ee
As noted in the previous section, for $f(u)=-1$ equations (\ref{e2}) reduce to (\ref{b2}), by construction.
Equation (\ref{e2b}) can be integrated once to obtain
\be
 \dot f=C~e^{-dA}
\label{e2e}\ee
where $C$ is an integration constant.

Using (\ref{e2a}) we may rewrite (\ref{e2c}) as
\be
(d-1)\left[\dot f\dot A+f(d\dot A^2+\ddot A)\right]+V=0
\label{e3}\ee
or equivalently
\be
(d-1){d\over du}\left(f\dot A~e^{dA}\right)=-V~e^{dA}
\label{e4}\ee
Similarly, (\ref{e2d}) can be written as
\be
{d\over du}\left(f\dot \phi~e^{dA}\right)=V'~e^{dA}
\label{e5}\ee

For $f(u)=1$, and $A(u)=u/\ell$, (\ref{e1}) corresponds to an AdS metric, with AdS radius $\ell$, whereas for a dS metric we have $f(u)=-1$ and
\be
A(u)=u H\sp  H={1\over \ell}~~\;,
\ee
 where again $\ell$ is the dS radius.

\subsection{The superpotential formalism}

We introduce the superpotential as usual, with
\eql{e26}{W(\phi) \equiv -2(d-1)\dot A (u)\sp \dot\phi=W'\;.}
We can also translate (\ref{e2b}) to an equation over $\phi$ using
\be
\pa_u=W'(\phi)\pa_{\phi}
\label{w29}\ee
It becomes
\be
W'f''+\left(W''-{d\over 2(d-1)}W\right)f'=0
\label{e28}\ee
which can be integrated once to obtain
\be\label{e28_a}
f'={1\over W'}\exp\left[{d\over 2(d-1)}\int d\phi {W\over W'}\right]
\ee

From (\ref{e26}) and (\ref{e2c}) we obtain

\eql{e29}{\bigg(\frac{d}{4 (d-1)}W^2(\phi)-\frac{(W'(\phi))^2}{2}\bigg)f(u)-\frac{\dot{f}(u)}{2} W(\phi)+V(\phi)=0.}
or equivalently
\be
{\bigg(\frac{d }{4 (d-1)}W^2(\phi)-\frac{(W'(\phi))^2}{2}\bigg)f(\phi)-{W'\over 2}f'(\phi) W(\phi)+V(\phi)=0.}
\label{e30}\ee

Equation (\ref{e2b}) can be rewritten as
\be
f'={C\over W'}e^{-dA}
\label{w29a}\ee
and (\ref{e30}) becomes
\be
{\bigg(\frac{d }{4 (d-1)}W^2(\phi)-\frac{(W'(\phi))^2}{2}\bigg)f(\phi)-{C\over 2} W(\phi)e^{-dA}+V(\phi)=0.}
\label{e31}\ee

Further, by integrating equation (\ref{e2b}), we obtain

\eql{e32}{f(u)=\int_{-\infty}^{u} du'\exp \bigg[ \frac{d}{2(d-1)} \int_{\phi_*}^{\phi(u')}d\phi' \frac{W(\phi')}{W'(\phi')} \bigg]+1,}
having set $f(u) \rightarrow 1$ as $u \rightarrow -\infty$, and $\phi_*$ is a constant of integration.

We can finally combine the equations to write a single equation for $W$ as

\be
b_0V+b_1V'+b_2V''=0
   \label{e33}\ee
with
\be
b_0=2 W'(W''(2(d-1)W''-dW)+d W'^2-2(d-1) W^{(3)}W')
   \label{e34}\ee
\be
b_1=d W^2 W'' - 4 (d-1) W'^2 W'' + W (-d W'^2 + 2 (d-1) W''^2
   \label{e35}\ee
   $$+ 2 (d-1) W' W^{(3)})$$

\be
b_2=2 (d-1) W'(W'^2 - W W'')
      \label{e36}\ee

      We also obtain
      \be
      f={2 V W'-WV' \over W'(W'^2-WW'')}
      \label{e37}\ee

The singular points of the equations are points where the coefficient of the leading derivative $W^{(3)}$ in (\ref{e33}) vanishes. This coefficient is $W'(2 V W'-WV')$. Points where $W'$ vanishes correspond as usual to extrema of the potential or bouncing points. When $2 V W'-WV'$ vanishes, $f$ from (\ref{e37}) vanishes and therefore such points are horizons.

The local analysis of solutions and their regularity can found in appendix \ref{flats}.

\subsection{The asymptotic solutions} \label{asymptotic}

To complete the description, in this section we shall study the asymptotic behaviour of solutions for two cases, where $f$ becomes asymptotically constant, as well as the more general case where this restriction is lifted.

\subsubsection{Constant $f$ solutions}

Consider a solution moving off to $\pm \infty$ in a directions that $V\to -\infty$ with $f>0$ constant\footnote{The case where the potential vanishes asymptotically is another possibility that may describe some moduli behavior in string theory. As in this paper we assume no Minkowski extrema, we do not consider such a case.}.
By rescaling $f$ to 1 we must find the asymptotic solutions of
\be
{d~W^2
\over 4(d-1)}-{W'^2\over 2}+V=0
\label{as1}\ee

The generic and dominant solution is
\be
W^+_{s}=e^{\sqrt{d\over 2(d-1)}\phi}\left[W_0-{1\over W_0}\int d\phi e^{-\sqrt{d\over 2(d-1)}\phi}V(\phi)+\cdots\right]
\label{e40}\ee
valid when the potential grows slower at infinity than $W_s$.
It is a one-parameter parameter of solutions and is ``singular" in the generalized (Gubser) sense, \cite{Gubser,thermo,cgkkm}.
For this solution we obtain
\be
\lim_{\phi\to\pm\infty}{\dot \phi\over \dot A}=-\sqrt{2d(d-1)}
\label{as3}\ee
Note that the leading part of this solution is independent of $V$ and it exists independent of the asymptotics of $V$.

There is another branch where the solution  is
\be
W^-_{s}=e^{-\sqrt{d\over 2(d-1)}\phi}\left[W_0+{1\over W_0}\int d\phi e^{\sqrt{d\over 2(d-1)}\phi}V(\phi)+\cdots\right]
\label{as4}\ee
and this is again singular in the Gubser sense.

The regular solution is obtained as follows. Set
\be
\kappa=\lim_{\phi_{\pm \infty}}{V'\over V}\sp \kappa\leq \sqrt{2d\over d-1}
\label{as5}\ee
where the inequality is required for the Gubser bound to be respected.
This is also valid for $\kappa\leq 0$.

In that case, the tuned (regular) solution reads
\be
W_0={\sqrt{-V}\over \sqrt{{d\over 4(d-1)}-{\kappa^2\over 8}}}+C\exp\left[{d\over d-1}\int d\phi~ {V\over V'}\right]+\cdots
\label{as6}\ee
We note that when the Gubser bound for the potential is violated, this solution does not exist.
For this solution we obtain
\be
\lim_{\phi\to\pm\infty}{\dot \phi\over \dot A}=-(d-1)\kappa
\label{as7}\ee

\bigskip

\subsubsection{The non-constant $f$ solutions}

\bigskip

We shall now consider the potential with an asymptotic form as in (\ref{as5})
which implies that asymptotically
\be
V\simeq -V_0~e^{\kappa\phi}+\cdots
\label{as7_1}\ee
and the asymptotic solution of the equations
\be
	2(d-1)\ddot{A}(u)+\dot{\phi}^2(u)=0\sp
\dot f=C~e^{-dA}
\label{as8}\ee
 and
\be
(d-1){d\over du}\left(f\dot A~e^{dA}\right)=-V~e^{dA}\sp
{d\over du}\left(f\dot \phi~e^{dA}\right)=V'~e^{dA}
\label{as12a}\ee

We also consider the asymptotic ansatz, justified in \cite{cgkkm},
\be
\dot A={a\over u-u_0}\sp \dot \phi={b\over u-u_0}
\label{as9}\ee
with $u\to u_0$ and $a>0$.
From (\ref{as9}) we obtain
\be
A=A_0+a\log(u-u_0)\sp \phi=\phi_0+b\log(u-u_0)
\label{as13}\ee
Equation (\ref{as8}) gives
\be
a={b^2\over 2(d-1)}\sp f=f_0+{C'\over 1-ad}(u-u_0)^{1-ad}
\label{as14}\ee

We must distinguish two cases:

\begin{itemize}

\item $1-ad>0\to ad<1$.

In that case $f\to f_0$ as $u\to u_0$ and we obtain from (\ref{as12a})
\be
b\kappa=-2\sp \-a(d-1)(da-1)f_0=V_0~e^{\kappa \phi_0}\sp b(da-1)f_0=-\kappa V_0 ~e^{\kappa \phi_0}
\label{as15}\ee
and finally
\be
a={2\over (d-1)\kappa^2}\sp b=-{2\over \kappa}\sp {f_0\over V_0}={\kappa^2\over 2}~{e^{\kappa \phi_0}\over {2d\over (d-1)\kappa^2}-1}
\label{as16}\ee

We now check that
\be
ad={2d\over (d-1)\kappa^2}\geq 1
\label{as17}\ee
because of the Gubser criterion in (\ref{as5})
Therefore,  this solution is not self-consistent.

\item $1-ad<0\to ad>1$.In that case $f\to \infty$ as $u\to u_0$ and we obtain from (\ref{as12a})
the same solution as above, but now the Gubser bound is valid.

\end{itemize}

The solution (\ref{as16}) is in that language
\be
W=-{4\over \kappa^2}~e^{-{\kappa\over 2}(\phi-\phi_0)}+\cdots\sp f=f_0+{(d-1)\kappa^2~C'\over (d-1)\kappa^2-2d}~e^{-{(d-1)\kappa^2-2d\over 2(d-1)\kappa}(\phi-\phi_0)}+\cdots
\label{as17_1}\ee

To check if there are other solutions,  we study the superpotential equation

\be
{\bigg(\frac{d }{4 (d-1)}W^2(\phi)-\frac{(W'(\phi))^2}{2}\bigg)f(\phi)-{W'\over 2}f'(\phi) W(\phi)+V(\phi)=0.}
\label{as18}\ee
\be
W'f''+\left(W''-{d\over 2(d-1)}W\right)f'=0~~~\to~~~~
f'={C\over W'}e^{-dA}
\label{w34}\ee
and examine the previous solutions with constant $f$.
In the solution (\ref{e40}) we find that
\be
e^{dA}\sim {1\over W}
\label{w35}\ee
and therefore
\be
f \sim f_0+{C e^{-dA_0}\over QW_0}\phi\sp Q=\sqrt{d\over 2(d-1)}
\label{w36}\ee
However, this is not a consistent solution of the superpotential equations, because the linear correction to $f$ is leading as $\phi\to \infty$ instead of subleading.
{To summarize, the asymptotic behaviour of the solution generically, for $f$ not approaching a constant value, is given by (\ref{as17_1}).}

\subsection{Curvature invariants and regularity of solutions} \label{cr_invariants_flat}

For the metric (\ref{e1}), we calculate some invariants in terms of $A,f,\phi$ and their derivatives. The scalar curvature is given by

\be\label{e39}
R=-(2d+1)\dot{A}\dot{f} -2df\bigg( \frac{d+1}{2} \dot{A}^2+\ddot{A}\bigg)-\ddot{f}\;.
\ee
Using the equations of motion (\ref{e2a})-(\ref{e2d}) we can simplify
\be\label{e39_1}
R={d+1\over d-1}V+{f\over 2}\dot\phi^2={d+1\over d-1}V+{fW'^2\over 2}
\ee

Therefore, finite Ricci scalar implies that $fW'^2$ must be finite.
Further, we have

\be\label{e41}
R_{\mu \nu} R^{\mu \nu}=\frac{1}{4}\Big\{ 4(d-1)\left[\dot{A} \dot{f}+f (d\dot{A}^2+\ddot{A})\right]^2
+\left[(d+2)\dot{A} \dot{f} +2df(\dot{A}^2+\ddot{A})+\ddot{f}\right]^2
\ee
$$
+\left[(d+2)\dot{A} \dot{f}+2 f (d\dot{A}^2+\ddot{A})+\ddot{f}\right]^2\Big\}=
$$
\be
={(d+1)\over (d-1)^2}V^2+{V\over d-1}f\dot\phi^2+{f^2\over 4}\dot\phi^4={(d+1)\over (d-1)^2}V^2+{V\over d-1}fW'^2+{(fW'^2)^2\over 4}
\ee
where in the last step we used the equations of motion.

We also have,
\be \label{e42}
\partial_\mu \phi \partial^\mu \phi=f \dot{\phi}^2=fW'^2\;.
\ee
Finally,
\begin{align} \label{e43}
R_{\mu \nu \rho \sigma}R^{\mu \nu \rho \sigma}&=(7+2d)\dot{A}^2 \dot{f}^2+2df^2\left[(d+1) \dot{A}^4+4 \dot{A}^2 \ddot{A}+2\ddot{A}^2\right]\\ \nonumber
&+6 \dot{A} \dot{f} \ddot{f}+\ddot{f}^2+4f\Big\{\dot{A}\dot{f}\left[(2d+1)\dot{A}^2+(d+2)\ddot{A}\right] + (\dot{A}^2+\ddot{A})\ddot{f}\Big\}=
\end{align}
$$
={d^2-4d+7\over (d-1)^2}V^2+{fV\over d-1}\left(2(d-1)^2(d-2)\dot A^2-(d-3)\dot\phi^2\right)+
$$
$$+
{f^2\over 4(d-1)}\left[4d(d-1)^3(d-2)\dot A^4-4(d-1)^2(d-2)\dot A^2\dot\phi^2+(d+1)\dot\phi^4\right]
$$
\be
={d^2-4d+7\over (d-1)^2}V^2+{fV\over d-1}\left[{(d-2)\over 2}W^2-(d-3)W'^2\right]+
\label{w37}\ee
$$+
{f^2\over 4(d-1)}\left[{d(d-2)\over 4(d-1)}W^4-(d-2)W^2W'^2+(d+1)W'^4\right]
$$

We therefore observe that the invariants are polynomials of $V$, $f W'^2$ and $f W^2$.
For constant $f$, it was shown in \cite{Bourdier13} that the curvature invariants diverge only when either $V(\phi)$ or $W(\phi)$ diverge. $V(\phi)$ can only diverge at $\phi=\pm\infty$. $W(\phi)$ cannot diverge without $V(\phi)$ diverging. Therefore,  any divergence happens at $\phi\to \pm \infty$.

When $f$ is not constant then the system of equations (\ref{e28}), (\ref{e30}) is regular everywhere except at points when $W'=0$ or $f=0$. The first of these cases was studied for $f=constant$ and corresponds to bounces when $V'\not=0$, that are regular.
This is further analyzed in appendix \ref{flats}.

It also corresponds to extrema of the potential, $V'=0$. In that case, the minus solutions around minima (AdS) or maxima (dS) are singular when $f$ is not constant.

The second corresponds to horizons and in that case the two local solutions of the equations degenerate to one.
Therefore, even when $f$ is not constant, both $W,W'$ and $f$ are regular at regular values of $\phi$. We conclude that curvature singularities can only happen
at $\phi\to \pm \infty$ or at appropriate extrema of the potential.

\bigskip

\subsection{Solutions near maxima and minima of the potential} \label{local}

\bigskip

Before we begin the study near the extrema of the potential properly, we look at the local behaviour of $f(\phi)$ near these points. Here, as above, we can set these points at $\phi=0$ by a shift of the scalar field. Therefore, we look at a region where the potential is expanded as

\be
V(\phi)\simeq -{d(d-1)\over \ell^2}+{m^2\over 2}\phi^2+\mathcal{O}(\phi^3)
\label{e47}\ee

First, we set
\be
W(\phi) \simeq W_0+W_2 \phi^2+{\cal O}(\phi^3)+\cdots
\ee
 We require $W_1=0$ because we are interested in flows that either start or end at the critical points of the potential.

Then, by (\ref{e32}), changing variables to $\phi$ we obtain for $f$ in terms of the superpotential,

\eql{e44}{f(\phi)=\int d\phi \exp \bigg[ \frac{d}{2(d-1)} \int d\phi \frac{W(\phi)}{W'(\phi)} \bigg]\frac{1}{W'(\phi)}.}

Upon performing the integral, we have,

\eql{e45}{f(\phi) = C_1+ C_2 \phi^{\frac{d W_0}{4(d-1)W_2}} \bigg( \frac{4(d-1)W_2}{d W_0}+ \mathcal{O}(\phi) \bigg),}
where $C_1, C_2$ are constants of integration. The ratio $W_0/W_2$ can be determined from equation (\ref{e33}). Indeed, setting $W(\phi) \simeq W_0+W_2 \phi^2+\mathcal{O}(\phi^3)$ one acquires, for the cases (\ref{e47}), (\ref{e49}), (\ref{e50}), (\ref{e51}) (listed individually in the following subsections):

\be
\frac{d W_0}{4(d-1)W_2}=\frac{d}{\Delta_\pm},
\label{w44}\ee

where $\Delta_\pm$ is defined as in (\ref{i23}).

Therefore, $f$ can be written as

\eql{e46}{f(\phi) = 1+ C_2 \phi^{\frac{d}{\Delta_\pm}} \bigg( \frac{\Delta_\pm}{d}+ \mathcal{O}(\phi) \bigg),}

We first turn our attention to the cases of maxima with negative potential, and minima with positive potential. In these cases $\Delta_\pm>0$, and $f$ is well-behaved at the point $\phi=0$.

\subsubsection{Maximum in the AdS regime}

First we study the potential (\ref{e47}) in the case of a maximum, that is

\be
V(\phi)\simeq -{d(d-1)\over \ell^2}+{m^2\over 2}\phi^2+\cdots\sp m^2<0.
\label{ee47}\ee

From equation (\ref{e30}), we have that $W(0)=\frac{2(d-1)}{\ell}$, by taking a positive superpotential. Following the discussion of the previous section, we determine $W_2=\frac{\Delta_\pm}{2 \ell}$, as expected.

We now look for possible subleading irregular terms in $W$. To do this, suppose that $W(\phi)$ is a solution of (\ref{e30}), and add a perturbation $W \rightarrow W + \delta W$. Then to linear order in $\delta W$, (\ref{e30}) becomes

\be
\delta W' \bigg[ W' f + \frac{1}{2} W f'\bigg]-\delta W \bigg[ \frac{d W}{2(d-1)} f - \frac{1}{2} f' W' \bigg]=0,
\label{e48}\ee
whose solution is found to be

\be
\delta W_- \sim \phi^{\frac{d}{\Delta_-}},
\label{w45}\sp
\delta W_+ \sim 1,
\ee
to leading order.

The situation here is essentially the same as in the AdS RG flows; only the perturbation $\delta W_-$ is subleading with respect to the unperturbed solution. Therefore, once again, the solutions are the same as (\ref{i22_a}) and (\ref{i22_b}), namely

\begin{subequations}			\label{e49}
\begin{align}
	W_+(\phi)
			=&{1\over \ell}
			\le[
				{2(d-1)} + {\Delta_+\over2}\phi^2+\cO(\phi^3)
			\ri], 			\label{e49a}\\
	W_-(\phi)
			=&{1\over \ell}
			\le[
				{2(d-1)} + {\Delta_-\over2}\phi^2+\cO(\phi^3)
			\ri]+C|\phi|^{d/\Delta_-}\le[1 +\cO(\phi)\ri]
			+\cO(C^{2})
			\label{e49b},\\
	\Delta_\pm =&\ha \le( d \pm \sqrt{d^2 +4 m^2\ell^2}  \ri) \quad \text{with}\quad -{d^2\over 4\ell^2}<m^2<0,
	\label{e49c}
\end{align}
\end{subequations}

Hence, $\phi(u)$ utilizing (\ref{e26}) is

\begin{subequations}\label{e50}
	\begin{align}\label{e50a}
		&\dot\phi(u)=W_+'(\phi)
			\implies
		\phi(u)=\phi_+ e^{\Delta_+ u/\ell}+\cdots\\
		&\dot\phi(u)=W_-'(\phi)
			\implies
		\phi(u)=\phi_- e^{\Delta_- u/\ell}
			+{
				d~C\ell
					\over
				\Delta_-(2\Delta_+-d)}~\phi_-^{\Delta_+/\Delta_-}~
			e^{\Delta_+u \ell}+\cdots\label{e50b}
	\end{align}
\end{subequations}
where $\phi_+$ and $\phi_-$ are integration constants. Similarly $A(u)$ is given by

\begin{subequations}\label{e51}
	\begin{align}
		&\dot A(u)=-{1\over 2(d-1)}W_+(\phi)
			\implies
		A(u)=-{u-u_*\over \ell}+{\phi^2_+ \over 8(d-1)} e^{2\Delta_+ u/\ell}
		+\cO\le( e^{3\Delta_+u/\ell}\ri)
		\label{e51a}\\
		&\dot A(u)=-{1\over 2(d-1)}W_-(\phi)
			\implies
		A(u)=-{u-u_*\over \ell}+ {\phi^2_- \over 8(d-1)} e^{2\Delta_- u/\ell}
		+\cO\le(C e^{ud/\ell}\ri)
		\label{e51b}
	\end{align}
\end{subequations}

Again, we take $u \rightarrow -\infty$, because we are interested in small values of $\phi$. As before, the scale factor diverges. Since $f$ approaches unity in that limit, the metric describes an asymptotically AdS space-time near the boundary, and the solution describes a flow leaving a UV fixed point.

Hence, the case of maxima of a negative potential in these coordinates is identical the holographic RG flows, studied in Chapter 1.

\subsubsection{Minimum in the dS regime}

Here, we examine the case of a potential of the form

\be
V(\phi)\simeq {d(d-1)H^2}+{m^2\over 2}\phi^2+\cdots\sp m^2>0.
\label{e52}\ee

We now express $f(\phi)$ as

\eql{e53}{f(\phi) = -1 + C_2 \phi^{\frac{d}{\Delta_\pm}} \bigg( \frac{\Delta_\pm}{d}+ \mathcal{O}(\phi) \bigg),}
{where in (\ref{e45}) we have set $C_1=-1$, } so that indeed for $\phi \rightarrow 0$, the space is asymptotically dS with a canonically normalized time coordinate.

Now from equation (\ref{e30}), we have that $W(0)=-2(d-1)H$, by choosing a negative superpotential. It follows that $W_2=-\frac{\Delta_\pm}{2}H$.

In order to look for allowable non-regular deformations of the solutions we follow the standard procedure. Once again, it follows that the possible deformations are, to leading order in $\phi$, $\delta W_- \sim \phi^{\frac{d}{\Delta_-}}$ and $\delta W_+ \sim 1$, where $\Delta_\pm =\ha \le( d \pm \sqrt{d^2 - 4 m^2 H^2}  \ri)$. The only allowable deformation is $\delta W_-$.

Therefore, the form of the solutions is the same as previously studied, in the context of the cosmological evolution

\begin{subequations}			\label{e54}
\begin{align}
	W_+(\phi)
			=&-H
			\le[
				{2(d-1)} + {\Delta_+\over2}\phi^2+\cO(\phi^3)
			\ri], 			\label{e54a}\\
	W_-(\phi)
			=&-H
			\le[
				{2(d-1)} + {\Delta_-\over2}\phi^2+\cO(\phi^3)
			\ri]+C|\phi|^{d/\Delta_-}\le[1 +\cO(\phi)\ri]
			+\cO(C^{2}).
			\label{e54b}
\end{align}
\end{subequations}

By (\ref{e26}) we obtain

\begin{subequations}\label{e55}
	\begin{align}\label{e55a}
		&\dot\phi(u)=W_+'(\phi)
			\implies
		\phi(u)=\phi_+ e^{-\Delta_+ H u}+\cdots\\
		&\dot\phi(u)=W_-'(\phi)
			\implies
		\phi(u)=\phi_- e^{-\Delta_- H u}
			+{
				d~C
					\over
				H\Delta_-(2\Delta_+-d)}~\phi_-^{\Delta_+/\Delta_-}~
			e^{-\Delta_+ H u}+\cdots\label{e55b}
	\end{align}
\end{subequations}
where $\phi_+$ and $\phi_-$ are integration constants. Notice that here $u$ is a timelike coordinate, and so equations (\ref{e55}) are the same with equations (\ref{field}).

For the scale factor,

\begin{subequations}\label{e56}
	\begin{align}
		&\dot A(u)=-{1\over 2(d-1)}W_+(\phi)
			\implies
		A(u)=H(u-u_*)+{\phi^2_+ \over 8(d-1)} e^{-2\Delta_+ H u}
		+\cO\le( e^{-3\Delta_+ H u}\ri)
		\label{e56a}\\
		&\dot A(u)=-{1\over 2(d-1)}W_-(\phi)
			\implies
		A(u)=H(u-u_*)+ {\phi^2_- \over 8(d-1)} e^{-2\Delta_- H u}
		+\cO\le(C e^{-u d H}\ri)
		\label{e56b}
	\end{align}
\end{subequations}

where $u_*$ is an integration constant. For small $\phi$ we must take $u \rightarrow + \infty$, therefore,  the very first term of $A(u)$ dominates, leading the black hole metric to be asymptotically dS. This situation again describes a flow leaving a ``UV fixed point", using the intuition provided by RG flows.

Once again, we see that the discussion of this extremum in this coordinate system, is essentially not different from the one with the asymptotically dS ansatz, when we studied the cosmological evolution.

\subsubsection{Minimum in the AdS regime}
Now we turn our attention to the potential
\be
V(\phi)\simeq -{d(d-1)\over \ell^2}+{m^2\over 2}\phi^2+\cdots\sp m^2>0.
\label{e57}\ee

As mentioned before, in this case we have

\be
\frac{d}{\Delta_-}<0 \sp 0<\frac{d}{\Delta_+}<1.
\label{w48}\ee

From (\ref{e45}), the solution for $f$  with $d / \Delta_-$ diverges. In that case, the solution is no longer asymptotically AdS, unless we fix the constant $C_2=0$. This makes $f$ globally constant.
We conclude that the minus solution at an AdS minimum is regular only if $f$ is constant.

On the other hand, for the plus solution, (\ref{e45}) and (\ref{w48}) indicate that $f$ is regular.

Again, we have two possible deformations of the $W$ solution, $\delta W_- \sim \phi^{\frac{d}{\Delta_-}}$ and $\delta W_+ \sim 1$. However, here none is subleading with respect to the regular part of $W$, and so is not an allowable deformation. The solutions therefore are

\begin{subequations}\label{31}
\begin{align}
	W_\pm(\phi)
			=&{1\over \ell}
			\le[
				{2(d-1)} + {\Delta_\pm\over2}\phi^2+\cO(\phi^3)
			\ri],
			\label{e58}\\
\Delta_\pm =&\ha \le( d \pm \sqrt{d^2 +4 m^2\ell^2} \ri)\quad \text{with}\quad m^2>0. \label{e59}
\end{align}
\end{subequations}

Notice that the $W_+$ solution has a minimum at $\phi=0$ while $W_-$ has a maximum.

Integrating equations (\ref{e26}) we obtain

\begin{subequations} \label{e60}
\begin{align}
		&\dot\phi(u)=W_+'(\phi)
			\implies
		\phi(u)=\phi_+ e^{\Delta_+ u/\ell}+... \label{e60a}\\
		&\dot\phi(u)=W_-'(\phi)
			\implies
		\phi(u)=\phi_- e^{\Delta_- u/\ell}
			+...  \label{e60b}\\
	&A_\pm(u)=-{u-u_*\over \ell}
				- {1\over 8(d-1)}{\phi_\pm^2}e^{2\Delta_\pm u/ \ell}+...
\label{w49}\end{align}
\end{subequations}
where $\phi_\pm$ and $u_*$ are integration constants.

For small $\phi$, we take $u \rightarrow - \infty$ for the $W_+$ solution, while we take $u \rightarrow + \infty$ for the $W_-$ solution.
Therefore, the $W_+$ solution has a scale factor that diverges as one approches the extremum (and the solution is regular) while the $W_-$ solution has a vanishing scale factor, and a divergent $f$ function (if it is not constant), and the solution is singular.

\subsubsection{Maximum in the dS regime}
Our final case is a potential of the form
\be
V(\phi)\simeq {d(d-1)H^2}+{m^2\over 2}\phi^2+\cdots\sp m^2<0.
\label{e61}\ee

We treat this in the same way we treated (\ref{e57}). We therefore find that the possible solutions are (choosing a negative W)

\begin{subequations}\label{e62}
\begin{align}
	W_\pm(\phi)
			=&-H
			\le[
				{2(d-1)} + {\Delta_\pm\over 2}\phi^2+\cO(\phi^3)
			\ri],
			\label{e62a}\\
\Delta_\pm =&\ha \le( d \pm \sqrt{d^2 -4 m^2\ell^2} \ri)\quad \text{with}\quad m^2<0. \label{e62b}
\end{align}
\end{subequations}

Where $W_+$ has a minimum and $W_-$ has a maximum, at $\phi=0$
The backness function $f$ is still given by  (\ref{e45}).

For $\phi$ and $A$ we have

\begin{subequations} \label{e63}
\begin{align}
		&\dot\phi(u)=W_+'(\phi)
			\implies
		\phi(u)=\phi_+ e^{-\Delta_+ H u}+... \label{e63a}\\
		&\dot\phi(u)=W_-'(\phi)
			\implies
		\phi(u)=\phi_- e^{-\Delta_- H u}
			+...  \label{e63b}\\
	&A_\pm(u)=H(u-u_*)
				- {1\over 8(d-1)}{\phi_\pm^2}e^{-2\Delta_\pm H u}+...
\label{w50}\end{align}
\end{subequations}

For small $\phi$, we take $u \rightarrow + \infty$ for $W_+$ and $u \rightarrow -\infty$ for $W_-$. In the former case, the plus solution has a diverging scale factor (that corresponds to a ``large" universe), while the minus solution  correspond to a scale factor vanishing (a universe undergoing a ``big-bang"). In this second case, if $f$ is not constant, there is a curvature singularity.

\subsection{Basic properties of interpolating solutions} \label{properties}

We shall use the convention that we take as the direction of the flows to be the direction in which the potential $V$ decreases (or  $W$ decreases).
Our goal here is to describe flows that start in the dS regime and eventually flow to an extremum of the AdS regime.

We first investigate when the flow of the scalar $\phi$ ends. Recall that $\phi$ satisfies a second order equation and therefore,  for $\phi$ to stop, both $\ddot\phi$ and $\dot \phi$ should be zero at the same point.
Looking at the equation for $\phi$ (\ref{e2d}), we observe that for $f\not=0$ it is a regular second order equation. Therefore,  away from a horizon,  for the $\phi$ flow to stop,  both $\ddot\phi$ and $\dot \phi$ should be zero at the same point.
For this to happen , $\ddot A=0$ from (\ref{e2a}) which means that near this point the scale factor is that of AdS (linear in u) and $V'=0$ from (\ref{e2d}).\footnote{Note that $\dot \phi=W'=0$ near bounces, but in such cases, $\ddot\phi\not=0$ and the flow does not stop.}

The case where we have a horizon, is being  discussed in appendix \ref{structure}. In that case, the $\phi$ flow does not stop, except if the horizon happens at an extremum of the potential.
Therefore, in all cases, flows stop only at extrema of the potential, or go off to infinity.

It is important to recall that the scale factor $e^A$  behaves as follows for $W_-$ flows

\begin{enumerate}

\item  $u\to -\infty$, $e^A\to 0$ at a maximum in dS regime (no tuning parameter).

\item  $u\to +\infty$, $e^A\to +\infty$ at a minimum in dS regime.

\item   $u\to +\infty$, $e^A\to 0$ at a minimum  in AdS regime (no tuning parameter).

\item  $u\to -\infty$, $e^A\to +\infty$ at a maximum in AdS regime.

\end{enumerate}

For the + flows we have instead\footnote{We shall not eventually consider here the $W_+$ flows further, as we know that they need special potentials in order for them  to be regular.}

\begin{enumerate}

\item[5] $u\to +\infty$, $e^A\to +\infty$ at a maximum in dS regime (no tuning parameter).

\item[6] $u\to +\infty$, $e^A\to +\infty$ at a minimum in dS regime (no tuning parameter).

\item[7]  $u\to -\infty$, $e^A\to +\infty$ at a minimum  in AdS regime (no tuning parameter).

\item[8] $u\to -\infty$, $e^A\to +\infty$ at a maximum in AdS regime (no tuning parameter).

\end{enumerate}

We have noted that {near dS extrema,  $\dot A>0$ and $f<0$} where we take $u$ to increase as the potential decreases.

 { Near AdS extrema, $\dot A<0$ and $f>0$,  where we take $u$ to increase as the potential decreases.}

Therefore, in a flow that interpolates between dS and AdS, we must have $\dot A$ vanish somewhere along the flow. This is a point where also the superpotential will vanish because of (\ref{e26}). Also $f$ must vanish somewhere along the flow so that coordinate $u$ turns from timelike to spacelike.

At a point where $\dot A=0$ (\ref{e2b}) implies that $\ddot f=0$ and (\ref{e2c}) implies that $2V=f\dot \phi^2$ so that $fV>0$. Therefore,  if this happens in the dS regime, $V>0$ then $f>0$ while if it happens in the AdS regime, $V<0$ then $f<0$.

This implies that if $\dot A=0$ happens in the dS regime ($V>0$) then $f$ must have changed sign earlier (as it started negative at the dS extremum from which the flow originates), and therefore there was a horizon ($f=0$) earlier in the flow.
If this transition happens in the AdS regime $V<0$ then $f$ must vanish later in the flow.

From (\ref{e2b}) we learn that $\dot f$ is either positive or negative along the whole flow, (or identically zero). This depends on the sign of the integration constant $C$. Therefore, $f$ is monotonic along the flow and it can vanish at most once during a flow.

At a horizon (\ref{e15}) gives that
\be
\dot \phi \dot f\sim V'~~~~{\rm and}~~~~\dot A\dot f\sim -V\;.
\label{ae1}   \ee
   Therefore,
\be
{d\phi\over dA}\sim -{V'\over V}\;.
\label{ae2}
\ee

We also learn from (\ref{e2b}) that at extrema where $e^A\to 0$ (this includes AdS minus minima, and dS  minus maxima), for regularity we must have $C=0$ and therefore flows in their neighbourhood must have $f$=constant.
Therefore, flows that arrive or depart at dS maxima or AdS minima as minus flows, must have $f$ constant and therefore cannot cross between the two regimes.

For a flow to cross between the two regimes, for generic potentials\footnote{Note that the plus solutions are regular only for special tuned potentials, \cite{exotic}.}, it must start at a dS  minimum and end in an AdS maximum.

\begin{figure}[h]
\centering
\includegraphics[ height=10cm, width=15cm]{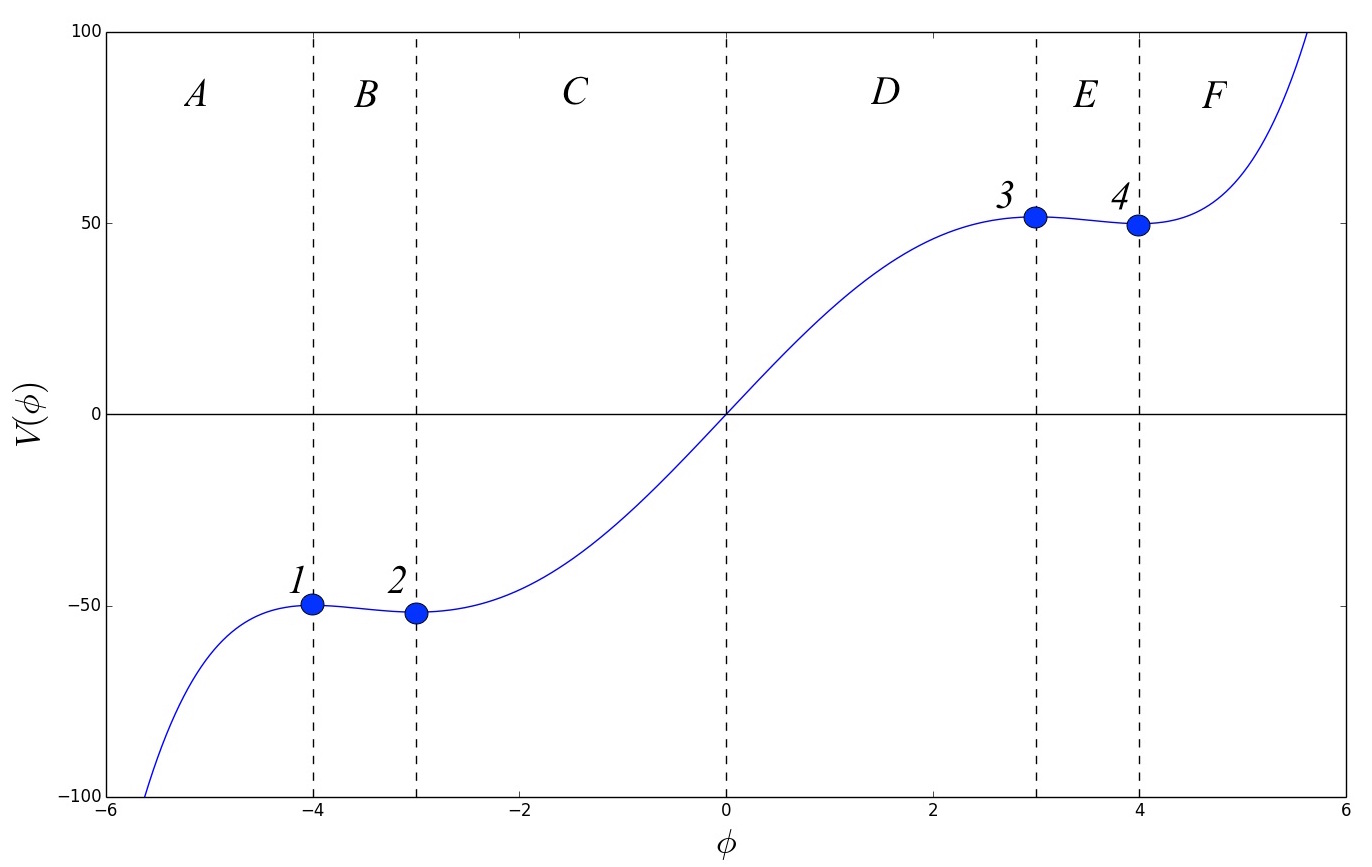}
\caption{A generic potential that contains minima and maxima in both the AdS and dS regimes. The four extrema divide our region of study into six subregions, ranging from $A$ to $F$, with respect to the placing of the extrema. The numbers $1$ to $4$ label each of the extrema.}
\label{plot1}
\end{figure}

This implies that we shall not be able to find a flow starting from a dS minimum and ending in a AdS minimum  with a minus solution (because the latter admits only an $f$=constant regular solution in its neighborhood).
It therefore seems that generic cross-regime flows,  if they exist,  must start from dS minima and end in AdS maxima.

This conclusion agrees also with the counting of parameters.
In a purely AdS regime, the solution starts at a maximum where there is a one-parameter family of solutions and this parameter is tuned so that the solution is regular everywhere and it eventually ends in a minimum.
The situation of pure dS regime solutions is similar, although here the solutions starts at a minimum with a single parameter family of solutions and ends at a maximum (where the ``big-bang" is).
However, to cross from dS to AdS, a horizon in between is necessary and this provides a further tuning condition as shown in appendix \ref{structure}.
Therefore,  we need an extra parameter in the solutions. This is the case when the solution starts at a dS minimum and ends in a AdS maximum.

Consider a flow coming from dS to AdS, and assume that $\dot A=0$ at a point in the AdS regime. In that case, as shown above,  the horizon must be earlier in the flow (we shall assume that the flow starts at point 4 in figure \ref{plot1} at $u=-\infty$ and ends in point 1 in the same figure as $u\to\infty$).

There are two cases to consider here:
\begin{itemize}

\item The horizon is at some point in the C region in figure  \ref{plot1}. In that case, as $\dot A$ has changed sign before, $\dot A$ at C is negative.
From the horizon relations      $\dot \phi \dot f\sim V'$  and $\dot A\dot f\sim -V$ we deduce that, $\dot \phi<0$ and $\dot f<0$.
The sign of $\dot f$ is compatible with the expectation that it will decrease along the flow until it will become negative.
 On the other hand $\dot \phi<0$ is incompatible with the direction of the flow.

\item The horizon is at some point in the B region in figure  \ref{plot1}. In that case, as $\dot A$ has changed sign before, $\dot A$ at B is negative.
From the horizon relations      $\dot \phi \dot f\sim V'$  and $\dot A\dot f\sim -V$ we deduce that, $\dot \phi>0$ and $\dot f<0$.
The sign of $\dot f$ is compatible with the expectation that it will decrease along the flow until it will become negative.
 Also $\dot \phi>0$ is compatible with the direction of the flow. This case therefore seems possible

\end{itemize}

Assume now that $\dot A=0$ in the dS regime. In that case, as shown above the horizon must be later  in the flow (we shall assume that the flow starts at point 1 in figure  \ref{plot1} at $u=-\infty$ and ends in point 4 in the same figure as $u\to\infty$).

There are again two cases to consider:
\begin{itemize}

\item The horizon is at some point in the D region in figure  \ref{plot1}. In that case, as $\dot A$ has not yet changed sign, $\dot A$ at D is positive.
From the horizon relations      $\dot \phi \dot f\sim V'$  and $\dot A\dot f\sim -V$ we deduce that, $\dot \phi<0$ and $\dot f<0$.
The sign of $\dot f$ is compatible with the expectation that it will decrease along the flow until it will become negative.
 On the other hand $\dot \phi<0$ is incompatible with the direction of the flow.

\item The horizon is at some point in the E region in figure  \ref{plot1}.In that case, as $\dot A$ has not yet changed sign, $\dot A$ at E is positive.
From the horizon relations      $\dot \phi \dot f\sim V'$  and $\dot A\dot f\sim -V$ we deduce that, $\dot \phi>0$ and $\dot f<0$.
The sign of $\dot f$ is compatible with the expectation that it will decrease along the flow until it will become negative.
 Also $\dot \phi>0$ is compatible with the direction of the flow. This case therefore seems also possible.

\end{itemize}

Finally there is a third possibility: that $\dot A=0$ exactly where $V=0$. In that case the horizon must also be at this same point.
Now the relation $\dot \phi\dot f\sim V'>0$ implies that $\dot f>0$ since $\dot \phi>0$ and this is impossible for this type of flow.

We conclude that the horizon can be either in the B region or the E region.

\bigskip

We now summarize what we have shown so far. We want to find generic flows that start from the dS regime, corresponding to $V>0$, and ending in an extremum in  AdS, where $V<0$. We have shown that if this is to generically happen, then the flow must start from a dS minimum, and end at an AdS maximum.

Recall that as we approach a dS minimum, the scale factor $e^A$ diverges as $u \rightarrow + \infty$ and $f$ is negative in a neighborhood of the dS minimum. Similarly, as we approach an AdS maximum, then $e^A$ diverges as $u \rightarrow - \infty$, and $f$ is positive in a neighborhood of the AdS maximum. Hence, if we have flows that start from a dS minimum and end at an AdS maximum, then we must have a horizon where $f=0$ along the flow, since $f$ must change sign. Using as an example the potential shown in figure \ref{plot1}, we have seen that a horizon can only exist in region B or region E in the figure.

\begin{figure}[h]
\centering
\includegraphics[ height=10cm, width=15cm]{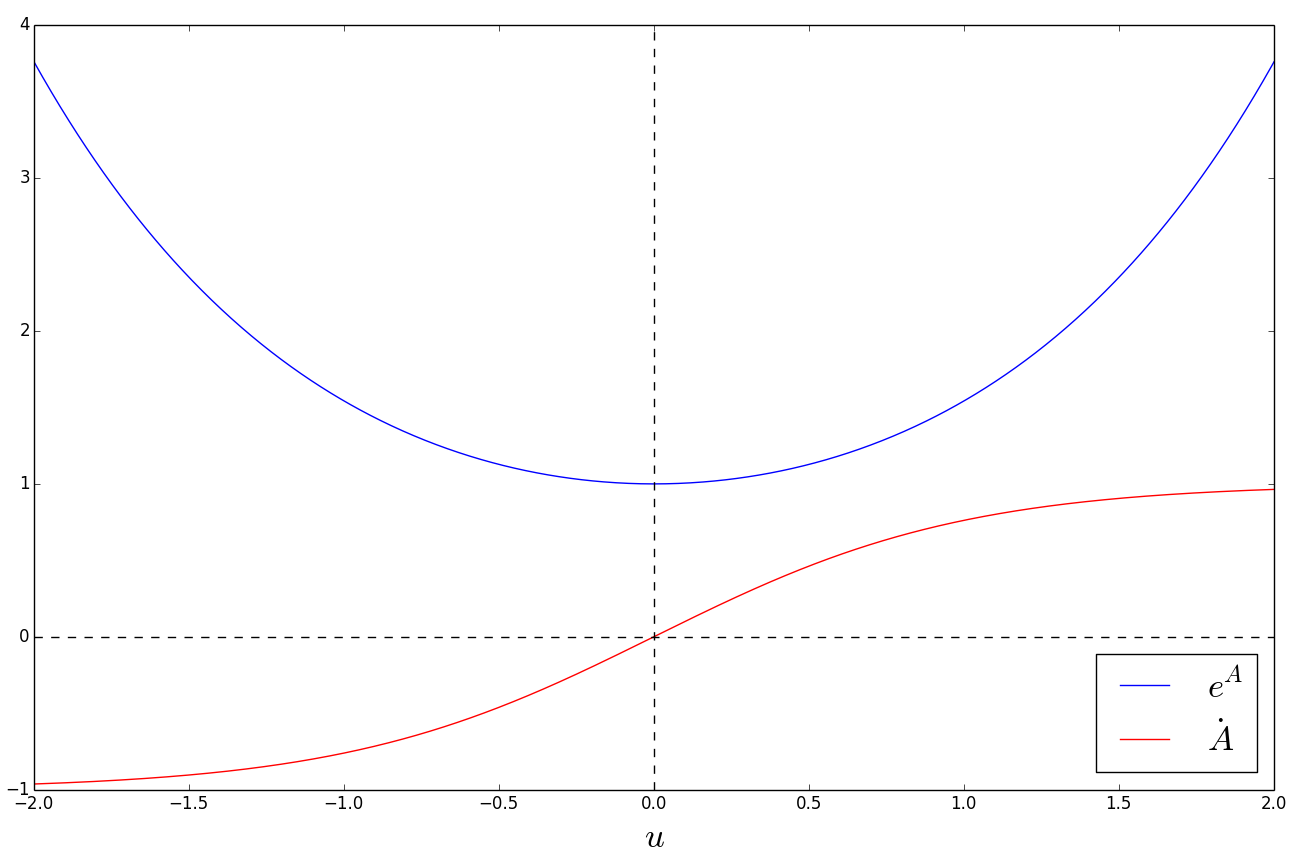}
\caption{Example of the expected qualitative behaviour of the scale factor (in blue) and $\dot{A}$ (in red), in case of a flow originating from a dS minimum at $u= +\infty$, and ending in an AdS maximum at $u=-\infty$. In this figure, we have shifted the extremum of the scale factor to $u=0$. Notice that $\dot{A}$ is increasing, in direct contradiction with equation (\ref{e2b}).}
\label{fig:plot1_a}
\end{figure}

Notice now that any such interpolating flow must have the scale factor, $e^A$, diverge as $u \rightarrow \pm \infty$, since in the vicinity of the dS minimum/AdS maximum the scale factor diverges as $e^{\pm {u \over \ell} }$. Further, we can only have \emph{one} minimum of $e^A$ along the flow, where $\dot{A}=0$, since $\ddot{A} \leq 0$ along the whole flow. The qualitative behaviour of the scale factor for interpolating flows, is shown in figure \ref{fig:plot1_a}, in blue. In red, we see the corresponding behaviour of $\dot{A}$. Such a flow, as seen in the figure, must have $\dot{A}$ increasing, which contradicts equation (\ref{e2a}).

We have therefore shown that, generically, \emph{there cannot exist flows that start from the dS regime and end at the AdS regime, that do not require a fine- tuned potential}.

\begin{figure}[h]
\centering
\includegraphics[ height=10cm, width=15cm]{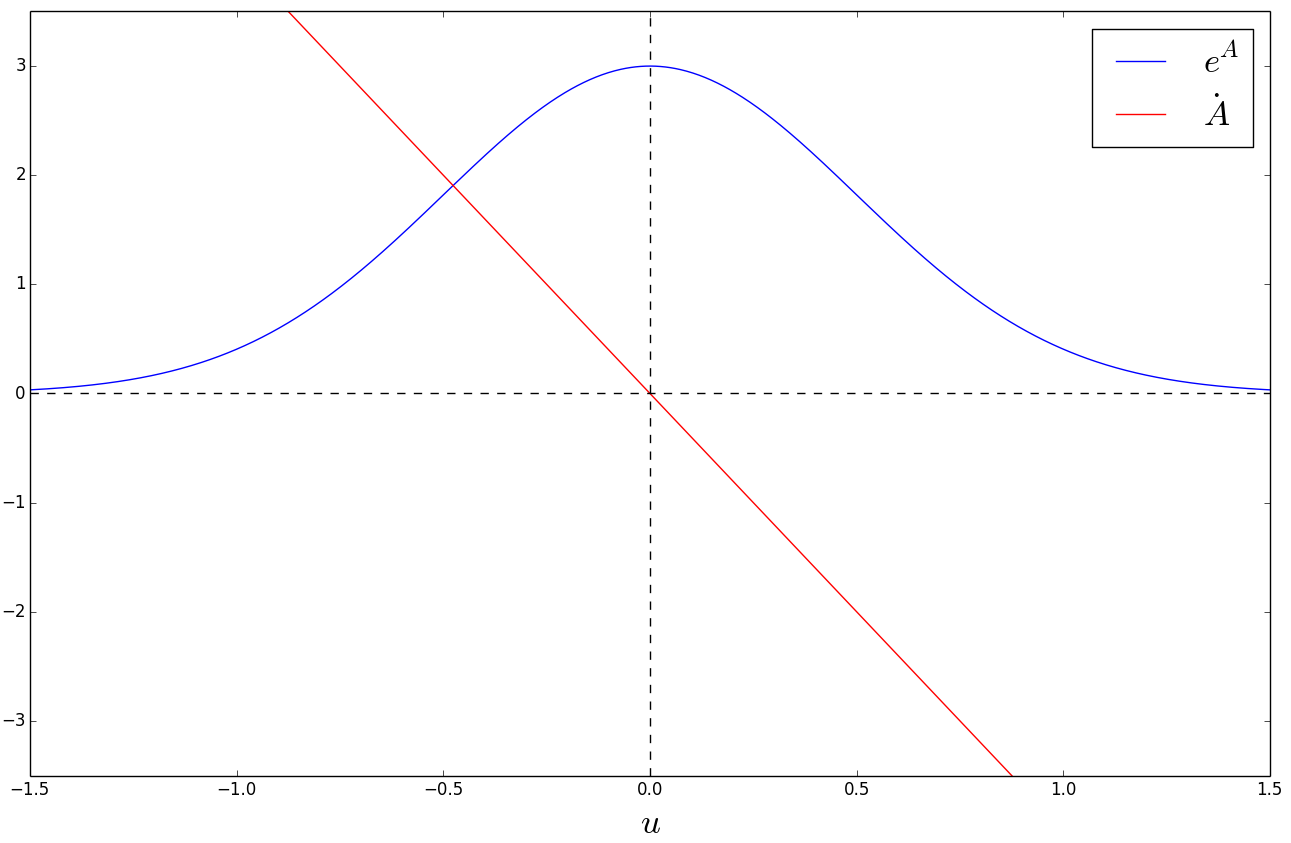}
\caption{Example of the expected qualitative behaviour of the scale factor (in blue) and $\dot{A}$ (in red), in case of a flow originating from a dS maximum at $u= +\infty$, and ending in an AdS minimum at $u=-\infty$. In this figure, we have shifted the extremum of the scale factor to $u=0$. Notice that $\dot{A}$ is decreasing, in accordance with equation (\ref{e2b}). Such a flow is permissible by the equations of motion, but it needs to be fine tuned.}
\label{fig:plot1_b}
\end{figure}

The only behavior for the scalae factor  that is permissible by equation (\ref{e2b}), is shown in figure \ref{fig:plot1_b}. Notice that, as $u\rightarrow \pm \infty$, the scale factor vanishes and there is no boundary.
A vanishing scale factor in the dS regime, can only be obtained near a maximum of the potential with a minus solution. However such solutions, with non-constant $f$ (necessary for the solution to interpolate to AdS) are all singular.
We therefore conclude that interpolating solutions from dS to AdS in our flat-sliced ansatz do not exist for {\it any potential}.

There can be solutions on the other hand that start and end in the AdS regime while they delve a bit in between in the dS regime.
In appendix \ref{exam} we present a specially prepared tuned potential and solution that realizes the behavior in figure \ref{fig:plot1_b} in the AdS regime. This is an AdS like solution without a boundary but two Poincar\'e horizons. This is not allowed if the potential is strictly negative, but it is allowed if the potential has a positive excursion in between the two extrema.

\subsection{Examples of numerical flows}

{Having developed formal arguments regarding the behaviour of flows through a flat slicing in the previous sections, we shall demonstrate here some numerical solutions, in order to visualise the global behaviour of these flows in concrete examples.}

In particular, we solve the equations (\ref{e2a})-(\ref{e2c}) for various cases : flows starting from an AdS minimum, flows passing through a horizon situated in the dS regime, as well as in the AdS regime. This system of equations is second order in $\phi$, second order in $f$, and first order in $A$. We therefore need 5 initial conditions. In general, we need $\phi(0),\dot\phi(0)$, $f(0),\dot f(0)$ and $A(0)$ at a given point of the flow.

We can however simplify the space of physically distinct solutions as follows;
by independent rescalings of $\vec x$, $t$ and a shift of $u$ we may set $A(0)=0$, $f(0)=\pm 1$ and by a rescaling of $u$ we may set $\dot f(0)=\pm 1$ if it is non-zero.
We are therefore left with two non-trivial boundary conditions, $\phi(0)$ and $\dot\phi(0)$.
If the point $u=0$ is a minimum in AdS regime, or a maximum in dS regime, or a horizon, then there is one relation among the 5 initial conditions that fixes completely one of them: $\dot\phi$.  In that case we are left with a single remaining non-trivial initial condition associated to $\phi(0)$.

Before we proceed, we must choose a potential in order to solve the Einstein equations numerically. We construct one with the condition that it admits a maximum and a minimum in both AdS and dS regimes, as well as exponential behaviour at $\phi \rightarrow \pm \infty$. We require this asymptotic behaviour since this is the generic behavior in supergravity and string theory. We therefore start with the condition

\be
V'(\phi)=k(e^{a_0 \phi}+e^{-a_0 \phi})(\phi-a_1)(\phi-a_2)(\phi+a_3)(\phi+a_4)
\label{d1a}\ee
 We shall choose:
 \be
 a_0={1\over 8}\sp a_1=a_4=4\sp a_2=a_3=3
 \label{d1b}\ee
 so that
 \be
 V(\phi)=\frac{4}{5} \left[e^{-{\phi\over 8}} \left(-\phi^4-32 \phi^3-743 \phi^2-11888 \phi-95248\right)+\right.
\label{d1c}\ee
$$+\left.
e^{\phi\over 8} \left(\phi^4-32 \phi^3+743
   \phi^2-11888 \phi+95248\right)\right]
$$

This potential is shown in figure  \ref{plot1}.

\subsubsection{Flows from an AdS minimum}

\begin{figure}[t]
\centering
\includegraphics[ height=10cm, width=15cm]{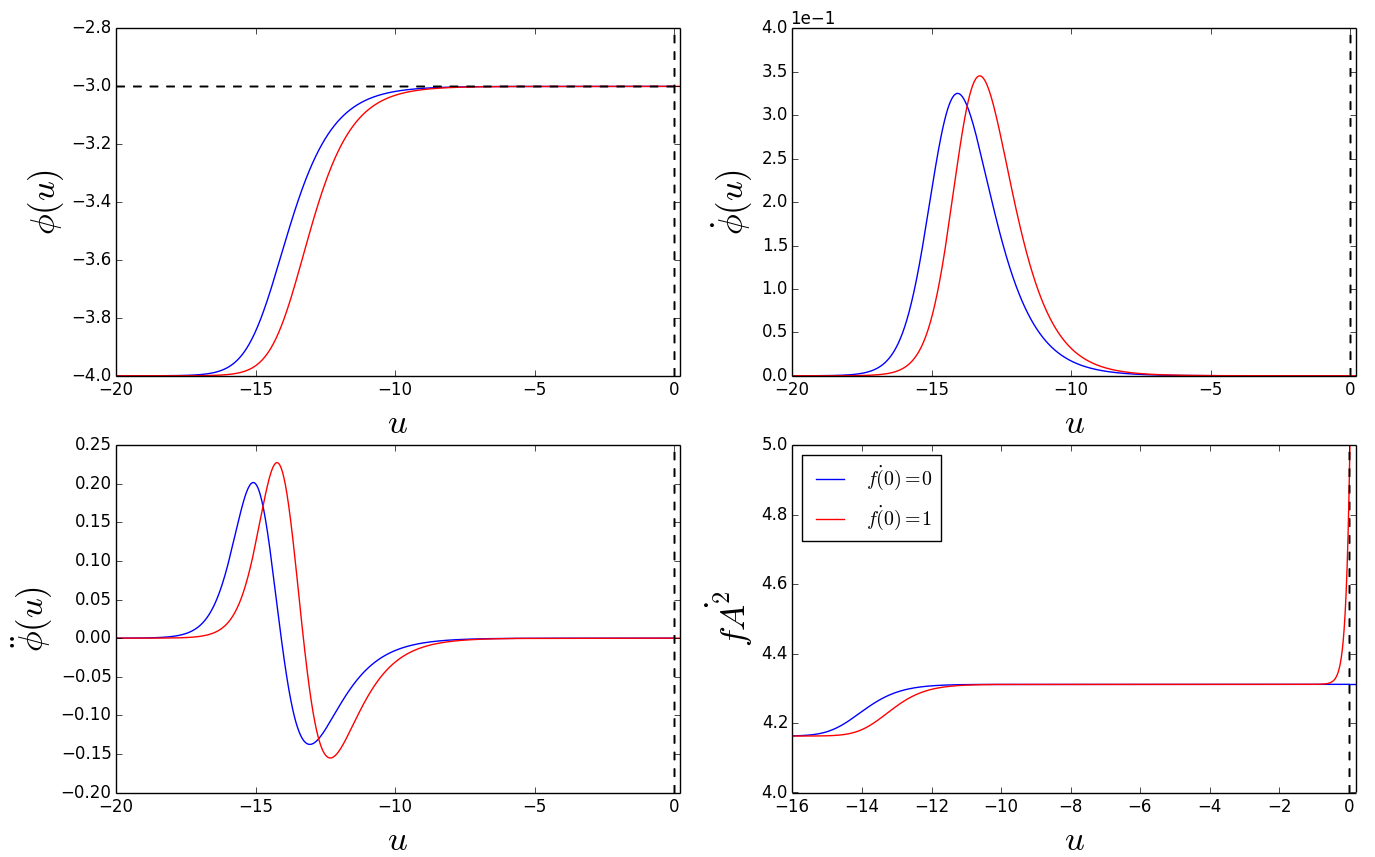}
\caption{{The solution to the Einstein equations with the initial conditions (\ref{initial1}) for $\phi$, $\dot{\phi}$,  and $\ddot{\phi}$ as well as $f \dot{A}^2$. Both cases $\dot{f}(0)=0$ and  $\dot{f}(0)=1$ are considered. The $f=\text{constant}$ solution (in blue) describes a flow that starts from an AdS maximum (situated at $\phi=-4$) and terminates at an AdS minimum (situated at $\phi=-3$). The $f \neq \text{constant}$ solution (in red) describes a flow that starts from an AdS maximum (situated at $\phi=-4$), also terminates at $\phi=-3$ but is, however, singular. This can be seen from the divergence of  $f \dot{A}^2$ or, equivalently, $f W^2$. This solution does not have a horizon. The vertical dashed line corresponds to $u=0$, and the horizontal dashed line in the upper left figure corresponds to the point of the AdS minimum, $\phi=-3$.}}
\label{fig:plot3}
\end{figure}

\begin{figure}[t]
\centering
\includegraphics[ height=10cm, width=15cm]{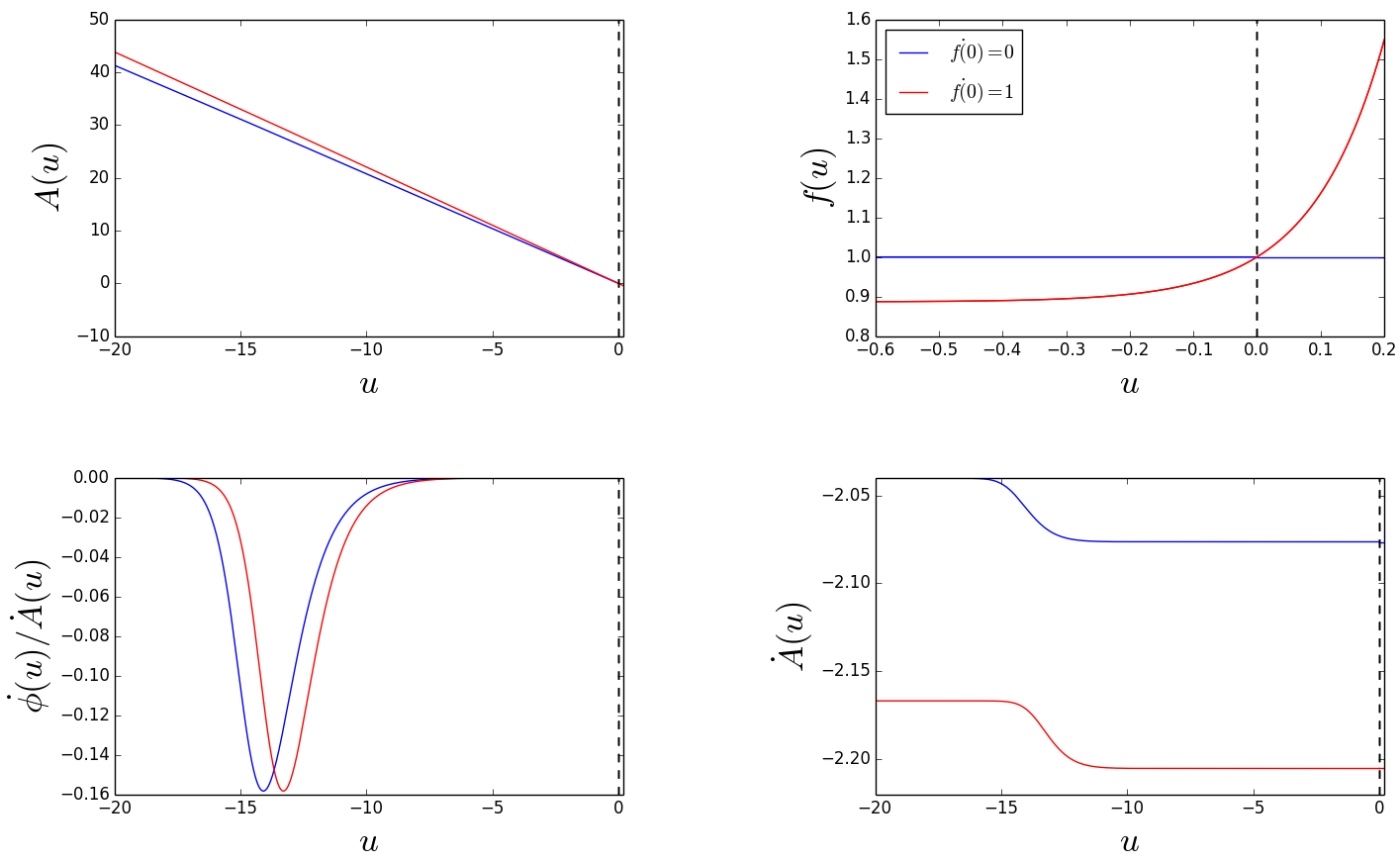}
\caption{{The solution to the Einstein equations with the initial conditions (\ref{initial1}) for $A$, $\dot{A}$,  and $f$ and $\dot{\phi} / \dot{A}$. Both cases $\dot{f}(0)=0$ and  $\dot{f}(0)=1$ are considered. The $f=\text{constant}$ solution (in blue) describes a flow that starts from an AdS maximum (situated at $\phi=-4$) and terminates at an AdS minimum (situated at $\phi=-3$). The $f \neq \text{constant}$ solution (in red) describes a flow that starts from an AdS maximum (situated at $\phi=-4$), also terminates at an AdS minimum, but is singular. Notice that $f$ diverges exponentially, as expected from (\ref{e2a}) and (\ref{e2b}) for points where both $\dot{\phi}$ and $\ddot{\phi}$ vanish. The vertical dashed line corresponds to $u=0$.}}
\label{fig:plot4}
\end{figure}

In this section, we shall consider flows that start close to an AdS minimum, here situated at $\phi_\text{min}=-3$. The initial conditions that we use are

\be \label{initial1}
f(0)=1\sp
\dot{f}(0)=
\begin{cases}
1, \\
0
\end{cases}\sp
\phi(0)=\phi_\text{min}+\epsilon
\ee
\be
\dot{\phi}(0)=\Delta_-(\phi(0)-\phi_\text{min})=\Delta_- \epsilon  \sp
A(0)=0,
\ee

where $\epsilon$ is a small parameter with $|\epsilon| \ll 1$. Taking $\epsilon>0$ we start the solution directly to the right of the AdS minimum, whereas $\epsilon<0$ corresponds to the solution starting directly to the left of the minimum. The parameter $\Delta_-$ is defined as

\be
\Delta_-=\frac{1}{2} (d-\sqrt{d^2+4 V''(\phi_\text{min})})
\label{w51}\ee

for the case of AdS extrema (see (\ref{i33_b})). Therefore,  we must solve the Einstein equations using (\ref{initial1}), for both cases $\dot{f}(0)=0$ as well as  $\dot{f}(0)=1$. In figure \ref{fig:plot3} we plot $\phi$, $\dot{\phi}$, $\ddot{\phi}$, $f \dot{A}^2$, whereas in \ref{fig:plot4} we plot $A$, $\dot{A}$,  and $f$ and $\dot{\phi} / \dot{A}$.

{
From these figures, we can deduce the qualitative behaviour of the two solutions, one with $\dot{f}(0)=0$ and one with $\dot{f}(0)=1$. Both solutions start from an AdS maximum situated at $\phi=-4$, and flow towards the AdS minimum at $\phi=-3$. In both cases $\phi$ asymptotically approaches $\phi=-3$, which is the AdS minimum. For the case $\dot{f}(0)=0$ this behaviour is expected, from the analysis of section (\ref{section:2}) and corresponds to a standard AdS to AdS flow that is dual to an RG flow. This is a regular solution.

However, the solution with initial condition $\dot{f}(0)=1$, even though it also stops at the AdS minimum, is not a regular solution. This can be seen from figure \ref{fig:plot3} where $f \dot{A}^2$, and equivalently $f W^2$, diverges. Recall from section \ref{cr_invariants_flat} that the curvature invariants are polynomials of $V$, $f W^2$, and $f W'^2$. Hence divergence of $f W^2$ implies that the geometry of this solution is irregular.

This behaviour for the solution with nonconstant $f$ at an extremum of the potential is indeed expected; since $\phi$ obeys a second order equation, namely (\ref{e2d}), the flow stops when both $\dot{\phi}$ and $\ddot{\phi}$ vanish at a point, which from (\ref{e2d}) is an extremum of the potential since $V'=0$ there. In figure \ref{fig:plot3}, indeed $\dot{\phi} \sim 0$ and $\ddot{\phi} \sim 0$ as $u \rightarrow \infty$. From (\ref{e2a}) we expect that $\dot{A} \sim \text{constant}$, which implies that $A$ is linear, as is indeed the case as shown in figure \ref{fig:plot4}. From (\ref{e2e}) then, $\dot{f}$ and therefore $f$ must diverge exponentially as $u \rightarrow \infty$. This behaviour is shown in figure \ref{fig:plot4} for $f$.. We thus see here numerically what was also shown in the local analysis of section \ref{local} near extrema of the potential. That is, the only regular solutions that arrive at an AdS minimum have constant blackness function.
}

\subsubsection{Flows through a horizon between two dS extrema}

In this section we consider the solution of the Einstein equations for the initial conditions

\be\label{initial2}
f(0)=\epsilon \sp
\dot{f}(0)= 1 \sp
\phi(0)=\phi_h
\ee
\be
\dot{\phi}(0)=\frac{V'(\phi_h)}{\dot{f}(0)}  \sp
A(0)=0.
\ee

Because the horizon is a singular point of the equations, we must set our initial conditions at a point close to the horizon, where $f$ is nonzero yet very small. This is signified by a parameter $\epsilon$ such that $|\epsilon| \ll 1$. We have also set $u=0$ to correspond to the horizon (or, more precisely, to the point where $f = \epsilon $).

Recall that from appendix \ref{fho}, the expressions for $\dot{\phi}(0)$ as well as $\dot{A}(0)$ are fixed and therefore,  we also have
\be
\dot{A}(0)=-\frac{V(\phi_h)}{(d-1)\dot{f}(0)}.
\ee

Further, the dS extrema for our potential shown in figure \ref{plot1} are situated at $\phi=3$ and $\phi=4$ for the maximum and minimum respectively. We therefore allow for $3<\phi_h<4$, following the discussion of section (\ref{properties}).

\begin{figure}[t]
\centering
\includegraphics[ height=10cm, width=15cm]{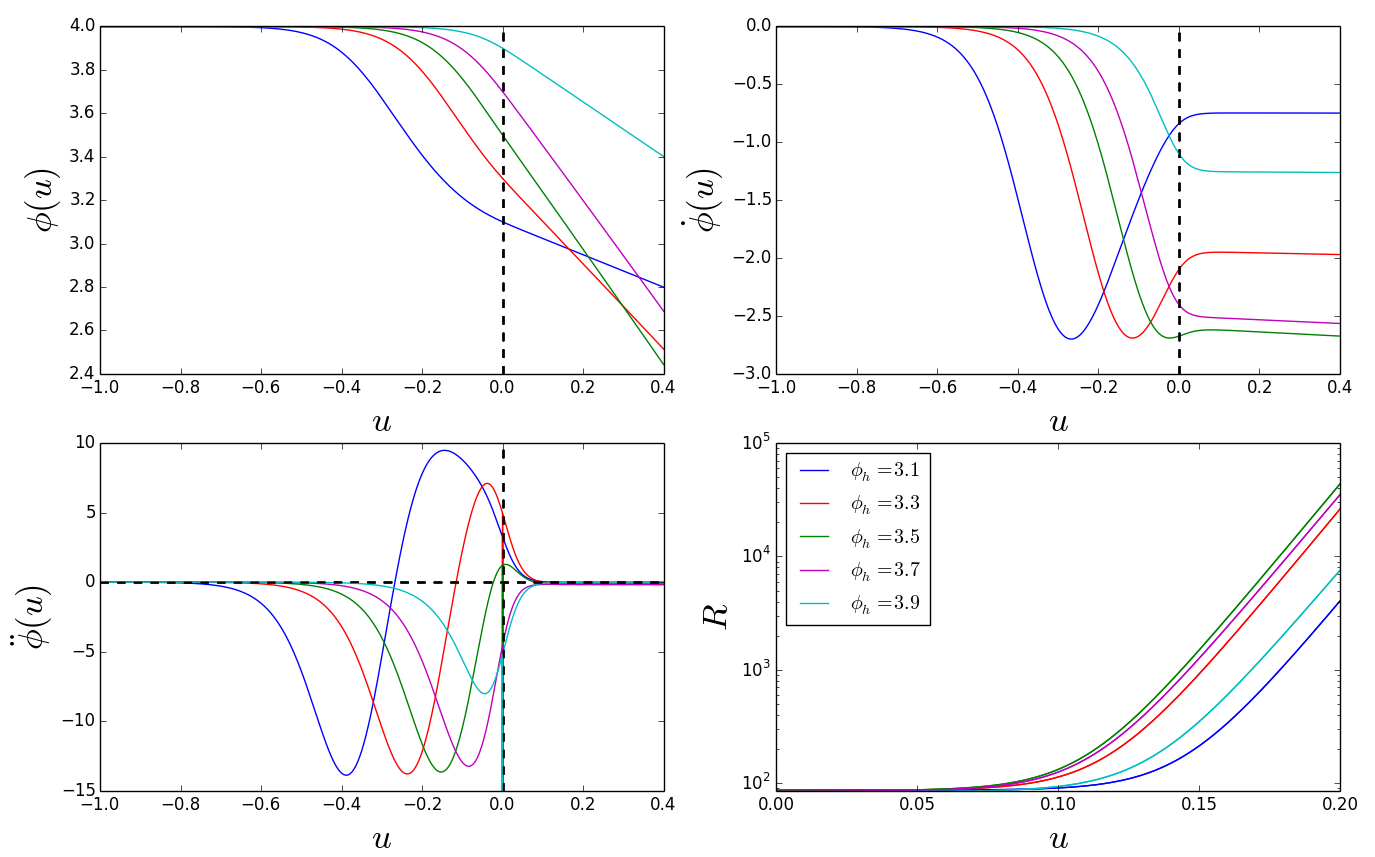}
\caption{The solution to the Einstein equations with the initial conditions (\ref{initial2}) for $\phi$, $\dot{\phi}$,  and $\ddot{\phi}$ as well as the scalar curvature, $R$, for various positions of the horizon between the two dS extrema of the potential. The vertical dashed line corresponds to the placement of the horizon. These solutions describe a flow that starts at a dS minimum situated at $\phi=4$, evolve through a horizon, and head to $\phi \rightarrow - \infty$ where the potential is divergent. All these solutions are singular.}
\label{fig:plot7}
\end{figure}

\begin{figure}[t]
\centering
\includegraphics[ height=10cm, width=15cm]{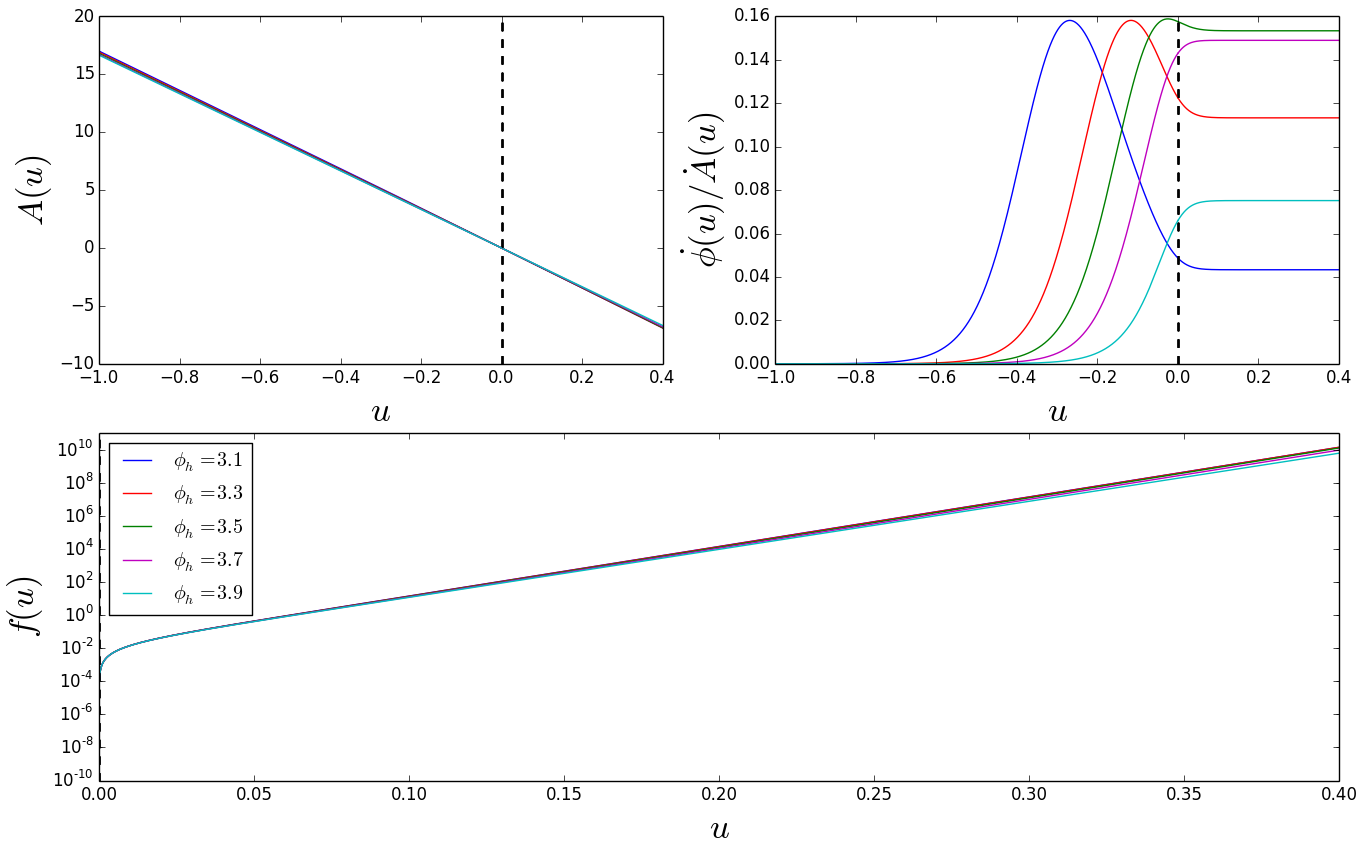}
\caption{The solution to the Einstein equations with the initial conditions (\ref{initial2}) for $A$,  $f$, and $\dot{\phi} / \dot{A}$. Notice how $f$ diverges as $u \rightarrow \infty$. As in figure \ref{fig:plot7}, these solutions describe a flow that starts at a dS minimum at $\phi=4$, evolve through a horizon, and head to $\phi \rightarrow - \infty$ where the potential is divergent. All these solutions are singular.}
\label{fig:plot8}
\end{figure}

In figure \ref{fig:plot7}, we plot the behaviour of $\phi, \dot{\phi}, \ddot{\phi}, R$, where $R$ is the Ricci scalar. The dashed vertical line corresponds to the horizon. In figure \ref{fig:plot8}, we plot the behaviour of $A$,  $f$, and $\dot{\phi} / \dot{A}$.

As $u \rightarrow + \infty$ we have $\phi \rightarrow -\infty$, and $f$ as well as $R$ diverge. The solution is irregular, and hits a singularity where $V(\phi) \rightarrow -\infty$.

Considering the above solutions, we conclude that the general behaviour is the following : flows start from the dS minimum where $f<0$, they flow through a horizon situated between the minimum and the maximum, and then hit the singularity at $\phi \rightarrow -\infty$, passing through all extrema of the potential.

The cosmological interpretation of such solutions near the minimum is that of an expanding large universe. Going back in time, the universe shrinks until a horizon forms hiding the analogue of the big-bang singularity.

\subsubsection{Flows through a horizon between two AdS extrema}

In this section we consider the solution of the Einstein equations for the initial conditions (they are in form identical to (\ref{initial2}), but we repeat them for convenience):

\be \label{initial3}
f(0)=\epsilon \sp
\dot{f}(0)= 1 \sp
\phi(0)=\phi_h
\ee
\be
\dot{\phi}(0)=\frac{V'(\phi_h)}{\dot{f}(0)}  \sp
A(0)=0.
\ee

As in the previous subsection, we take

\be
\dot{A}(0)=-\frac{V(\phi_h)}{(d-1)\dot{f}(0)}.
\ee

Further, the AdS extrema for our potential shown in figure ref{plot1}  are situated at $\phi=-3$ and $\phi=-4$ for the minimum and maximum respectively. We therefore allow for $-3>\phi_h>-4$.

\begin{figure}[h]
\centering
\includegraphics[ height=10cm, width=15cm]{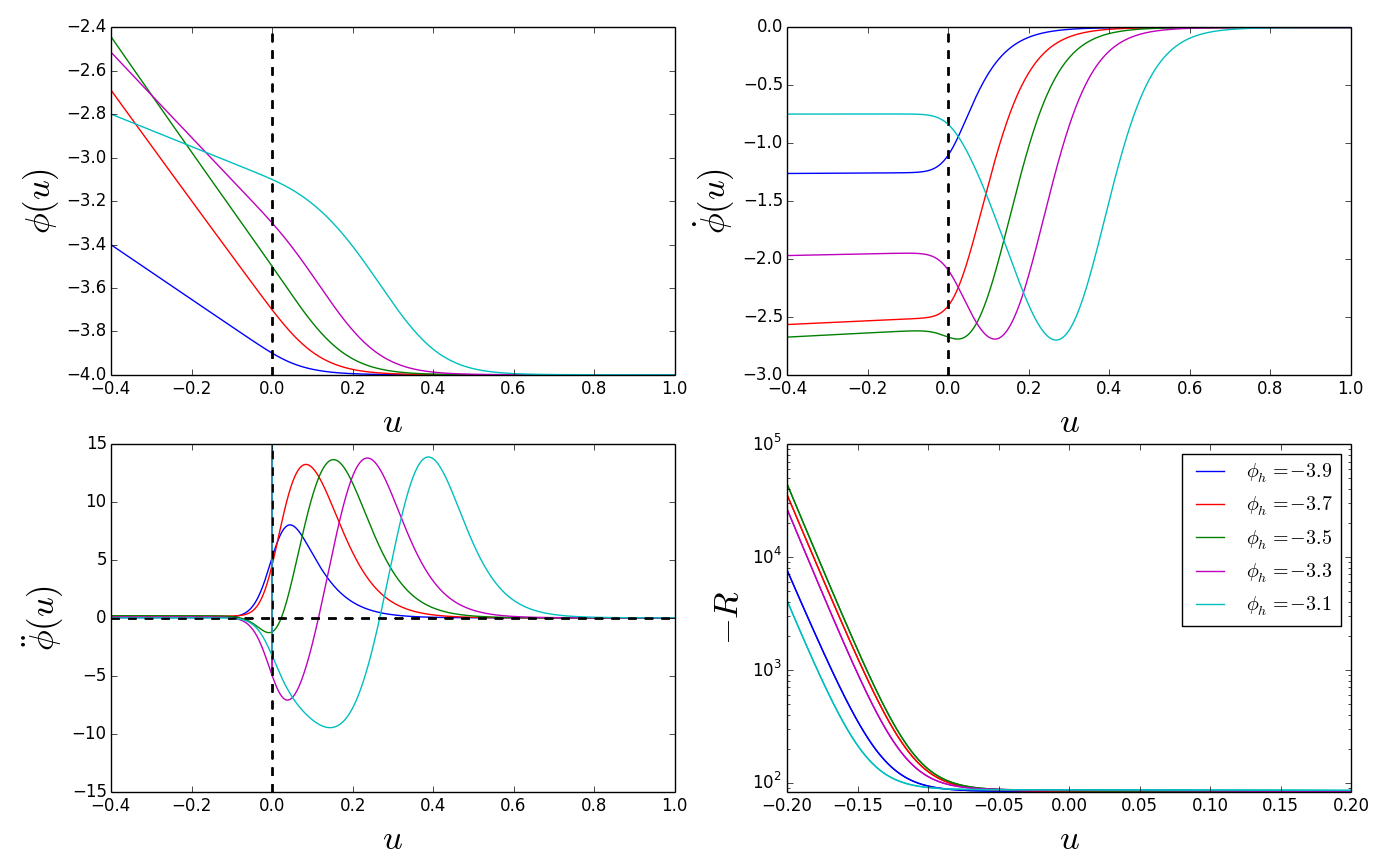}
\caption{The solution of the Einstein equations with the initial conditions (\ref{initial3}) for $\phi$, $\dot{\phi}$,  and $\ddot{\phi}$ as well as the minus scalar curvature, $R$ (the reason for the minus sign is that $R$ is negative, and itself cannot be plotted logarithmically). Notice that as $u\rightarrow - \infty$, the scalar curvature diverges, and $\phi \rightarrow + \infty$. Different colours correspond to different placements of the horizon. These solutions describe flows that come from $\phi \rightarrow + \infty$, evolve through a horizon placed in the AdS regime, and terminate at an AdS maximum situated at $\phi=-4$.}
\label{fig:plot9}
\end{figure}

\begin{figure}[h]
\centering
\includegraphics[ height=10cm, width=15cm]{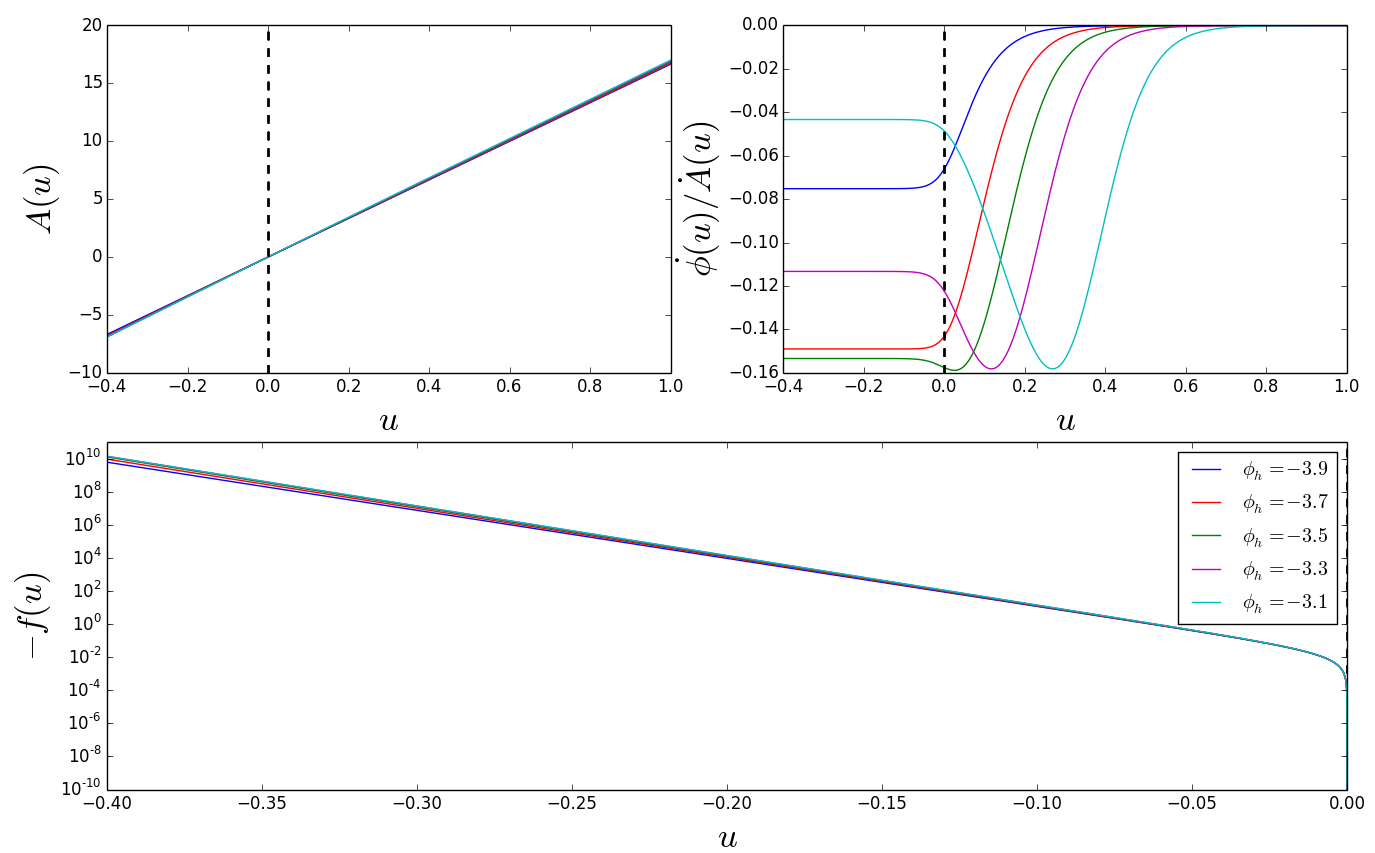}
\caption{The solution of the Einstein equations with the initial conditions (\ref{initial3}) for $A$,  $-f$, and $\dot{\phi} / \dot{A}$. As usual, the vertical dashed line corresponds to the position of the horizon. These solutions describe flows that come from $\phi \rightarrow + \infty$, evolve through a horizon placed in the AdS regime, and terminate at an AdS maximum  situated at $\phi=-4$, as shown in figure \ref{fig:plot9}.}
\label{fig:plot10}
\end{figure}

In figures \ref{fig:plot9} and \ref{fig:plot10} we show the numerical solutions for the flows starting from an AdS maximum and flowing through a horizon.

As $u \rightarrow - \infty$ we have $\phi \rightarrow +\infty$, and $f$ as well as $R$ diverge. The solution hits a singularity where $V(\phi) \rightarrow +\infty$.
Again,  considering the above numerical solutions, we see that the general behaviour is the following : flows start from the AdS maximum where $f>0$, they flow through a horizon situated between the minimum and the maximum, and then hit the singularity at $\phi \rightarrow +\infty$, passing through all other extrema of the potential. Their interpretation are as the solution describing
the thermal state of the standard AdS flows.

\section*{Acknowledgements}
\addcontentsline{toc}{section}{Acknowledgements}

We would like to thank David Andriot, Dionysios Anninos, Diego Hofman, Francesco Nitti, Lukas Witkowski, and Timm Wrase for useful discussions.
We also thank Dionysions Anninos, Francesco Nitti and Lukas Witkowski for a critical reading of the manuscript.

This work was supported in part by European Union's Seventh Framework Programme under grant agreements (FP7-REGPOT-2012-2013-1) no 316165 and the Advanced
ERC grant SM-grav, No 669288.



\newpage
\appendix
\renewcommand{\theequation}{\thesection.\arabic{equation}}
\addcontentsline{toc}{section}{Appendix}
\section*{Appendices}

\section{Review of coordinate systems in de Sitter and Anti de Sitter\label{a}}

In this appendix we review briefly some commonly used coordinate systems in dS as well as AdS. For further details see, for example \cite{anninos}, \cite{adsreview}.

\subsection{Coordinate systems in de Sitter}

The (d+1)-dimensional de Sitter space is defined as the manifold:
\be
-X_0^2+\sum_{i=1}^{d+1}X_i^2={1\over H^2}
\label{a24}\ee

embedded in $\mathbb{R}^{d+2}$. Below, we list some commonly used coordinate systems:

\begin{itemize}

\item {\bf Global coordinates}

\be
ds^2=-d\tau^2+{\cosh^2(H\tau)\over H^2}d\Omega_{d}^2
\label{a25}\ee

where $\tau\in R$.
de Sitter space has two timelike boundaries ${\cal I}^+$ and ${\cal I}^+$.
Here, $\tau\to\infty$ is the future boundary ${\cal I}^+$ while $\tau\to -\infty$ is the past boundary ${\cal I}^-$
The picture of de Sitter in global coordinates is that of a spatial sphere with variable radius. In the infinite past, the sphere has infinite radius and this is the ${\cal I}^-$ past boundary. The sphere radius subsequently decreases as times goes on, and shrinks to a minimum size at $\tau=0$. Finally it starts re-expanding for $\tau>0$ and eventually becomes of infinite size in the far future where the future boundary is ${\cal I}^+$.
The analytic continuation $\tau\to i\tau$ turns de Sitter into a $(d+1)$-dimensional sphere.

\item {\bf Poincar\'e coordinates}

\be
ds^2=-dT^2+e^{2HT}d\vec x^2={1\over (Ht)^2}(-dt^2+d\vec x^2)
\label{a26}
\ee

where $T\to\infty$ is ${\cal I}^+$ while $T\to-\infty$ is a single point on ${\cal I}^-$.
The same applies to $t\to \pm \infty$
The $T=-\infty$ point looks like a big-bang singularity point as space shrinks to a point. However, in reality this is a coordinate singularity similar to the coordinate singularity at the origin of spherical coordinates. Indeed, the dS curvature is everywhere fine and constant.

\item {\bf Static patch coordinates}.
\be
ds^2=-\left(1-H^2r^2\right)dt^2+{dr^2\over (1-H^2r^2)}+r^2~d\Omega_{d-1}^2
\label{a27}\ee
There are two horizons ar $r=\pm {1\over H}$.
$r\to\infty$ is ${\cal I}^+$ while  $r\to-\infty$ is ${\cal I}^-$
We may do another radial coordinate change to $Hr=\sin(Hu)$ so that the metric above becomes
\be
ds^2=-\cos^2(Hu)dt^2+du^2+{\sin^2(Hu)\over H^2}~d\Omega_{d-1}^2
\label{a31}\ee
and describes the sector between the two horizons. For $(Hr)^2>1$ we change as
\be
Hr=\pm\cosh(Hu)
\label{a32}\ee
so that the two boundaries are at $u\to \infty$.
The metric becomes
\be
ds^2=\sinh^2(Hu)dt^2-du^2+{\cosh^2(Hu)\over H^2}d\Omega_{d-1}^2
\label{a33}\ee

\item {\bf AdS slice coordinates}.
\be
ds^2=-dt^2+{\sinh^2(Ht)\over H^2}dH_{d}^2=-dt^2+{\sinh^2(Ht)\over H^2}\left(dR^2+\sinh^2 R d{\Omega}^2_{d-1}\right)
\label{a28}\ee
where $dH^2_d$ is the metric of unit radius Euclidean AdS$_d$.
$t\to\infty$ is ${\cal I}^+$ while $t\to 0$ is a big-bang like singularity which is however a coordinate singularity (as dS has constant curvature everywhere).
The analytic continuation to a sphere is $t\to it$.

\item {\bf dS slice coordinates}.
\be
ds^2=dw^2+\sin^2(Hw)\left(-dt^2+{\cosh^2(Ht)\over H^2}d\Omega_{d-1}^2\right)
\label{a29}\ee
Now $w\in [0,\pi]$. There is one point of ${\cal I}^+$ at $w=\pi$ and a a single point of ${\cal I}^-$ at $w=0$.
The analytic continuation to a sphere is $t\to it+{\pi\over 2H}$.

\end{itemize}

\subsection{Coordinate systems in anti de Sitter}

\begin{itemize}

\item (d+1)-dimensional Minkowski anti de Sitter is defined as the manifold:
\be
X_0^2+X_1^2-\sum_{i=1}^{d}X_i^2=\ell^2
\label{a30}\ee

\item {\bf Global coordinates}
\be
ds^2=\ell^2\left(-\cosh^2\rho~d\tau^2+d\rho^2+\sinh^2(\rho)d\Omega_{d-1}^2\right)
\label{a30_1}\ee
The universal cover has $\tau\in R$. $\rho\geq 0$.
AdS$_{d+1}$ has a single spacelike boundary with topology $R\times S^{d-1}$.
The boundary of AdS$_{d+1}$ is at $\rho\to\infty$, while $\rho=0$ is the center of AdS$_{d+1}$

By introducing $\tan\theta=\sinh\rho$ we can make the radial coordinate compact and the metric is
\be
ds^2={\ell^2\over \cos^2\theta}\left(-d\tau^2+d\theta^2+\sin^2\theta d\Omega_{d-1}^2\right)={\ell^2\over \cos^2\theta}\left(-d\tau^2+d\Omega_{d}^2\right) \sp \theta\in \left[0,{\pi\over 2}\right)
\label{a34}\ee
In this coordinate system it is clear that the time it takes a null radial geodesic to arrive from the center to the boundary is $\pi/2$.

\item {\bf Poincar\'e coordinates}.
\be
ds^2=-du^2+e^{-2{u\over \ell}}(-dt^2+\vec x^2)={\ell^2\over z^2}(dz^2-dt^2+d\vec x^2)
\label{a34_1}
\ee
$u\to-\infty$ is the AdS boundary  while $u\to \infty$ is a single point on the opposite side of the boundary.

\item {\bf Static patch coordinates}.
\be
ds^2=-\left(1+{r^2\over \ell^2}\right)dt^2+{dr^2\over \left(1+{r^2\over \ell^2}\right)}+r^2~d\Omega_{d-1}^2
\label{a35}\ee
The AdS boundary is at $r\to \infty$.

\item {\bf AdS slice coordinates}.
\be
ds^2=du^2+\ell^2\cosh^2{u\over \ell}dAdS_{d}^2=du^2+\cosh^2{u\over \ell}\left(dR^2+\sinh^2 R d{\Omega}^2_{d-1}\right)
\label{a36}\ee
where $dAdS^2_d$ is the metric of unit radius Euclidean AdS$_d$.
$u\to\infty$ is the boundary.

\item {\bf dS slice coordinates}.
\be
ds^2=dw^2+\ell^2\sinh^2{w\over \ell}dS_{d}^2=dw^2+\sinh^2{w\over \ell}\left(-dt^2+{\cosh^2(Ht)\over H^2}d\Omega_{d-1}^2\right)
\label{a37}\ee
Now $w\in (-\infty,0]$. $u\to-\infty$ is the AdS boundary.

\end{itemize}

\section{Solving the linearized scalar equation near global  AdS$_5$\label{linear}}

We consider here the case of AdS in global coordinates, and we turn on the scalar near the UV and the IR of the geometry.
There are three related coordinate systems that describe global AdS.
The first is the u-system that is adapted to our ansatz in (\ref{e1}) and in which the global AdS metric we reproduce here
\be
A={u\over \ell}+A_0\sp f=1+{\ell^2\over R^2}e^{-2A} \sp V=-{12\over \ell^2}
\label{C32}\ee
Here the boundary is at $u\to +\infty$ and the center at $u\to -\infty$.

There is also the static patch  coordinates in (\ref{a35}) with
\be
{r\over \ell}=e^{u\over \ell}
\label{C33}\ee
where the boundary is at $r\to \infty$ and the center at $r=0$.

Then there are the standard global coordinates in (\ref{a30_1}) with
\be
\ell \sinh\rho=r
\label{C34}\ee
and finally the conformally related coordinates in (\ref{a34}) with
\be
\tan\theta=\sinh\rho
\label{C35}\ee
We would like now to write the solution of the linearized equation for the scalar near the UV and IR of the potential. The equation is
\be
(\square- m^2)\phi=0
\label{C36}\ee
where as usual $m^2$ is given by the second derivative of the potential.
The general solution of this linearized equation is well known and can be found for example in appendix L.4 of \cite{book}. For AdS$_5$, the general solution in the coordinate system\footnote{Note that the AdS length is $L$ and the principal $S^3$ quantum number $\ell$ in \cite{book}.} (\ref{a34}) is composed of the modes,
\be
\phi_{\omega,j,\pm}=e^{i\omega\tau}~F(\theta) ~Y_{j}(\Omega_3)\;\;,
\label{C37}\ee
with $Y_{j}(\Omega_3)$ the spherical harmonic on $S^3$, an eigenstate of the
Laplacian on $S^3$ with eigenvalue $j(j+2)$,  $j=0,1,2,\cdots$, and
 \be
F(\theta)=(\sin \theta)^{j}(\cos\theta)^{\Delta_{\pm}}~_2F_1(a,b,c;\sin^2
\theta)\eqc
\label{C38}\ee
\be
a={1\over 2}(j+\Delta_{\pm}-\omega \ell)\sp
b={1\over 2}(j+\Delta_{\pm}+\omega \ell)\sp c=j +2\eqc
\label{C39}\ee
and
\be
\Delta_{\pm}=2\pm \sqrt{4+m^2L^2}\;\;,
\label{C40}
\ee
As we are interested in time independent solutions that are constant on the $S^3$ we take $j=\omega=0$ to obtain
\be
\phi_{\pm}=(\cos\theta)^{\Delta_{\pm}}~_2F_1({\Delta_{\pm}\over 2},{\Delta_{\pm}\over 2},2;\sin^2\theta)\eqc
\label{C41}\ee
To convert this solution to our coordinate system in (\ref{e1}) we use (\ref{C33})-(\ref{C35}),
\be
\tan\theta=\sinh\rho={r\over \ell}=e^{u\over \ell}
\label{C42}\ee to obtain
\be
\phi_{\pm}=\left(\sqrt{1+e^{2u\over \ell}}\right)^{-\Delta_{\pm}}~_2F_1\left({\Delta_{\pm}\over 2},{\Delta_{\pm}\over 2},2;{e^{2u\over \ell}\over 1+e^{2u\over \ell}}\right)\eqc
\label{C43}\ee

Near the boundary, $u\to\infty$. To obtain the leading behavior of the solution we need to transform that hypergeometric function using
\be
_2F_1\left({\Delta_{\pm}\over 2},{\Delta_{\pm}\over 2},2;x\right)={\Gamma\left(2\right)\Gamma\left(2-\Delta_{\pm}\right)\over \Gamma^2\left({\Delta_{\mp}\over 2}\right)}~ _2F_1\left({\Delta_{\pm}\over 2},{\Delta_{\pm}\over 2},\Delta_{\pm}-1;1-x\right)+
\label{C44}\ee
$$
+{\Gamma\left(2\right)\Gamma\left(\Delta_{\pm}-2\right)\over \Gamma^2\left({\Delta_{\pm}\over 2}\right)}~(1-x)^{\Delta_{\mp}\over 2}~ _2F_1\left({\Delta_{\mp}\over 2},{\Delta_{\mp}\over 2},\Delta_{\mp}-1;1-x\right)
$$
We obtain
\be
\phi_{\pm}=\left(\sqrt{1+e^{2u\over \ell}}\right)^{-\Delta_{\pm}}\left[
{\Gamma\left(2\right)\Gamma\left(2-\Delta_{\pm}\right)\over \Gamma^2\left({\Delta_{\mp}\over 2}\right)}~ _2F_1\left({\Delta_{\pm}\over 2},{\Delta_{\pm}\over 2},\Delta_{\pm}-1;{1\over 1+e^{2u\over \ell}}\right)+
\right.
\label{C45}\ee
$$
+\left.{\Gamma\left(2\right)\Gamma\left(\Delta_{\pm}-2\right)\over \Gamma^2\left({\Delta_{\pm}\over 2}\right)}~\left(1+e^{2u\over \ell}\right)^{-{\Delta_{\mp}\over 2}}~ _2F_1\left({\Delta_{\mp}\over 2},{\Delta_{\mp}\over 2},\Delta_{\mp}-1;{1\over 1+e^{2u\over \ell}}\right)\right]
$$
\be
={\Gamma\left(2\right)\Gamma\left(2-\Delta_{\pm}\right)\over \Gamma^2\left({\Delta_{\mp}\over 2}\right)}~\left(1+e^{2u\over \ell}\right)^{-{\Delta_{\pm}\over 2}}~ _2F_1\left({\Delta_{\pm}\over 2},{\Delta_{\pm}\over 2},\Delta_{\pm}-1;{1\over 1+e^{2u\over \ell}}\right)+
\label{C46}\ee
$$
+{\Gamma\left(2\right)\Gamma\left(\Delta_{\pm}-2\right)\over \Gamma^2\left({\Delta_{\pm}\over 2}\right)}~\left(1+e^{2u\over \ell}\right)^{-2}~ _2F_1\left({\Delta_{\mp}\over 2},{\Delta_{\mp}\over 2},\Delta_{\mp}-1;{1\over 1+e^{2u\over \ell}}\right)
$$
$$
\simeq {\Gamma\left(2\right)\Gamma\left(2-\Delta_{\pm}\right)\over \Gamma^2\left({\Delta_{\mp}\over 2}\right)}~e^{-\Delta_{\pm}{u\over \ell}}\left[1+{\cal O}\left(e^{-2{u\over \ell}}\right)\right]+{\Gamma\left(2\right)\Gamma\left(\Delta_{\pm}-2\right)\over \Gamma^2\left({\Delta_{\pm}\over 2}\right)}~e^{-4{u\over \ell}}\left[1+{\cal O}\left(e^{-2{u\over \ell}}\right)\right]
$$
The general therefore solution is a linear combination of the two and can be written as
\be
\phi_{UV}\simeq C_-\left[~e^{-\Delta_{-}{u\over \ell}}\left(1+{\cal O}\left(e^{-2{u\over \ell}}\right)\right)+{\Gamma^2\left({\Delta_{+}\over 2}\right)\over
\Gamma^2\left({\Delta_{-}\over 2}\right)}{}{\Gamma\left(\Delta_{-}-2\right)\over \Gamma\left(\Delta_{+}-2\right) }~e^{-4{u\over \ell}}\left(1+{\cal O}\left(e^{-2{u\over \ell}}\right)\right)\right]+
\label{C47}\ee
$$+
C_+\left[~e^{-\Delta_{+}{u\over \ell}}\left(1+{\cal O}\left(e^{-2{u\over \ell}}\right)\right)+{\Gamma^2\left({\Delta_{-}\over 2}\right)\over
\Gamma^2\left({\Delta_{+}\over 2}\right)}{}{\Gamma\left(\Delta_{+}-2\right)\over \Gamma\left(\Delta_{-}-2\right) }~e^{-4{u\over \ell}}\left(1+{\cal O}\left(e^{-2{u\over \ell}}\right)\right)\right]
$$
We can contrast this solution with the one in Poincar\'e coordinates that reads
\be
\phi_{\pm}=e^{-\Delta_{\pm}{u\over \ell}}
\label{C48}\ee

From the solution (\ref{C47}) we can reconstruct the leading part of the superpotential $W$ by matching to the first order equation (\ref{i9}).
We obtain a superpotential that has the structure indicated in  (\ref{C26}) with $\Delta_-$ appearing in the quadratic term  and all other $\Delta$-dependent exponents.
If instead we consider the solution with the source $\phi_-=0$, then the superpotential we obtain has $\Delta_+$ in the quadratic term and all $\Delta$-dependent exponents as well as $\hat C=0$.
This structure matches what is known in flat-sliced AdS, \cite{exotic}, with the only difference being the terms that reflect the presence  of the curvature of the $S^3$ proportional to $C$ in  (\ref{C26}).

Near the center, $u\to -\infty$ and we obtain instead a regular power series expansion in $e^{2{u\over \ell}}$,
\be
\phi_{\pm}=1+{\Delta_{\pm}(\Delta_{\pm}-8)\over 16}e^{2u\over \ell}+{\cal O}\left(e^{4u\over \ell}\right)
\label{C49}\ee
From this we can make a solution that asymptotes to the minimum ($\phi\to0$)as
\be
\phi_{IR}=C(\phi_+-\phi_-)\simeq {\cal O}\left({e^{2u\over \ell}}\right)
\label{C50}\ee

By contrast, the related solution in Poincar\'e coordinates is still (\ref{C47}) which is radically  different, as expected.
In this case, going again through the first order equations, we can reconstruct the superpotential and we obtain a regular power series expansion in $\phi$. The constants $C=\hat C=0$ in  (\ref{C26}) for IR regularity.

\section{Comparing AdS RG Flows and dS cosmologies\label{comparison2}}

In this appendix we collect the most important formulae and properties of near critical AdS and dS solutions for purposes of comparison.

\subsection{AdS Maxima of the potential}

\be
	V=-{d(d-1)\over \ell^2}+{m^2\over 2}\phi^2 + \cO(\phi^3), \qquad m^2 <0 \label{c1}
\ee

\be
	W_{\pm}(\phi)
			={1\over \ell}
			\le[
				{2(d-1)} + {\Delta_{\pm}\over2}\phi^2+\cO(\phi^3)
			\ri], 			\label{c2}
\ee
	\be
	\Delta_\pm =\ha \le( d \pm \sqrt{d^2 +4 m^2\ell^2}  \ri) \quad \text{with}\quad -{d^2\over 4\ell^2}<m^2<0,\sp \Delta_{\pm}\geq 0
	\label{c3}
\ee
The relation above can be written as
\be
\Delta_{+}\Delta_{-}=-m^2\ell^2=d(d-1){V''(0)\over V(0)}
\label{d1}\ee
and the BF bound
\be
{V''(0)\over V(0)}\leq {d\over 4(d-1)}
\label{BFads}\ee

\begin{subequations}\label{c4}
	\begin{align}\label{c5}
		&
		\phi_+(u)=\phi_+ e^{\Delta_+ u/\ell}+\cdots\\
		&
		\phi_-(u)=\phi_- e^{\Delta_- u/\ell}
			+{
				d~C\ell
					\over
				\Delta_-(2\Delta_+-d)}~\phi_-^{\Delta_+/\Delta_-}~
			e^{\Delta_+u \ell}+\cdots\label{c6}
	\end{align}
\end{subequations}
where $\phi_+$ and $\phi_-$ are integration constants.

The scale factors are found by integrating equation (\ref{i8}):
\begin{subequations}
	\begin{align}&
		A_+(u)=-{u-u_*\over \ell}+{\phi^2_+ \over 8(d-1)} e^{2\Delta_+ u/\ell}
		+\cO\le( e^{3\Delta_+u/\ell}\ri)
		\label{a8}\\
		&
		A_-(u)=-{u-u_*\over \ell}+ {\phi^2_- \over 8(d-1)} e^{2\Delta_- u/\ell}
		+\cO\le(C e^{ud/\ell}\ri)
		\label{a9}
	\end{align}
\end{subequations}
where $u_*$ is an integration constant.

The $+$ solution corresponds to an AdS Boundary and is tuned (no free parameters).
The $-$ solution corresponds also to an AdS boundary but it is not tuned (there is a free parameter $C$).
{\it In all cases $W>0$, and $\dot A<0$.}

\subsection{AdS minima of the potential}
\begin{subequations}\label{W_min}
\begin{align}
	W_\pm(\phi)
			=&{1\over \ell}
			\le[
				{2(d-1)} + {\Delta_\pm\over2}\phi^2+\cO(\phi^3)
			\ri],
			\label{c10}\\
\Delta_\pm =&\ha \le( d \pm \sqrt{d^2 +4 m^2\ell^2} \ri)\quad \text{with}\quad m^2>0. \label{c11}
\end{align}
\end{subequations}
Note that now we have necessarily $\Delta_-<0$  and $\Delta_+>0$.

The $W_+$ solution has a local minimum at $\phi=0$ while  $W_-$ has a local maximum. This implies a different geometrical and holographic interpretation.
\begin{subequations} \label{c12}
\begin{align}
		&
		\phi_+(u)=\phi_+ e^{\Delta_+ u/\ell}+... \label{c13}\\
		&
		\phi_-(u)=\phi_- e^{\Delta_- u/\ell}
			+...  \label{c14}\\
	&A_\pm(u)=-{u-u_*\over \ell}
				- {1\over 8(d-1)}{\phi_\pm^2}e^{2\Delta_\pm u/ \ell}+...
\end{align}
\end{subequations}
where $\phi_\pm$ and $u_*$ are integration constants.

The $+$ solution corresponds to an AdS Boundary (UV) where the scale factor diverges and is tuned (no free parameters).
The $-$ solution corresponds to an AdS interior (IR) where the scale factor vanishes,  and is  also tuned.
Therefore,  the regular solution near AdS minima corresponds to superpotentials with no free parameters.
{\it In all cases $W>0$, and $\dot A<0$.}

We conclude that in the AdS regime, only for the minus solution around maxima, has a free adjustable parameter.

\subsection{dS minima of the potential}
The potential has the following series expansion around a local minimum:
\be
	V={d(d-1)H^2}+{m^2\over 2}\phi^2 + \cO(\phi^3), \qquad 0<m^2<\frac{d^2 H^2}{4}
\label{c15}
\ee
The equation \eqref{b6_1} has solutions:
\begin{subequations}			\label{d18}
\begin{align}
	W_+(\phi)
			=&-H
			\le[
				{2(d-1)} + {\Delta_+\over2}\phi^2+\cO(\phi^3)
			\ri], 			\label{c16}\\
	W_-(\phi)
			=&-H
			\le[
				{2(d-1)} + {\Delta_-\over2}\phi^2+\cO(\phi^3)
			\ri]+C|\phi|^{d/\Delta_-}\le[1 +\cO(\phi)\ri]
			+\cO(C^{2})
			\label{c17},\\
	\Delta_\pm =&\ha \le( d \pm \sqrt{d^2 -4 {m^2\over H^2}}  \ri) \quad \text{with}\quad {d^2 H^2 \over 4}>m^2>0\sp \Delta_{\pm}\geq 0
	\label{c18}
\end{align}
\end{subequations}
\be
\Delta_{+}\Delta_{-}={m^2\over H^2}=d(d-1){V''(0)\over V(0)}
\ee
and the BF bound
\be
{V''(0)\over V(0)}\leq {d\over 4(d-1)}
\label{BFds}\ee

\begin{subequations}\label{field}
	\begin{align}\label{c19}
		&
		\phi_+(t)=\phi_+ e^{-\Delta_+ H t}+\cdots\\
		&
		\phi_-(u)=\phi_- e^{-\Delta_- H t}
			+{
				d~C
					\over
				H\Delta_-(2\Delta_+-d)}~\phi_-^{\Delta_+/\Delta_-}~
			e^{-\Delta_+ H t}+\cdots\label{c20}
	\end{align}
\end{subequations}
where $\phi_+$ and $\phi_-$ are integration constants.

The scale factors are found by integrating equation (\ref{b5}):
\begin{subequations}\label{c21}
	\begin{align}
		&
		A_+(t)=H(t-t_*)+{\phi^2_+ \over 8(d-1)} e^{-2\Delta_+ H t}
		+\cO\le( e^{-3\Delta_+ H t}\ri)
		\label{c22}\\
		&
		A_-(t)=H(t-t_*)+ {\phi^2_- \over 8(d-1)} e^{-2\Delta_- H t}
		+\cO\le(C e^{-t d H}\ri)
		\label{c23}
	\end{align}
\end{subequations}
where $t_*$ is an integration constant and the sub-leading terms come from the $\cO(\phi^2)$ terms in $W_\pm$. We also have $\dot A_{\pm}>0$.

The expressions above are valid for small $\phi$, so we must take $t \to +\infty$.

The $+$ solution corresponds to the ${\cal I}^+$ boundary of deSitter ($t\to +\infty$, $e^A\to \infty$) and is tuned (no free parameters).

The $-$ solution corresponds to the ${\cal I}^+$ boundary of deSitter ($t\to +\infty$, $e^A\to \infty$) and is not tuned ( there is a single free parameter $C$).
{\it In all cases $W<0$, and $\dot A>0$.}

\subsection{dS  maxima of the potential}

We now assume the potential has the form \eqref{d18},  but  with $m^2<0$. The two solutions of the superpotential are:
\begin{subequations}\label{d22}
\begin{align}
	W_\pm(\phi)
			=&-H
			\le[
				{2(d-1)} + {\Delta_\pm\over2}\phi^2+\cO(\phi^3)
			\ri],
			\label{d22a}\\
\Delta_\pm =&\ha \le( d \pm \sqrt{d^2 -4 {m^2\over H^2}} \ri)\quad \text{with}\quad m^2<0. \label{d22b}
\end{align}
\end{subequations}
Now, we have $\Delta_-<0$  and $\Delta_+>0$.

The $W_+$ solution has a local maximum at $\phi=0$ while  $W_-$ has a local minimum. Solving for the field and scale factor in each case, we have:
\begin{subequations} \label{d23}
\begin{align}
		&
		\phi_+(t)=\phi_+ e^{-\Delta_+ H t}+... \label{d23a}\\
		&
		\phi_-(t)=\phi_- e^{-\Delta_- H t}
			+...  \label{d23b}\\
	&A_\pm(t)=H(t-t_*)
				- {1\over 8(d-1)}{\phi_\pm^2}e^{-2\Delta_\pm H t}+...\sp \dot A_{\pm}>0
\end{align}
\end{subequations}
where $\phi_\pm$ and $t_*$ are integration constants.

The $+$ solution corresponds to the ${\cal I}^+$ boundary of deSitter ($t\to +\infty$, $e^A\to \infty$) and is tuned.

$-$ solution corresponds to the ${\cal I}^-$ boundary of deSitter ($t\to -\infty$, $e^A\to 0$) and is also tuned.
{\it In all cases $W<0$, and $\dot A>0$.}

We conclude that in the dS regime, only for the minus solution around minima, has a free adjustable parameter.

{
\section{The local structure of solutions} \label{structure}

In this appendix we  study the local behaviour of the equations (\ref{e28}), (\ref{e30}),  for the superpotential $W(\phi)$ and $f(\phi)$ for  a flat slicing respectively, about a generic point.

\subsection{The flat sliced  case\label{flats}}

Instead of solving (\ref{e28}), (\ref{e30}) we may solve the equivalent third order equation (\ref{e33}) and then determine $f$ from (\ref{e37}).

Around a generic point (that we shift so that it is at $\phi=0$ we may expand the potential and $W$ into a regular power series as follows
\be
V(\phi)=\sum_{n=0}^{\infty}{V_n\over n!}\phi^n\sp
W(\phi)=\sum_{n=0}^{\infty}{W_n\over n!}\phi^n
\label{C1}\ee
The function $f$ has also a similar expansion.

As equation  (\ref{e33}) is of third order, $W_0,W_1,W_2$ remain undetermined (they are in one-to-one correspondence with the constants of integration, while the rest of the coefficients are determined by the differential equation  in  (\ref{e33}). We give the first nontrivial one

\be
W_3={d W_1^2 - d W_0 W_2 + 2 (d-1) W_2^2\over (d-1) ( 2 V_0 W_1-V_1 W_0)}
V_0- {W_1^2 - W_0 W_2\over 2 V_0 W_1-V_1 W_0}V_2+
\label{C2}\ee
$$
+{
 d W_0(W_0W_2-W_1^2)+2(d-1)W_2(W_0W_2-2W_1^2)\over
 2 (d-1) W_1 ( 2 V_0 W_1-V_1 W_0)}V_1+
 $$
We may also compute
\be
f_0={2 V_0 W_1-V_1 W_0\over W_1(W_1^2 - W_0W_2)}
\label{C3}\ee
and
\be
f_1=-{d V_1 W_0^2 - 2 d V_0 W_0 W_1 - 2 (d-1) V_1 W_1^2 +
  4 (d-1) V_0 W_1 W_2\over 2(d-1)W^2_1(W_1^2 - W_0W_2)}
\label{C4}\ee

From this expansion we can conclude that the singular points of the equation are of two types:
\begin{enumerate}

\item Points where $W'\equiv W_1$ vanishes. These are the singular points of the equations in the case that $f$ is constant, studied in detail in \cite{exotic}.

    They are of two kinds: the first corresponds to $W_1$ vanishing when the potential is extremal $V_1=0$. Such points have been studied in detail in section \ref{local}.
What was found there is that near a dS minimum or an AdS maximum such solutions are regular.

Near a dS maximum (plus) solution or an AdS minimum (plus solution) the solutions are again regular if we tune one of the constants of integration.
However, near a dS maximum (minus) solution or an AdS minimum (minus solution) such solutions are singular if and only if the blackness function is non-constant.

The second possibility concerns points where $V_1\not= 0$ but $W_1=0$.
In the constant $f$ case such points were called bounces and were studied in detail in \cite{exotic}.  In the case with a non-trivial $f$ function in the AdS regime they have been studied in \cite{gur}.

In this case, the indicial equation near a point $\phi=0$ with $W_1=0$ but $V_1\not=0$ gives the following expansion\footnote{Depending on $\phi>0$ or $\phi<0$ the expansion can also be written in terms of $(-\phi)^{n\over 2}$, see \cite{exotic}.}
\be
W=W_0+\sum_{n=3}^{\infty} ~W_{n/2}~\phi^{n\over 2}\sp f=\sum_{n=0}^{\infty}f_{n/2}~\phi^{n\over 2}
\label{C5}\ee
Solving the equation (\ref{e33}) we find that $W_0,W_{3\over 2}$ are arbitrary, while  we obtain recursively
\be
W_{2}=-{3\over 2}{W_{3/2}^2\over W_0}{V_0\over V_1}-{d\over 6(d-1)}W_0
\label{C6}\ee
\be
W_{5/2}= {39\over 20}{W_{3/2}^3\over W_0^2}{V_0^2\over V_1^2}+{13\over 30}{d\over d-1}{W_{3/2}}{V_0\over V_1}+{3\over 20}W_{3/2}{V_2\over V_1}+{d^2\over 135(d-1)^2}{W_0^2\over W_{3/2}}
 \label{C7}     \ee

      and
      \be
      f_0={8\over 9}{V_1\over W_{3/2}^2}\sp f_{1/2}={8\over 9}{V_0\over W_{3/2}W_0}+{16d\over 27(d-1)}{V_1W_0\over W_{3/2}^3}
      \label{C8}\ee
      These solutions are the generalization of the bounces of \cite{exotic} to our ansatz.
      As shown in \cite{exotic,gur} and as explained in section \ref{bounce}, curvature invariants that are contractions of the Riemann tensor are polynomials of $fW^2$ and $fW'^2$ that are finite at a bounce.
Covariant derivatives of the Riemann tensor can have additional, $ \pa_{u}$ derivatives that are translated to $\pa_{u}=W'(\phi)\pa_{\phi}$.
Although $\pa_{\phi}$ derivatives of the superpotential starting with $W''$ are divergent at a bounce, they always come multiplied with $W'$ that is vanishing at the bounce.

Therefore,  both $\pa_{u}^n(fW^2)$ and $\pa_{u}^n(fW'^2)$ are finite at the bounce. This is equivalent to the fact (shown in \cite{exotic}) that $A(u)$ and $\phi(u)$ have absolutely regular expansions in $u$ around the bounce.

\item The other class of singular points appear when  $2 V_0 W_1-V_1 W_0=0$. At such points $f_0=0$ and they correspond to horizons.
    These points and the structure of the solutions are analysed in subsection \ref{fho}. As is well known, they signal a reduction of independent constants but the solutions and the curvature invariants are regular there.

\end{enumerate}

Therefore,  putting together the analysis of the previous works and the one here we can ascertain that the only singularities that appear in the class of flows described by the flat-sliced ansatz can happen only when the solutions end up at $\phi\to \pm\infty$ or at dS maxima and AdS minima with minus solutions and non-constant $f$.

\subsection{Solutions near a flat horizon\label{fho}}

As we have seen, one of the points where solutions can be singular, if not tuned, is a horizon, $f=0$.
We therefore parametrize a horizon at $u=0$ as
\be
f=f_0 u^{s}\left[1+f_1 u+{\cal O}(u^2)\right]
\label{n3}\ee
We shall also first assume that $s\not=1$.

From (\ref{e2e})  we obtain
\be
A=-{1\over d}\left[(s-1)\log u+{\cal O}(u^0)\right]\sp \dot A=\left[{1-s\over d u}-{f_1(s+1)\over d s}+{\cal O}(u)\right]
\label{n4}\ee
We also obtain from (\ref{e2a})
\be
\phi=\phi_0\pm \sqrt{2(1-s)(d-1)\over d}\log u+\cdots
\label{n5}\ee
which implies that $0<s< 1$.
We also obtain that
\be
u\sim e^{\pm{\phi\over \sqrt{2(1-s)(d-1)\over d}}}~~~{\rm with}~~~\phi\to\mp\infty
\label{n6}\ee
Therefore, for $s\not= 1$ a horizon can appear only at the boundaries of the scalar space $\phi\to \pm\infty$.

From (\ref{e2c}) we obtain
\be
(d-1)\left(\dot f\dot A+f\left[d\dot A^2+\ddot A\right]\right)=-2{( d-1)\over d s}f_0f_1~u^{s-1}+\cdots =-V
\label{n7}\ee
If we assume that as $\phi\to \mp\infty$, and parametrize the asymptotic form of the potential as
\be
V\sim e^{\mp a\phi}\;,
\label{n1}\ee
 then
 \be
 1-s={2(d-1)\over d}a^2
\label{n2} \ee
 If $a$ varies up to the Gubser bound,
 \be
 0<a^2<{d\over d-1}
 \label{n8}\ee
 we obtain
 \be
 0<1-s<2~~~\to~~~-1\leq s< 1
\label{n9} \ee
We conclude that horizons with $s\not=1$ may appear only at $\phi\to \mp \infty$ if the asymptotics of the potential are as in (\ref{n1}) with $a$ given in (\ref{n2}).

{In order to check regularity, we need only check the asymptotic behaviour of $f \dot{A}^2$ and $f \dot{\phi}^2$, as the curvature invariants are polynomials of these two quantities. We have

\be
f \dot{A}^2=u^{s-2} \left[ \frac{f_0 (s-1)^2}{d^2} + \mathcal{O}(u) \right],
\label{n10}\ee

\be
f \dot{\phi}^2=u^{s-2} \left[ -\frac{2f_0 (s-1)(d-1)}{d} + \mathcal{O}\left(u\right) \right].
\label{n11}\ee

Since $s<1$, we see that as $u \rightarrow 0$ the curvature invariants diverge and the solutions are irregular at the horizon. Yet, if $s=1$ this is no longer the case, as the $\mathcal{O}(1)$ terms in equations (\ref{n10}) and (\ref{n11}) will vanish, and the invariants become regular.

}

We shall now study the local behaviour of solutions of (\ref{e2}) near a horizon with $s=1$ in (\ref{n3}), assuming regularity. Consider a horizon at $u=u_h$ and write the expansion of relevant functions,

\be
f(u)=f_1{(u-u_h)\over \ell}+{f_2\over 2}{(u-u_h)^2\over \ell^2}+{f_3\over 3!}{(u-u_h)^3\over \ell^3}+{\cal O}\left((u-u_h)^4\right)
\label{e9}\ee
\be
A(u)=A_h+A_1{(u-u_h)\over \ell}+{A_2\over 2}{(u-u_h)^2\over \ell^2}+{A_3\over 3!}{(u-u_h)^3\over \ell^3}+{\cal O}\left((u-u_h)^4\right)
\label{e10}\ee
\be
\phi(u)=\phi_h+\phi_1{(u-u_h)\over \ell}+{\phi_2\over 2}{(u-u_h)^2\over \ell^2}+{\phi_3\over 3!}{(u-u_h)^3\over \ell^3}+{\cal O}\left((u-u_h)^4\right)
\label{e11}\ee

Note that all coefficients $f_i,A_i,\phi_i$ are defined, using the length scale $\ell$ so that they are dimensionless.

Equations (\ref{e2a})-(\ref{e2d}) imply
\be
(d-1)A_1f_1+\ell^2~V_h=0\sp f_1\phi_1-\ell^2~V'_h=0
\label{e12}\ee
together with
\be
2f_1\phi_2=\phi_1~\ell^2~V''_h\sp A_2=-{\phi_1^2\over 2(d-1)}\sp f_2=-{d}f_1A_1
\label{e13}\ee
where
\be
V_h\equiv V(\phi_h)\sp V'_h\equiv V'(\phi_h)
\label{e14}\ee
and so on.

For a regular horizon, $f_1\not =0$ and therefore
\be
A_1=-{\ell^2~V_h\over (d-1)f_1}\sp \phi_1={\ell^2~V'_h\over f_1}
\label{e15}\ee
\be
\phi_2={\ell^4~V'_hV''_h\over 2f_1^2}\sp A_2=-{\ell^4~V_h'^2\over 2(d-1)f_1^2}\sp f_2={d\over (d-1)}\ell^2~V_h
\label{e16}\ee
\be
\phi_3=\ell^6~{V'_h(d(V'_h)^2-dV_hV''_h+(d-1)(V''_h)^2+2(d-1)V'_hV'''_h)\over 6(d-1)f_1^3}
\label{e17}\ee
\be
A_3=-\ell^6~{(V'_h)^2V''_h\over 2(d-1)f_1^3}\sp f_3=\ell^4~{d(2dV_h^2+(d-1)(V'_h)^2)\over 2(d-1)^2f_1}
\label{e18}\ee
and so on.

This solution is also valid when the horizon is at a point where $V(\phi_h)=0$.

If the horizon is at an extremum of the potential, $V'(\phi_h)=0$, then this can happen only if $\phi=constant$ and $A$ is linear in $u$, therefore,  the scale factor is that  of $AdS$ and the final solution is the AdS-Schwarzschild black hole (whose horizon can be anywhere)

If we expand the equations around a generic point, then we need as initial conditions 5 parameters that we can take to be the values of $A,f,\phi$ at that point plus two first derivatives (say, $\dot f, \dot\phi$)

We have seen above, that near a horizon $A_h,f_h=0,\phi_h$ are free parameters, but $f_1,\phi_1$ are correlated by equation (\ref{e12}). {Therefore,  a horizon enforces a codimension-one tuning of the initial conditions.}
An immediate corollary is that if at the horizon $V_h=0$, then (\ref{e15}) implies that $A_1\sim \dot A=0$ there.

Consider now the case where the horizon is at an extremum of the potential: $V'(\phi_h)=0$.
Then the equations imply
\be
A_1=-{V(\phi_h)\over (d-1)f_1}\sp A_2=A_3=\cdots=0\sp \phi_1=\phi_2=\cdots=0
\label{w25}\ee
\be
f_2={d\over d-1}V(\phi_h)\sp f_3={d^2\over (d-1)^2}{V^2(\phi_h)\over f_1}
\label{w26}\ee
which resums to
\be
f(u)={f_1\over dA_1}\left[1-e^{-dA_1 u}\right]\;.
\label{w27}\ee
In this case , $\phi$ is constant and the horizon can be anywhere.

{We now have deduced the local behaviour of solutions close to a horizon, which will,  in due time,  allow us to impose initial conditions in solving the Einstein equations numerically, as well as asserting some global properties.

\section{A tuned potential example} \label{exam}

In this appendix we  illustrate how specially tuned potentials may allow nontrivial flows that start and end on minima of the potential, and therefore,  do not have arbitrary integration constants but are regular. This will provide an example of the behavior indicated in figure \ref{fig:plot1_b}.
It will be achieved by the implementation of the inverse method explained in appendix F of \cite{exotic}.

We shall use our flat-sliced ansatz (\ref{e1}) with the blackness function $f=1$.
The equations to be solved become
\be
 \ddot{\f}-d\dot{A}\dot{\f}=V'
\label{bb1}\ee
\be
2(d-1) \ddot{A} + \dot{\f}^2 =0 \, ,
\label{bb2}
\ee
\be
\label{bb3} d(d-1) \dot{A}^2 - \frac{1}{2} \dot{\f}^2 + V=0 \, ,
\ee
We would like to consider a flow between an extremum of the potential at $\f=\f_i$ and another extremum  of the potential at $\f_f$ with
\be
V(\f_i)=-{d(d-1)\over \ell_i^2}\sp V(\f_f)=-{d(d-1)\over \ell_f^2}
\label{bb4}\ee

We start from the function
\be
\phi(u)=\bar \f+{\Delta\f\over 2}\tanh{u-u_0\over L}\sp \Delta\f=\f_f-\f_i\sp \bar \f\equiv {\f_i+\f_f\over 2}
\label{bb9}\ee
where $L$ is a fiducial length scale that we shall adjust later on.
This function interpolates monotonically between $\f_i$ and $\f_f$ as advertised.

We must now use (\ref{bb1})-(\ref{bb3}) to find the associated $A$ and $V$ that is compatible with (\ref{bb9}).
From (\ref{bb9}) we obtain
\be
\dot \f={\Delta\f\over 2L}~{1\over \cosh^2{u-u_0\over L}}
\label{bb10}\ee
Next we  integrate once (\ref{b2}) to obtain
\be
\dot A={a\over L}-{\Delta\f^2\over 24(d-1)L}\tanh{u-u_0\over L}\left(2+{1\over \cosh^2{u-u_0\over L}}\right)
\label{bb12}\ee
and once more
\be
A(u)=A_0+a{u\over L}-{\Delta\f^2\over 24(d-1)}\left[2\log\left(\cosh{u-u_0\over L}\right)-{1\over 2}{1\over \cosh^2{u-u_0\over L}}\right]
\label{bb13}\ee

We can evaluate the Kretchmann invariant as
\be
R_{\m\n;\r\s}R^{\m\n;\r\s}=4d(\ddot A+\dot A^2)^2+2d(d-1)\dot A^4={d\over (d-1)^2}\left(2(d-1)\dot A^2-\dot\f^2\right)^2+2d(d-1)\dot A^4
\label{b36}\ee
and it is clear that it is regular for the flow discussed above.

Now we use (\ref{bb3}) to calculate the potential
\be
V={\dot \f^2\over 2}-d(d-1)\dot A^2=
\label{bb15}\ee
\be
={1\over 2}\left({\Delta\f\over 2L}~{1\over \cosh^2{u-u_0\over L}}\right)^2-d(d-1)\left[{a\over L}-{\Delta\f^2\over 24(d-1)L}\tanh{u-u_0\over L}\left(2+{1\over \cosh^2{u-u_0\over L}}\right)\right]^2
\label{bb16}\ee
$$
=-d(d-1){a^2\over L^2}+{d\Delta\f^2\over 12L}{a\over L}\tanh{u-u_0\over L}\left(2+{1\over \cosh^2{u-u_0\over L}}\right)-
$$
$$
-{d\Delta\f^4\over 24^2(d-1)L^2}\tanh^2{u-u_0\over L}\left(2+{1\over \cosh^2{u-u_0\over L}}\right)^2
+{\Delta\f^2\over 8L^2}~{1\over \cosh^4{u-u_0\over L}}
$$

In order to write it in terms of $\f$ we need to invert (\ref{bb9}) as
\be
\tanh{u-u_0\over L}={2\over \Delta\f}(\f-\bar\f)
\label{bb17}\ee
and rewrite
\be
{1\over \cosh^2{u-u_0\over L}}=1-\tanh^2{u-u_0\over L}=1-{4(\f-\bar\f)^2\over \Delta\f^2}
\label{bb18}\ee
Therefore,
\be
V=-d(d-1){a^2\over L^2}+{d\Delta\f a\over 6L^2}(\f-\bar\f)\left(3-{4(\f-\bar\f)^2\over \Delta\f^2}\right)-
\label{bb19}\ee
$$
-{d\Delta\f^2\over 12^2(d-1)L^2}(\f-\bar\f)^2\left(3-{4(\f-\bar\f)^2\over \Delta\f^2}\right)^2
+{\Delta\f^2\over 8L^2}~\left(1-{4(\f-\bar\f)^2\over \Delta\f^2}\right)^2
$$
$$
=-d(d-1){a^2\over L^2}+{\Delta\f^2\over 8L^2}+{d\Delta\f a\over 2L^2}(\f-\bar\f)-\left(1+{d\Delta\f^2\over 16(d-1)}\right){(\f-\bar \f)^2\over L^2}-{2dA_1\over 3L \Delta\f}(\f-\bar \f)^3+
$$
$$
+\left({2\over \Delta\f^2}+{d\over 12(d-1)}\right){(\f-\bar \f)^4\over L^2}-{d\over 9(d-1)}{(\f-\bar \f)^6\over L^2\Delta\f^2}
$$

and finally
\be
L^2~V(\f)
=-d(d-1)a^2+{\Delta\f^2\over 8}+{da\Delta\f\over 2}(\f-\bar\f)-\left(1+{d\Delta\f^2\over 16(d-1)}\right){(\f-\bar \f)^2}-{2d a\over 3\Delta\f}(\f-\bar \f)^3+
\label{bb20}\ee
$$
+\left({2\over \Delta\f^2}+{d\over 12(d-1)}\right){(\f-\bar \f)^4}-{d\over 9(d-1)}{(\f-\bar \f)^6\over \Delta\f^2}
$$

Note that a generic sextic potential depends on 7 real parameters, but the potential above is special as it  depends on 4 arbitrary parameters.

We can compute
\be
L^2~V''(\f)
=-2\left(1+{d\Delta\f^2\over 16(d-1)}\right)-{4d a\over \Delta\f}(\f-\bar \f)+
\label{bb24c}\ee
$$
+\left({24\over \Delta\f^2}+{d\over (d-1)}\right){(\f-\bar \f)^2}-{10d\over 3(d-1)}{(\f-\bar \f)^4\over \Delta\f^2}
$$

We rewrite (\ref{bb10}) as
\be
\dot\f={\Delta \f\over 2L}\left(1-{4(\f-\bar\f)^2\over \Delta\f^2}\right)\equiv W'
\label{bb22}\ee
from which we can compute the superpotential as
\be
W=-2(d-1){a\over L}+{\Delta \f\over 2L}(\f-\bar\f)\left(1-{4(\f-\bar\f)^2\over 3 \Delta\f^2}\right)
\label{bb23}\ee
This is compatible with the other flow equation $W=-2(d-1)\dot A$.
It is also compatible with the superpotential equation
\be
V(\f)={W'^2\over 2}-{d\over 4(d-1)}W^2
\label{bb24}\ee
It describes the flow from $\phi_i$ to $\phi_f$.

The extrema of the potential are given by the solutions of either of the two equations
\be
W'=0\sp W''={d\over 2(d-1)}W
\label{bb25}\ee
From the first one we obtain
\be
\f_{\pm}=\bar\f\pm {\Delta\f\over 2}\sp \f_+=\f_f\sp \f_-=\f_i
\label{bb26}\ee
and from (\ref{bb24c}) we find
\be
L^2V''_{\pm}=4+{d\left(\Delta\f^2\mp 12(d-1)a\right)\over 6(d-1)}\sp L^2V_{\pm}=-{d\over 144(d-1)}\left(\Delta\f^2\mp12(d-1)a\right)^2
\label{bb27}\ee
We note that the extrema are always AdS extrema.
We also observe that whether the extrema correspond to maxima or minima depends on the values of $\Delta\f$ and $a$.
The associated AdS scales are
\be
\ell_{\pm}={12(d-1)L\over |\mp \Delta\f^2+12(d-1)a|}
\label{bb28}\ee

The associated scalar scaling dimensions are given as usual by
\be
\Delta_{\pm}={1\over 2}\left(d\pm\sqrt{d^2-4d(d-1){V''\over V}}\right)
\label{bb29}\ee
where the $\pm$ in the formula above do not refer to the $\pm$ of the two extrema in (\ref{b26}).
The BF bound  is always satisfied as the superpotential is real.

Expanding the superpotential around these two extrema we obtain
\be
LW=-{1\over 6}\left(\mp \Delta\f^2+ 12(d-1)a\right)-(\f-\f_{\pm})^2+{\cal O}((\f-\f_{\pm})^3)
\label{bb31}\ee

From (\ref{bb12}) we obtain
\be
L\dot A=a-{\Delta\f^2\over 24(d-1)}\tanh{u-u_0\over L}\left(2+{1\over \cosh^2{u-u_0\over L}}\right)
\label{bb32}\ee
which indicates that
\be
L\dot A=\left\{ \begin{array}{lll}
\displaystyle {12(d-1)a-\Delta\f^2\over 12(d-1)},&\phantom{aa} & u\to+\infty     ,\\ \\
\displaystyle {12(d-1)a+\Delta\f^2\over 12(d-1)},&\phantom{aa}&  u\to -\infty       .
\end{array}\right.\sp
\f=\left\{ \begin{array}{lll}
\displaystyle \f_f,&\phantom{aa} & u\to +\infty     ,\\ \\
\displaystyle \f_i,&\phantom{aa}&  u\to -\infty       .
\end{array}\right.
\label{bb33}\ee

 \begin{figure}[t]
\centering
\includegraphics[width=0.45\textwidth]{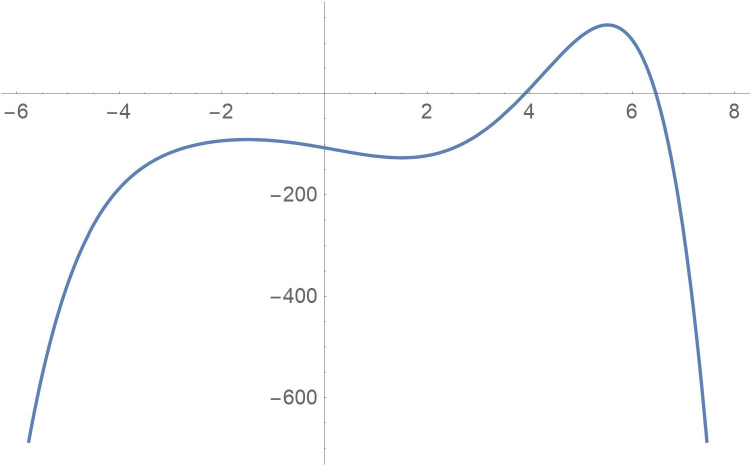} \includegraphics[width=0.45\textwidth]{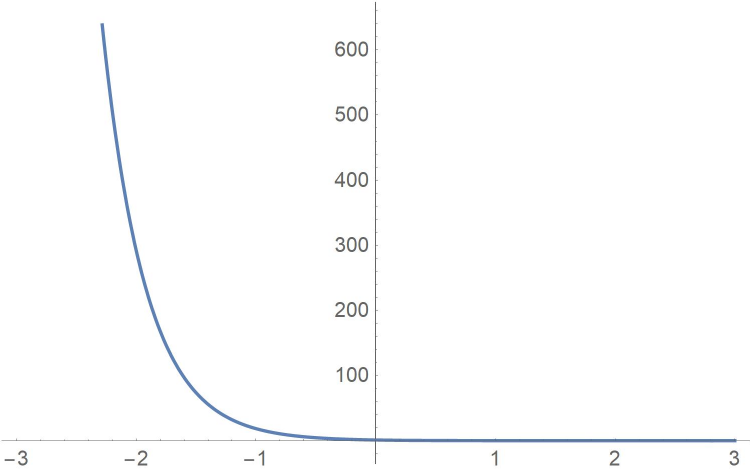}
\caption{Left figure: The potential $\e^2V(\f)$ versus $\f$, for case 1, with $\bar\f=0$, $d=4$, $a=-3$ and $\Delta\f=3$. The flow runs between $\phi_-=-3/2$ and $\f_+=3/2$. This is the subcase where the flow is from a maximum to a minimum. We have $\ell_{+}\simeq 0.31$ and $\ell_{-}\simeq 0.36$.Right figure: the scale factor $e^A$ as a function of $u$.}
\label{fig:1}
\end{figure}

 \begin{figure}[b]
\centering
\includegraphics[width=0.45\textwidth]{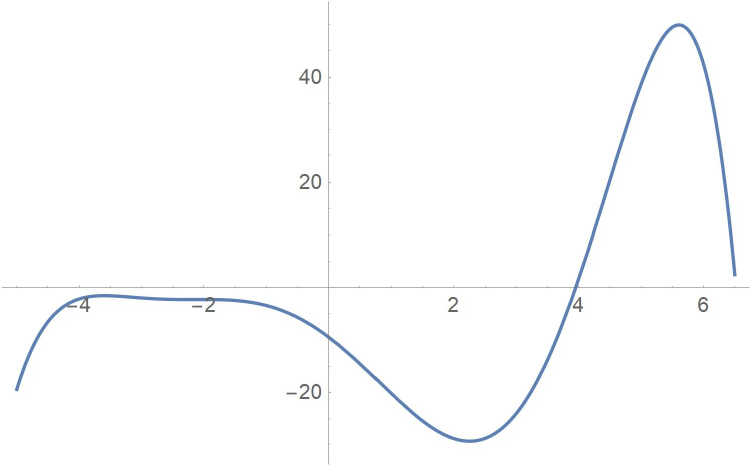} \includegraphics[width=0.45\textwidth]{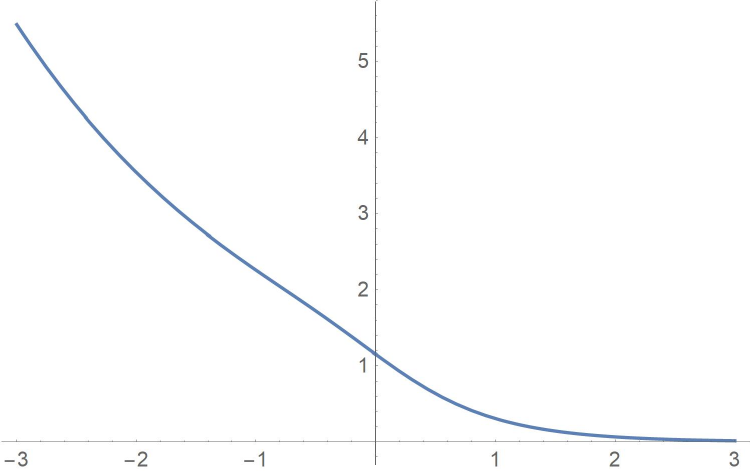}
\caption{Left figure: The potential $\e^2V(\f)$ versus $\f$, for case 1, with $\bar\f=0$, $d=4$, $a=-1$ and $\Delta\f=4.5$. The flow runs between $\phi_-=-2.25$ and $\f_+=2.25$. This is the subcase where the flow is from a minimum to a minimum. We have $\ell_{+}\simeq 0.64$ and $\ell_{-}\simeq 2.28$.Right figure: the scale factor $e^A$ as a function of $u$.}
\label{fig:2}
\end{figure}

We can distinguish three cases:
\begin{enumerate}

\item $12(d-1)a+\Delta\f^2<0$ which implies that $12(d-1)a<-\Delta\f^2$.
Note that here we must have $a<0$ necessarily.

In this case from (\ref{bb33}) we observe that at $u\to -\infty$ we have an AdS boundary, and at $u\to\infty$ we have a Poincar\'e horizon.
The flow is from $\f_i\to \f_f$ as $u$ increases from $-\infty$ to $+\infty$. We also have from (\ref{bb28}) that
\be
\ell_{IR}\equiv \ell_+=\ell_f~~~<~~~\ell_{UV}\equiv \ell_-=\ell_i
 \label{bb28b}\ee
 in accordance with the holographic C-theorem which states that $V(\f_{f})<V(\f_i)$ .
Note also that from (\ref{bb32}) $W$ remains positive for the whole flow.

The + extremum at $\f=\f_f$ is always a minimum. The other ones depends on parameters:

$\bullet$ If $12(d-1)a+\Delta\f^2<-24{d-1\over d}$ then the - extremum at $\f=\f_i$ is a maximum, and the flow is a conventional AdS to AdS flow.

$\bullet$ If $12(d-1)a+\Delta\f^2>-24{d-1\over d}$ then the - extremum at $\f=\f_i$ is a minimum, and the flow is an exotic minimum to minimum AdS flow, \cite{exotic}.

\item $12(d-1)a+\Delta\f^2>0$ and $12(d-1)a-\Delta\f^2<0$. In this case the scale factor vanishes at both end-points of the flow, at $\pm\infty$.
    It reaches a maximum in between. In particular, there is an intermediate point where $\dot A=0$.

In this case $W$ starts negative at $\f_i$ and ends positive at $\f_f$.
The change of sign follows the vanishing of $\dot A$ along the flow and is in agreement with the C-theorem, \cite{exotic}, ${dW\over du}\geq 0$.
Also since $W$ is bounded away from zero in the purely AdS case, it means that in this case the potential is positive somewhere in between $\f_i$ and $\f_f$.

In this case both $\f=\f_f$ and $\f=\f_i$ are always minima. Also, there is no unique ordering between $\ell_+$ and $\ell_-$ despite the relation ${dW\over du}\geq 0$ because $W$ changes sign and this inequality is trivially satisfied. Therefore,  the C-theorem fails and this is also reflected in the fact that the scale factor is not monotonic along the flow.

 \begin{figure}[t]
\centering
\includegraphics[width=0.45\textwidth]{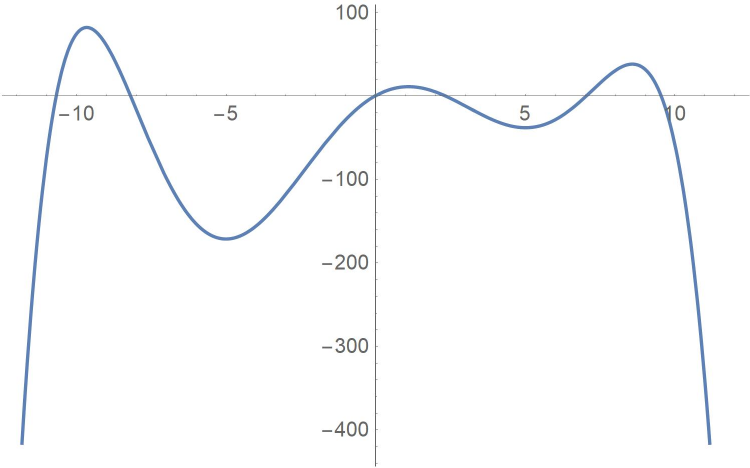} \includegraphics[width=0.45\textwidth]{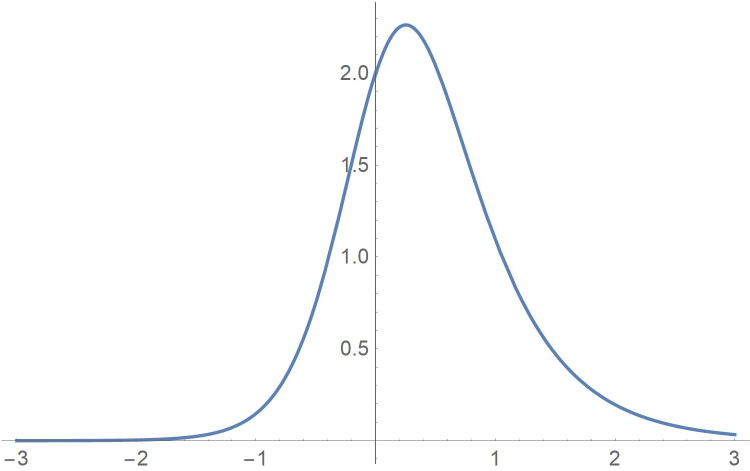}
\caption{Left figure: The potential $\e^2V(\f)$ versus $\f$, for case 2, with $\bar\f=0$, $d=4$, $a=1$ and $\Delta\f=10$. The flow runs between $\phi_-=-5$ and $\f_+=5$. We have $\ell_{+}\simeq 0.56$ and $\ell_{-}\simeq 0.24$. Right figure: the scale factor $e^A$ as a function of $u$.}
\label{fig:3}
\end{figure}

\item $12(d-1)a+\Delta\f^2>0$ and $12(d-1)a-\Delta\f^2>0$. In this case, at $\f_f$ we have an AdS boundary and at $\f_i$ an AdS Poincar\'e horizon. At $\f_i$ we always have a minimum while at $\f_f$ we may have a maximum or a minimum.
    Upon changing the direction of the flow by $u\to- u$, this case maps to case 1. The superpotential changes sign in this case.

\end{enumerate}

\end{document}